\documentclass{aastex63}

\usepackage{graphicx}
\usepackage{amsmath}

\usepackage{epsfig}

\usepackage{subfigure}
\usepackage{longtable}
\usepackage{times}

\shorttitle{Disk perturbation in vicinity of Kerr and EGB black holes}
\shortauthors{O.Donmez}

\begin{document}


\title{Perturbing the Stable Accretion Disk in Kerr and 4-D Einstein-Gauss-Bonnet Gravities:
 Comprehensive Analysis of Instabilities and Dynamics} 

\author{Orhan Donmez}
\altaffiliation{College of Engineering and Technology, American
  University of the Middle East, Egaila 54200, Kuwait}

\begin{abstract}   
  The study of a disturbed accretion disk holds great significance in the realm of astrophysics, as such events play a crucial role in revealing the nature of disk structure, the release of energy, and the generation of shock waves. Consequently, they can help explain the causes of X-ray emissions observed in black hole accretion disk systems. In this paper, we perturb the stable disk formed by spherical accretion around Kerr and EGB black holes. This perturbation reveals one- and two-armed spiral shock waves around the black hole. We find a strong connection between these waves and the black hole spin parameter ($a/M$) and the EGB coupling constant ($\alpha$). Specifically, we found that as $\alpha$ increases in the negative direction, the dynamics of the disk and the waves become more chaotic. Additionally, we observe that the angular momentum of the perturbing matter significantly affects mass accretion and the oscillation of the arising shock waves. This allows us to observe changes in $QPO$ frequencies. Particularly, perturbations with angular momentum matching the observed $C-$  type low-frequency $QPOs$ of the  $GRS 1915 + 105$  source. Thus, we conclude that the possibility of the occurrence of shock waves within the vicinity of $GRS 1915 + 105$ is substantial.
\end{abstract}

  \keywords{numerical relativity --- Kerr black hole --- EGB gravity --- perturbed accretion disk --- QPOs
}


\section{Introduction}
\label{Introduction}

The perturbations of accretion disks around black holes have been conducted on vastly different scales, ranging from stellar-mass black holes in $X$-ray binaries to supermassive black holes in Active Galactic Nuclei (AGN). These investigations aim to understand the effects of various physical processes and interactions that may lead to perturbations in the accretion disks.

In $X$-ray binaries, where a stellar-mass black hole accretes matter from a companion star, observations have revealed fluctuations in the emitted radiation, broad emission lines, and Quasi-Periodic Oscillations ($QPOs$). These phenomena suggest the presence of perturbed accretion disks, which may be influenced by interactions with the companion star or other external sources. Several observed $X$-ray binaries have been identified as potential candidates with perturbed accretion disks. $GRS$ $1915+105$ was examined through observations of high-frequency $QPOs$ and rapid flux variations \citep{Sreehari2020MNRAS, Majumder2022MNRAS, Feng2022, Ricketts2023}. At the same time, low-frequency $QPOs$ have been observed from the source $GRS$ $1915+105$, with frequencies typically ranging from a few to tens of Hertz. The source of low-frequency $QPOs$ is not fully understood. However, there are suggested models. These include the instability of the disk around the black hole due to the strong gravitational field from the black hole-disk interaction, or the resonance condition reached as matter spirals into the black hole \citep{Tomsick2000HEAD, Pawar2015MNRAS, Ingram2015MNRAS, Ingram2019}. The intense radiation from $MAXI$ $J1820+070$ provided evidence of a perturbed accretion disk, possibly caused by instabilities or interactions with surrounding matter \citep{Zhang2023}. The findings from $GX$ $339-4$ were in line with the presence of a perturbed accretion disk \citep{Jin2023}, which showed variations in $X$-ray emission, spectral changes, and $QPOs$. The perturbation likely arises from interactions with the companion star or other sources of external perturbations. Cygnus $X-1$ has exhibited spectral variability and $QPOs$ due to the perturbation of the accretion disk, which could be influenced by interactions with a companion star or clumps of material in the disk \citep{Miller2012, Dong2022}. The presence of perturbations in the accretion disk, possibly caused by interactions with the companion star or transient structures in the disk, could cause strong $X$-ray variability and complex flaring behavior.

\citet{Remillard2006} investigated the properties and behavior of 20 X-ray binaries, with 17 of them confirmed as transient systems hosting dynamically active black holes. Over the past decade, these transient sources underwent intensive monitoring during their typical year-long outburst cycles, utilizing the large-area timing detector of the Rossi X-Ray Timing Explorer. The evolution of these sources exhibits complexity, but a comprehensive comparison of six selected systems reveals shared patterns in their behavior. Central to this comparison are three distinct X-ray states of accretion, meticulously examined and quantitatively defined. They explored various phenomena occurring in strong gravitational fields, such as relativistically broadened Fe lines, high-frequency QPOs (ranging from 100 to 450 Hz), and relativistic radio and X-ray jets. These observations shed light on the interactions between black holes and their surrounding environments, providing valuable insights that complement the knowledge gained from gravitational wave detectors.

AGNs consist of supermassive black holes (SMBHs) with masses ranging from $10^6 M_\odot$ to $10^{10} M_\odot$, located at the centers of galaxies. These black holes continuously accrete matter, leading to their highly luminous nature. The presence of such SMBHs in galaxies suggests that supermassive black hole binaries (SMBHBs) should be common in galactic nuclei \citep{Khan2016}. However, detecting these systems in close proximity to the central object is extremely challenging \citep{DOrazio2018}. In the era of multi-messenger astrophysics, understanding the significance of SMBHBs in close proximity to the central object goes beyond evolutionary processes. They are regarded as crucial targets for connecting gravitational wave detections with electromagnetic counterparts \citep{Bowen2019}. This possibility is promising since merging black holes could interact with the surrounding gas, potentially leading to electromagnetic counterparts \citep{Palenzuela2010}.

$J095036+512838$ is classified as an AGN. It was investigated using observations of Ly-alpha and C IV broad lines \citep{Runnoe2021, Mohammed2023}. It belongs to a category of galaxies characterized by an extraordinarily bright center, believed to be fueled by a supermassive black hole that attracts and absorbs surrounding matter. The remarkable luminosity of AGNs is attributed to the intense radiation emanating from the accretion disk encircling the central black hole.

Perturbation is an immensely significant physical phenomenon with the potential to provide valuable insights into the origin of X-ray emissions and QPOs observed in the aforementioned astrophysical systems \citep{Ingram2019, Rana2020}. The perturbation induces disturbances in the accretion disk, breaking its spherical symmetry and generating strong shock waves that may oscillate during the system's evolution, even after reaching a steady state. As a consequence of these disturbances or irregularities, the observed emission lines experience shifts from their expected positions. These shifts offer valuable insights into the properties of the accretion disk, the environment surrounding the supermassive black hole, and the physical processes taking place in its vicinity. Analyzing these shifts aids astrophysicists in understanding the dynamics and characteristics of black hole systems.

A numerical investigation of the accretion disk helps us comprehend its structure, instabilities, and the occurrence of quasi-periodic behavior near the black hole for different gravities. This deeper understanding contributes to unraveling the underlying physical mechanisms responsible for the observed X-ray frequencies and variations \citep{Chakrabarti2002}. \citet{Orhan2006} investigates the dynamic evolution of star-disk interaction in the vicinity of a massive black hole, where physical perturbation plays a dominant role over other processes, and the gravitational region exhibits strong effects. A numerical simulation is employed to model the accretion disk around the black hole when a star gets captured by it. When the accretion disk, whether in a steady state or not, undergoes perturbation due to the presence of the star, the interaction results in the destruction of the disk surrounding the black hole, giving rise to a spiral shock wave and leading to the loss of angular momentum.

\citet{Stone2001} conducted a comprehensive exploration of the vertical shear instability (VSI) in accretion disks using 3D numerical simulations. The VSI is a hydrodynamic instability that arises from vertical velocity shear within the accretion disk. The study delves into the growth and characteristics of the VSI in various regions of the disk, shedding light on its influence on the overall dynamics of the disk. Notably, the simulations show that the VSI can induce substantial turbulence and vertical mixing, potentially impacting the accretion process and angular momentum transport within the disk.

In this study, we investigate the dynamics of perturbed accretion disks using a numerical relativistic hydrodynamics code, which allows us to gain valuable insights into the behavior and characteristics of these disks in the presence of strong gravitational fields, particularly in the vicinity of Kerr and Einstein-Gauss-Bonnet (EGB) black holes. Our numerical simulations focus on modeling the dynamics of accretion disks under perturbations arising from interactions with companion objects. The results from our numerical simulations reveal that perturbations in accretion disks can give rise to the formation of shock waves, oscillations, and other complex phenomena. The generation of shock waves leads to the heating of matter within the accretion disk, resulting in the production of electromagnetic radiation. Consequently, the perturbation induces shock waves, which, in turn, impact various aspects of the system, such as the light curve, radiation spectrum (including QPOs), and outburst durations \citep{Zhilkin2021}.

The physical origins of QPO frequencies are not yet fully understood \citep{Ingram2019}. In this work, we explain some observed QPOs with the shock waves generated by perturbing a stable disk. We also examine the influence of various parameters, including the black hole spin ($a/M$), the strength of the perturbation, and the EGB coupling constant ($\alpha$), on the behavior of the disk. Through these simulations, we aim to shed light on the intricate interplay between the accretion disk and external perturbations, providing a better understanding of the physical processes occurring in these extreme gravitational environments. The findings contribute to our broader comprehension of accretion processes in the vicinity of black holes and their significance in astrophysical observations.

The structure of the rest of the paper is outlined as follows. In Section \ref{GRHE1}, we introduce the Kerr and 4D EGB black hole spacetime metrics, along with lapse functions and shift vectors. The equations for general relativistic hydrodynamics are presented in a conserved form suitable for high-resolution shock-capturing schemes. Section \ref{InitialBC} presents the initial conditions for the setup of the steady-state accretion disk and perturbations. The necessary boundary conditions to yield physical solutions are also defined in this section. To place numerical computations on a stronger scientific foundation, the astrophysical applications of numerical results are important. Section \ref{Astro_Mot} includes X-ray binary systems and AGN sources, which are thought to have perturbed disks observationally. In Section \ref{Results}, we conduct a numerical examination of the initial steady-state accretion disk, the structure and instabilities of the disk around Kerr and EGB black holes, as well as QPO frequencies that can serve as a comparison basis for various sources. The impact of using alternative gravity on the structure and instability of the perturbed accretion disk is discussed. Finally, in Section \ref{Conclusion}, we discuss and summarize the numerical results, and potential future research directions are provided. We use geometrized units with $c=G=1$ throughout the paper to simplify the equations.


\section{Governing Equations}
\label{GRHE1}


\subsection{Metrics for Kerr and Einstein-Gauss-Bonnet gravities}
\label{GRHE2}
Using different gravities, such as those in Kerr and 4-D EGB black holes, allows for modeling the behavior of the accretion disk in the strong gravitational field near black holes. This approach provides a more comprehensive understanding of disk dynamics, including shock wave formation, oscillations, and other complex behaviors. By comparing the numerical results obtained with various gravity models, researchers can gain new insights into the underlying physical processes, enhance their understanding of black hole-disk systems, and bridge theoretical and observational approaches.

The Kerr black hole in Boyer-Lindquist coordinates is described by the following metric equation \citep{Donmez6}:

\begin{eqnarray}
  ds^2 = -\left(1-\frac{2Mr}{\sum^2}\right)dt^2 - \frac{4Mra}{\sum^2}sin^2\theta dt d\phi
  + \frac{\sum^2}{\Delta_1}dr^2 + \sum^2 d\theta^2 + \frac{A}{\sum^2}sin^2\theta d\phi^2
\label{GREq2}
\end{eqnarray}

\noindent where $\Delta_1 = r^2 - 2Mr + a^2$, and $A = (r^2 + a^2)^2 - a^2\Delta \sin^2\theta$. In Boyer-Lindquist coordinates, the shift vector and lapse function of the Kerr metric are $\beta^i = (0,0,-2Mar/A)$ and $\tilde{\alpha} = (\sqrt{\Delta_1/A})$.

Near the Kerr black hole, the strong gravitational field causes spacetime to curve significantly, leading to the alteration of the physical characteristics of the geometry in its vicinity. This curvature is a consequence of the black hole mass and angular momentum, resulting in intriguing phenomena such as frame-dragging, time dilation, and the existence of an event horizon. 4D EGB gravity characterizes the curvature and physical properties of the spacetime in the vicinity of the rotating black hole, taking into account the gravitational effects predicted by the Einstein-Gauss-Bonnet theory. The metric for the 4D EGB rotating black hole is given by \citep{Donmez_EGB_Rot, Donmezetal2022}:

\begin{eqnarray}
  ds^2 &=& -\frac{\Delta_2 - a^2sin^2\theta}{\Sigma}dt^2 + \frac{\Sigma}{\Delta_2}dr^2 -
  2asin^2\theta\left(1- \frac{\Delta_2 - a^2sin^2\theta}{\Sigma}\right)dtd\phi + 
  \Sigma d\theta^2 + \nonumber \\
  && sin^2\theta\left[\Sigma +  a^2sin^2\theta \left(2- \frac{\Delta_2 -  
   a^2sin^2\theta}{\Sigma} \right)  \right]d\phi^2.
\label{GREq3}
\end{eqnarray}

\noindent
The expressions for $\Sigma$ and $\Delta_2$ are given by $\Sigma = r^2 + a^2 \cos^2\theta$ and $\Delta_2 = r^2 + a^2 + \frac{r^4}{2\alpha}\left(1 - \sqrt{1 + \frac{8 \alpha M}{r^3}} \right)$, respectively. Here, $a$, $\alpha$, and $M$ represent the black hole spin parameter, Gauss-Bonnet coupling constant, and mass of the black hole, respectively. The black hole horizon is determined by solving the equations $\Delta_1=0$ and $\Delta_2=0$. The lapse function $\tilde{\alpha}$ and the shift vectors of the 4D EGB metric are given as $\tilde{\alpha} = \sqrt{\frac{a^2(1-f(r))^2}{r^2+a^2(2-f(r))} + f(r)}$ and $\beta^i = \left(0, 0, \frac{a r^2}{2\pi \alpha}\left(1 - \sqrt{1 + \frac{8 \pi \alpha M}{r^3}}\right)\right)$, respectively. Here, $f(r) = 1 + \frac{r^2}{2\alpha}\left(1 - \sqrt{1 + \frac{8 \alpha M}{r^3}} \right)$.

The EGB black hole is a modified theory of gravity that generalizes the traditional black hole solutions, such as the Schwarzschild and Kerr solutions, by incorporating an additional term involving the Gauss-Bonnet curvature. This extra term introduces higher curvature corrections, offering a new perspective on the behavior of black holes in extreme conditions and higher-dimensional spacetime \citep{Fernandes2022}.

\subsection{General Relativistic Equations}
\label{GRHE3}

In the context of the perturbed accretion disk phenomenon, a fluid, which could be gas or dust originating from a companion star, undergoes gravitational attraction towards a massive object, such as a black hole, and subsequently perturbs the stable accretion disk. This process holds significance in gaining insights into the interaction between matter and black holes, as well as stable accretion disks. To investigate the perturbation of the disk process involving a perfect fluid, we explore the presence of rotating black holes, specifically the Kerr and EGB black holes, by solving the GRH equations in the curved background. The stress-energy-momentum tensor of the perfect fluid is taken into account to analyze the dynamical behavior and characteristics of the perturbed accretion disk in the vicinity of the rotating black holes.

\begin{eqnarray}
 T^{ab} = \rho h u^{a}u^{b} + P g^{ab},
\label{GREq1}
\end{eqnarray}

\noindent
where $\rho$ represents the rest-mass density, $p$ denotes the fluid pressure, $h$ indicates the specific enthalpy, $u^{a}$ represents the 4-velocity of the fluid, and $g^{ab}$ is the metric of the curved spacetime. The indices $a$, $b$, and $c$ range from $0$ to $3$. To facilitate a comprehensive comparison of the dynamic evolution of the accretion disk around the rotating black hole, we utilize two distinct coordinate systems. The first involves the Kerr black hole in Boyer-Lindquist coordinates, while the second employs the rotating black hole metric in 4D EGB gravity as described in Section \ref{GRHE2}.

\noindent
In order to perform numerical solutions for the GRH equations, it is essential to transform them into a conserved form \citep{Donmez1}. By employing advanced numerical methods, simulations of fluid dynamics in strong gravitational fields, such as those near spinning black holes, become more reliable and efficient, ensuring precise and trustworthy results.

\begin{eqnarray}
  \frac{\partial U}{\partial t} + \frac{\partial F^r}{\partial r} + \frac{\partial F^{\phi}}{\partial \phi}
  = S.
\label{GREq4}
\end{eqnarray}

\noindent
The conserved variables are denoted by the vectors $U$, $F^r$, $F^{\phi}$, and $S$, representing the conserved quantities, fluxes along the $r$ and $\phi$ directions, and sources, respectively. These conserved quantities are formulated in relation to the primitive variables, as shown below:

\begin{eqnarray}
  U =
  \begin{pmatrix}
    D \\
    S_j \\
    \tau
  \end{pmatrix}
  =
  \begin{pmatrix}
    \sqrt{\gamma}W\rho \\
    \sqrt{\gamma}h\rho W^2 v_j\\
    \sqrt{\gamma}(h\rho W^2 - P - W \rho)
    \end{pmatrix}.
\label{GREq5}
\end{eqnarray}

\noindent In the given equation, the terms are defined as follows: the Lorentz factor is represented by $W = (1 - \gamma_{a,b}v^i v^j)^{1/2}$. The enthalpy is denoted by $h = 1 + \epsilon + P/\rho$, where $\epsilon$ stands for the internal energy. The three-velocity of the fluid is given by $v^i = u^i/W + \beta^i$. The pressure of the fluid is determined using the ideal gas equation of state, $P = (\Gamma - 1)\rho\epsilon$. The three-metric is denoted as $\gamma_{i,j}$, and its determinant is $\gamma$, both calculated using the four-metric of the rotating black holes. Latin indices $i$ and $j$ range from $1$ to $3$. The flux and source terms can be computed for any metric using the following equations:

\begin{eqnarray}
  \vec{F}^i =
  \begin{pmatrix}
    \tilde{\alpha}\left(v^i - \frac{1}{\tilde{\alpha}\beta^i}\right)D \\
    \tilde{\alpha}\left(\left(v^i - \frac{1}{\tilde{\alpha}\beta^i}\right)S_j + \sqrt{\gamma}P\delta^i_j\right)\\
    \tilde{\alpha}\left(\left(v^i - \frac{1}{\tilde{\alpha}\beta^i}\right)\tau  + \sqrt{\gamma}P v^i\right)
    \end{pmatrix},
\label{GREq6}
\end{eqnarray}

\noindent and,

\begin{eqnarray}
  \vec{S} =
  \begin{pmatrix}
    0 \\
    \tilde{\alpha}\sqrt{\gamma}T^{ab}g_{bc}\Gamma^c_{aj} \\
    \tilde{\alpha}\sqrt{\gamma}\left(T^{a0}\partial_{a}\tilde{\alpha} - \tilde{\alpha}T^{ab}\Gamma^0_{ab}\right)
   \end{pmatrix},
\label{GREq7}
\end{eqnarray}

\noindent where $\Gamma^c_{ab}$ is the Christoffel symbol.

\section{Initial and Boundary Conditions}
\label{InitialBC}

To examine the perturbed accretion disk surrounding the rotating Gauss-Bonnet black hole and compare it with the Kerr black hole, the General Relativistic Hydrodynamical (GRH) equations are solved on the equatorial plane. The simulation is performed using the code described in previous works by \citet{Donmez1, Donmez5, Donmez2}. The pressure of the accreted matter is calculated using the standard $\Gamma$-law equation of state for a perfect fluid, given by $P = (\Gamma - 1)\rho\epsilon$, where $\Gamma = 4/3$. Here, $P$ represents the pressure, $\rho$ is the density of the fluid, and $\epsilon$ denotes the internal energy.

To conduct the numerical simulation, two distinct initial conditions are utilized. Initially, the initial conditions presented in \citet{Donmez2023} are employed to establish a stable and steady-state accretion disk around the black hole under spherical accretion conditions. The initial density and pressure profiles are carefully tuned to ensure that the speed of sound is equal to $C_{s} = 0.1$ at the outer boundary of the computational domain.

In order to achieve spherical accretion and perturbation of the steady-state accretion disk within short timescales, it is crucial to carefully choose the speed of sound. When the speed of sound falls within a specific range, spherical accretion and the formation of shock waves occur around the black hole. For the disk to undergo perturbation and for a shock wave to form around the black hole, matter must be drawn towards the black hole. This is feasible when the perturbation exhibits supersonic behavior, indicated by a Mach number greater than 1. If the infalling matter is subsonic, a shock wave may not form, or its formation time may surpass the simulation duration. Thus, we focus on scenarios where the speed of matter entering the outer boundary exceeds the speed of sound. This facilitates rapid interaction between the disk and the perturbation, leading to the creation of a shock wave around the black hole. In essence, in cases where $V_{\infty} > C_{s}$, as observed in numerical simulations, a shock wave forms rapidly.

At this critical speed of sound, both spherical accretion and the stability of the disk around the black hole are disrupted, paving the way for the emergence of new physical structures. Selecting a speed of sound that is either too low or too high could result in scenarios where matter either never falls towards the black hole or directly plunges into it. Such situations may disrupt the disk formation mechanism and prevent the formation of shock waves due to perturbations. Moreover, an improper speed of sound at the outer boundary may introduce unwanted oscillations, altering the underlying physical processes near the black hole. Consequently, undesired artificial oscillations may manifest on the disk. This adjustment is crucial to maintain appropriate conditions for the simulation and prevent unphysical behaviors at the boundaries.

In pursuit of this goal, a constant density value ($\rho = 10^{-4}$) is assigned, and the pressure is computed using the perfect fluid equation of state throughout the entire computational domain. This approach ensures uniform conditions and accurate representation of the fluid behavior across the simulation domain. Subsequently, numerical simulations are conducted on the equatorial plane, where gas is injected from the outer boundary with velocities $V^r = -0.01$, $V^{\phi}=0$, and density $\rho=1$. As seen in \citet{Donmez2023}, the disk reaches a steady state at approximately $t=5000M$. However, to ensure certainty, the first phase is run up to $t=6000M$.

In the second stage of the modeling, the perturbation of the accretion disk is achieved by the infall of matter originating from a disrupting star or other physical mechanisms, causing it to fall toward the black hole from the outer boundary of the stable accretion disk that has reached a steady state. Once more, the initial density and pressure profiles are meticulously adjusted to ensure that the speed of sound equals $C_s = 0.1$ at the outer boundary of the computational domain. This meticulous tuning guarantees appropriate conditions for the simulation, particularly at the domain's outer limits.

In order to introduce a physically acceptable perturbation, the numerical simulations are performed on the equatorial plane, with gas being injected from the outer boundary with velocities $V^r = -0.01$, $V^{\phi}= V_{\infty}\sqrt{\gamma^{\phi\phi}}\sin\phi$, and density $\rho=2$ for $0<\phi<0.4$ at the outer boundary. $V_{\infty}$ is the asymptotic velocity of the matter at the outer boundary of the computational domain, as given in Table \ref{Inital_Con}. This setup ensures the infall of the gas toward the black hole supersonically, generating a significant perturbation in the system, allowing for a more insightful analysis of its behavior and dynamics. The detailed descriptions and a more comprehensive understanding of the initial models for rotating black holes, including both Kerr and Gauss-Bonnet black holes, can be found in Table \ref{Inital_Con}.

\begin{table}
\footnotesize
\caption{Adopted initial model used in the numerical simulation for Kerr and Gauss-Bonnet black holes. $Model$ is the name of the model. $Type$ represents the types of black holes used in the numerical simulations. $\alpha$ denotes the Gauss-Bonnet coupling constant, and $a/M$ signifies the black hole rotation parameter. $nt/nt_{Kerr}$ indicates the ratio of the time steps required to reach the maximum time of $30000M$ to the time steps required for Kerr.
}
 \label{Inital_Con}
\begin{center}
  \begin{tabular}{cccccc}
    \hline
    \hline

    $Model$ & $type$ & $\alpha (M^2)$ & $a/M$ & $V_{\infty}$ & $nt/nt_{Kerr}$\\
    \hline
    $K028$ &  & $-$   & $0.28$ & 0.2 & 1\\
    $K09$  &  & $-$   & $0.9$  & 0.2 & 1\\
    \hline
    $K028\_EGB1$ &  & $-4.93$  & $0.28$ & 0.2 &  1.15061\\
    $K028\_EGB2$ &  & $-3.03$  & $0.28$ & 0.2 & 1.03582\\
    $K028\_EGB3$ & $Gauss-Bonnet$ & $-0.37$   & $0.28$ & 0.2 & 1.01832\\
    $K028\_EGB4$ &  & $0.096$  & $0.28$ & 0.2 & 1.01561\\
    $K028\_EGB5$ &  & $0.41$   & $0.28$ & 0.2 & 1.01375\\
    $K028\_EGB6$ &  & $0.68$   & $0.28$ & 0.2 & 1.01163\\
    \hline
    $K09\_EGB1$ &  & $-3.61$   & $0.9$ &  0.2 & 1.15832\\
    $K09\_EGB2$ &  & $-1.56$   & $0.9$ & 0.2 & 1.11819\\    
    $K09\_EGB3$ &  & $-0.98$   & $0.9$ & 0.2 & 1.11321\\
    $K09\_EGB4$ &  & $-0.443$   & $0.9$ & 0.2 & 1.10711\\
    $K09\_EGB5$ &  $Gauss-Bonnet$ & $-0.063$   & $0.9$ & 0.2 & 1.10408\\
    $K09\_EGB6$ &  & $-0.023$  & $0.9$ & 0.2 & 1.10301\\
    $K09\_EGB7$ &  & $0.000625$   & $0.9$ & 0.2 & 1.10337\\
    $K09\_EGB8$ &  & $0.05$   & $0.9$ & 0.2 & 1.10313\\
    $K09\_EGB9$ &  & $0.05$   & $0.9$ & 0 & 1.05249\\    
    \hline
    \hline
  \end{tabular}
\end{center}
\end{table}

Uniformly spaced zones are utilized to discretize the computational domain in both the radial and angular directions. Specifically, there are $N_r=1024$ zones in the radial direction and $N_{\phi}=256$ zones in the angular direction. In the radial direction, the inner and outer boundaries of the computational domain are located at $r_{\text{min}}=3.7M$ and $r_{\text{max}}=100M$, respectively. The angular boundaries are specified as $\phi_{\text{min}}=0$ and $\phi_{\text{max}}=2\pi$. The code execution time, with $t_{\text{max}} = 33000M$, is considerably longer than expected. Numerical verification has confirmed that the qualitative outcomes of the simulations, such as the presence of QPOs and instabilities, the location of shocks, and the behavior of the accretion rates, are mostly independent of variations in the grid resolution \citep{Donmez3, Donmez_EGB_Rot, Donmezetal2022}. Furthermore, it is believed that positioning the outer boundary far enough away from the region where artifacts may possibly be produced ensures that they do not impact the physical solution.

In this paper, we explore the behavior of matter as it falls towards a black hole through accretion, aiming to uncover the impacts of Kerr and EGB gravities on the vicinity of the black hole horizon. Thus, it is imperative for the inner boundary of the computational domain to be positioned as close to the black hole horizon as possible. However, the location of the horizon for EGB black holes varies considerably based on the $\alpha$ parameter. For instance, with $\alpha = 0.6$, the horizon is approximately at $r=1M$, whereas with $\alpha=-4.93$, it is situated at $r=3.5M$. Therefore, for a fair comparison, the inner radius of the computational domain is standardized to $r=3.7M$ across all models.

Ensuring accurate boundary treatment is crucial to prevent the emergence of unphysical solutions in numerical simulations. To address this, specific boundary conditions are implemented for different regions. An outflow boundary condition is applied to the inner radial boundary. This allows gas to flow into the black hole, employing a simple zeroth-order extrapolation technique. At the outer boundary, gas is continuously injected with initial velocities and density, as mentioned above, to account for the inflow of material until the disk reaches a steady state. Periodic boundary conditions are employed along the angular direction. This means that the fluid properties at one end of the domain are considered equivalent to the properties at the opposite end. By implementing these appropriate boundary conditions, the simulation aims to maintain the accuracy and reliability of the results, ensuring that unphysical artifacts are minimized or avoided.

\section{Astrophysical Motivation}
\label{Astro_Mot}

Observational evidence strongly suggests that accretion disks and electromagnetic radiation are closely related and that disks act as the driver for the radiation. There is a large, rich literature on the topic of astrophysical electromagnetic radiation and disks. The presence of perturbed disks has been observed in various astrophysical observations. These observations have revealed the existence of nonlinear physical mechanisms that can occur on the disk. For example, frequency fluctuations in the radiation observed in AGNs have been noted \citep{Vaughan2011, McHardy2010, Gaskell2013}. This is entirely due to perturbations caused by interactions with other stars around the accretion disk. On the other hand, while black holes have stable disks around them, they can be perturbed by matter coming from surrounding stars, resulting in the formation of QPOs \citep{Artymowicz1996, Nixon2011}. These are the consequences of disk instability.

Observations from various sources have provided evidence that a substantial number of systems exhibit emissions originating from perturbed accretion disks. For instance, SMBH J095036+512838 was studied through observations of Ly-alpha and C IV broad lines \citep{Runnoe2021, Mohammed2023}. Similarly, GRS 1915+105 was examined through observations of high-frequency QPOs \citep{Sreehari2020MNRAS, Majumder2022MNRAS, Feng2022, Ricketts2023, Dhaka2023}. The results from GX 339-4 are consistent with the presence of a perturbed accretion disk \citep{Jin2023}. Additionally, the bright radiation from MAXI J1820+070 indicated the presence of a perturbed accretion disk \citep{Zhang2023}. For the sources MAXI J1820+070 and MAXI J1535-571, \citet{Ding2023} observed cross and hypotenuse patterns, suggesting the presence of QPOs. It is plausible that these different types of QPOs could arise from similar underlying mechanisms.

Another observed system believed to originate from a perturbed accretion disk is Mrk 1018. Situated within a host galaxy currently experiencing a significant merger event, Mrk 1018 is among the earliest identified Changing-Look (CL) AGNs, transitioning from a Seyfert 1.9 to a Seyfert 1 nucleus over time \citep{Brogan2023}. Over a span of 40 years, the spectral type of Mrk 1018 underwent changes from Type 1.9 to Type 1, and subsequently reverted back to Type 1.9. By analyzing archival spectral and imaging data, \citet{Kim2018} identified two kinematically distinct broad-line components, namely blueshifted and redshifted components. The velocity offset curve of the broad line displayed a characteristic pattern over time. Perturbations in the accretion disk caused by pericentric passages are deemed plausible in explaining the AGN activity and spectral changes in Mrk 1018.

Specific low-frequency QPOs originate from the geometric structure of the disk during the interaction between the black hole and the disk \citep{Pawar2015MNRAS, Ingram2015MNRAS, Ingram2019}. Here, we modeled the structure of the disk around rotating black holes with different gravities, revealing how the disk structure and resulting shock waves depend on the black hole spin parameter and the EGB coupling constant. This allows us to explore the potential physical mechanisms behind the observed low-frequency QPOs, which will be further elaborated on in the following sections.

\section{The Numerical Simulation of the accretion toward 4D EGB Rotating and Kerr Black Holes}
\label{Results}

This section presents the results of our numerical simulation, which explores the dynamics of the accretion disk around a black hole. We investigate how the steady-state disk is perturbed by a donor star in the X-ray binary system, uncovering instabilities, the dynamics of spiral shock waves, and potential QPOs in the disk. These phenomena manifest after the disk reaches a quasi-equilibrium state.


\subsection{Formation of Initial Steady-State Accretion Disk}
\label{Initial_Disk}

In astrophysical systems, observing regular emissions requires identifying a consistent physical mechanism operating within the disk. In this article, we investigate the instabilities induced by perturbations to the disk. To achieve this, it is essential to effectively perturb the disk by utilizing analytically defined disks formed around black holes through various physical mechanisms. For instance, in our previous study \citep{Donmez4}, we examined the perturbation of tori analytically known in Schwarzschild and Kerr gravity around black holes. However, the analytic solutions of tori around non-rotating and rotating black holes are unknown in EGB gravity. Therefore, to perturb a different stable disk, we utilize the stable disks identified in our prior research \citep{Donmez2023}.

\begin{figure*}
  \vspace{1cm}
  \center
   \psfig{file=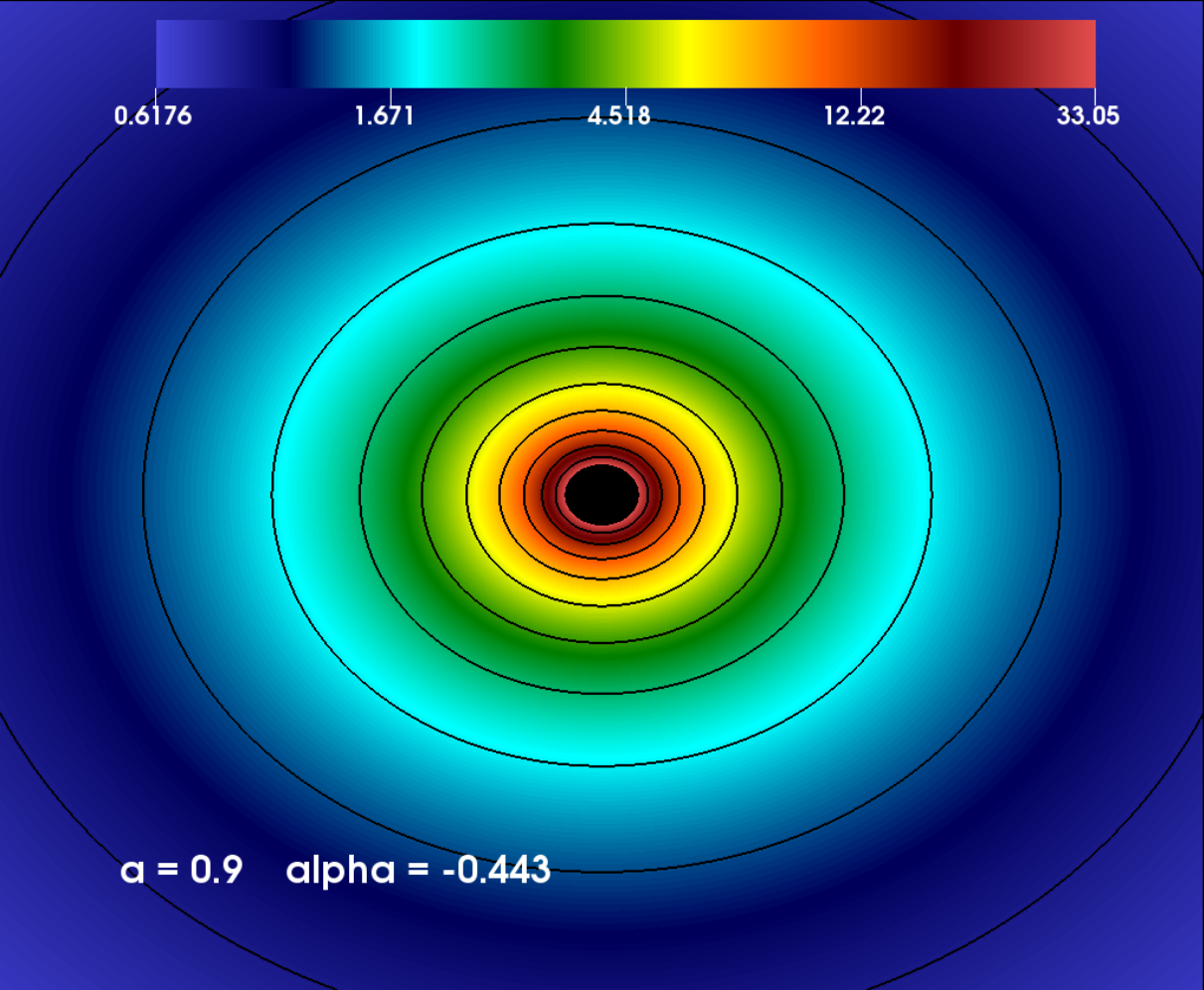,width=6.0cm}
   \psfig{file=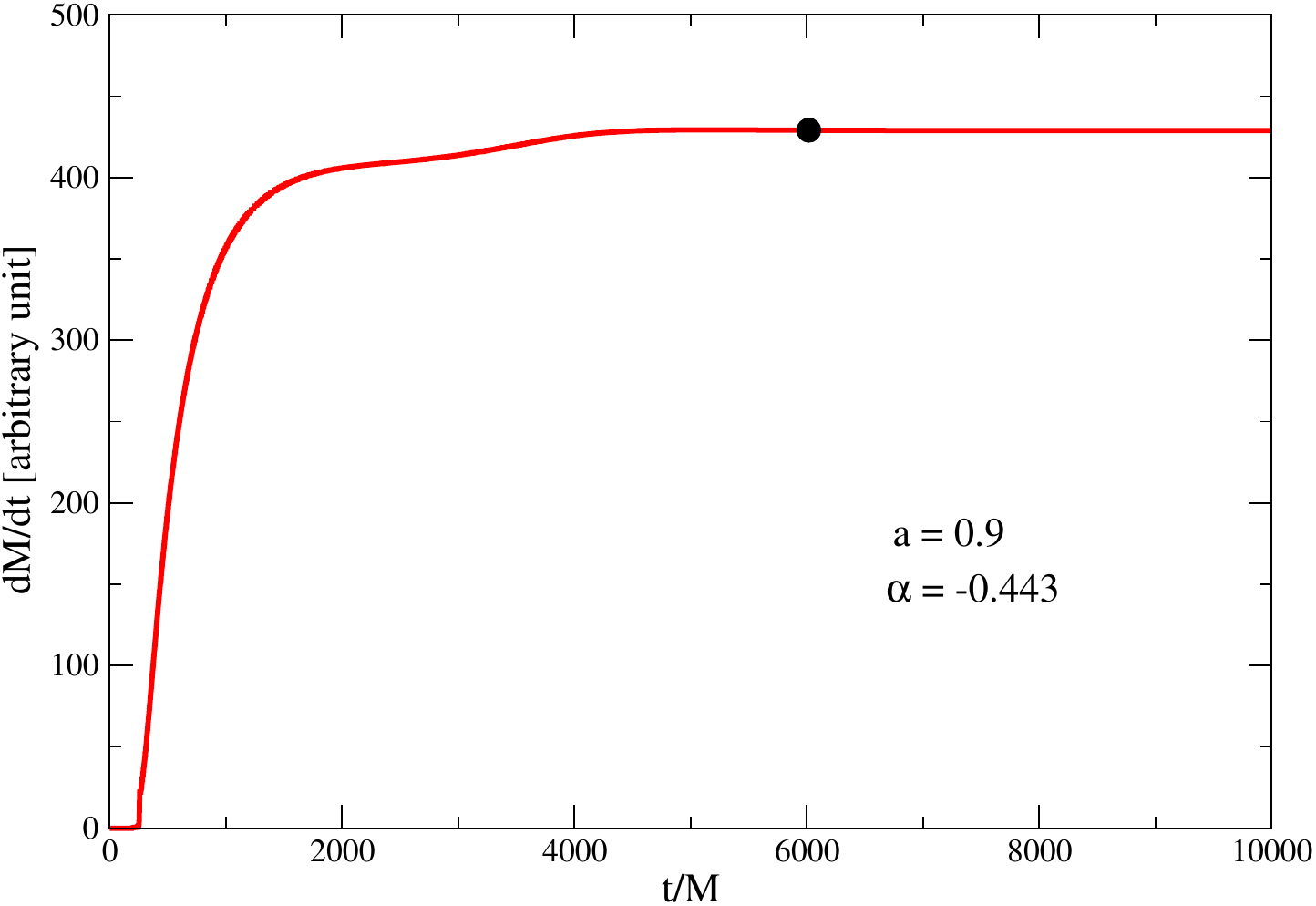,width=7.0cm} 
   \caption{
The logarithmic rest-mass density observed at $t = 6000M$ and the mass accretion rate plotted as a function of time demonstrate how this rate evolves over the simulated period. Both plots depict the formation of an initial steady-state accretion disk.
}
\label{Mass_acc}
\end{figure*}

The initial stable disk formed as a result of matter scattered around after a supernova explosion and falling back onto the newly formed black hole. This situation is explained in Section \ref{InitialBC} and detailed in \citep{Donmez2023, Donmez2023arXiv231108388D} concerning the stability status of the formation disk. In Fig. \ref{Mass_acc}, the logarithmic colored density of the disk, which has reached a steady state for a model used in \citep{Donmez2023}, is presented in the left panel. In the right panel of Fig. \ref{Mass_acc}, the stability behavior, varying over time, is derived from the calculation of the mass accretion rate at the inner boundary, at the point of the disk closest to the black hole. Particularly, the graph on the right shows that the disk reached a steady state around $t=5000M$ and always maintained its stable structure. It has also been reiterated in Section 5 of \citep{Donmez2023} that the disk does not oscillate after reaching a steady state. The solid circle depicted in the right panel of Fig. \ref{Mass_acc} represents the point in time when the disk experiences perturbation, following a certain period after achieving stability. This circumstance will be further elaborated in the subsequent sections.

\subsection{The Perturbed Accretion Disk around  the Rotating EGB  Black Hole}
\label{EGB_BH}

Modeling a perturbed stable accretion disk is crucial for understanding the behavior of matter around black holes, determining the mass accretion rate towards the black hole, unveiling the physical mechanism of shock waves formed on the disk \citep{Aoki2004}, variations in the disk light emission, and consequently explaining the observed $QPOs$ \citep{Abramowicz2013}. Here, we demonstrate how the stable disk around rotating black holes, as presented in Section \ref{Initial_Disk}, responds to perturbation. We illustrate the changes in the shock wave resulting from the perturbation and demonstrate how these changes depend on the parameters.

\begin{figure*}
  \vspace{1cm}
  \center
  \psfig{file=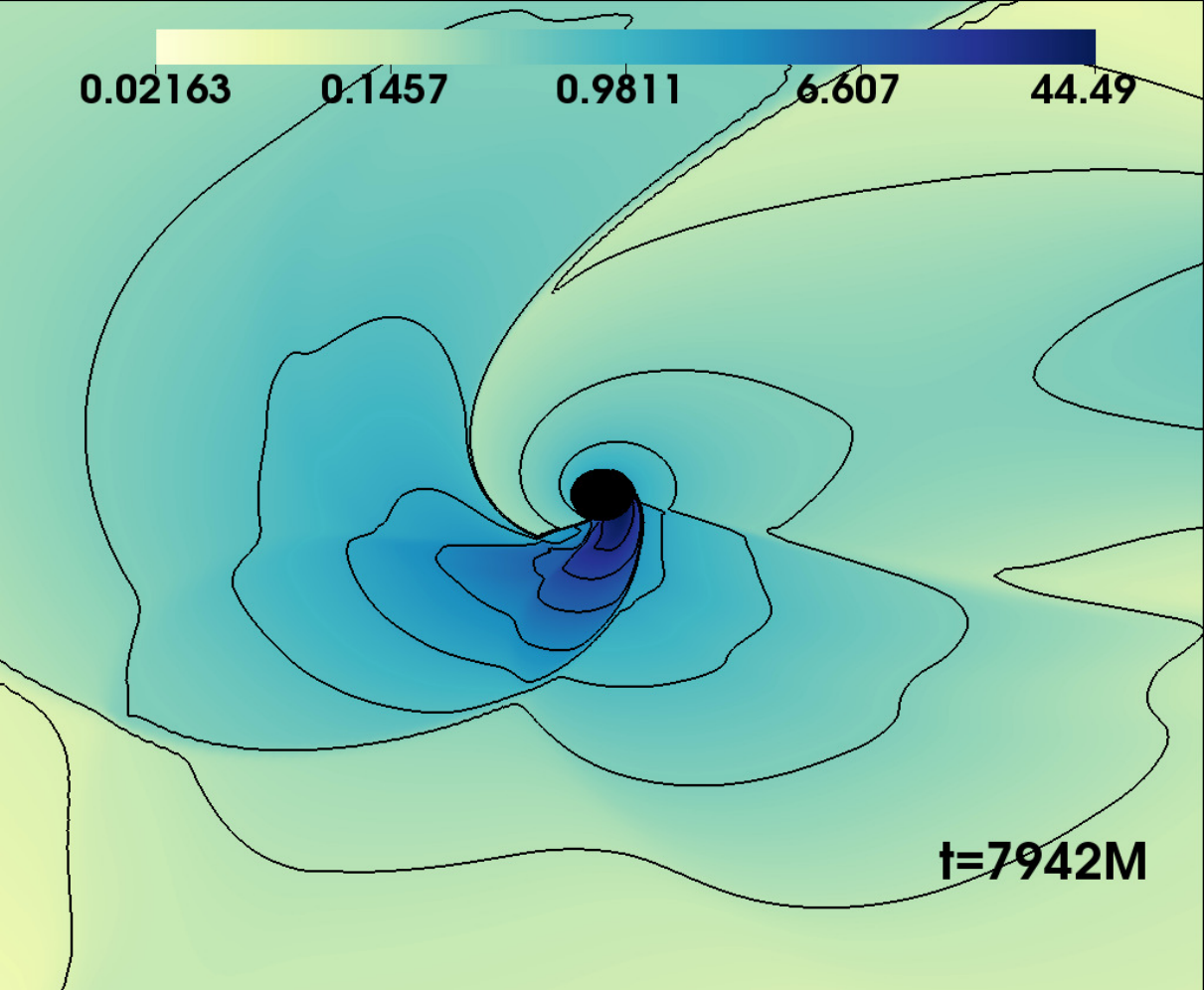,width=2.9cm,height=3.0cm}
  \psfig{file=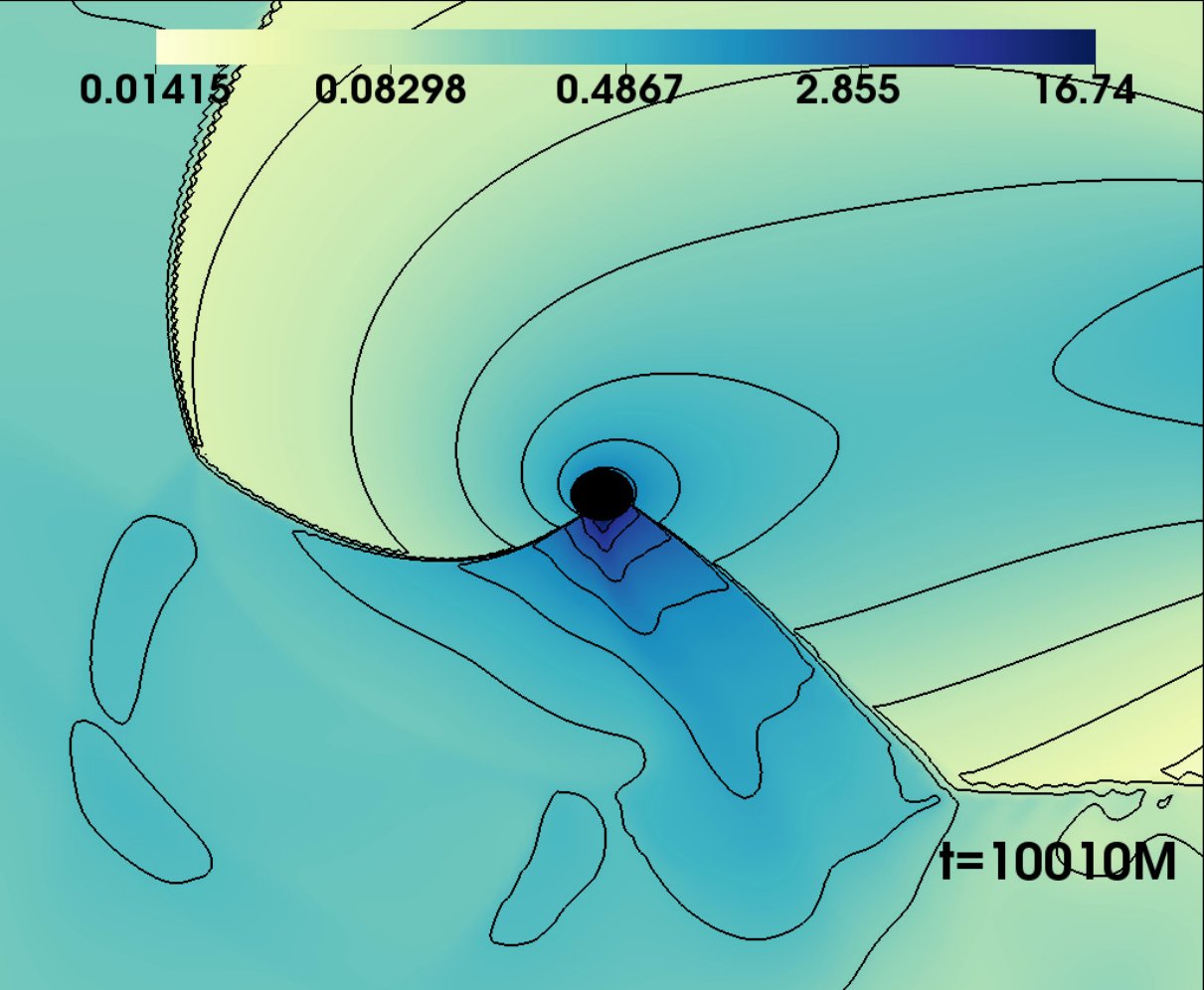,width=2.9cm,height=3.0cm}
  \psfig{file=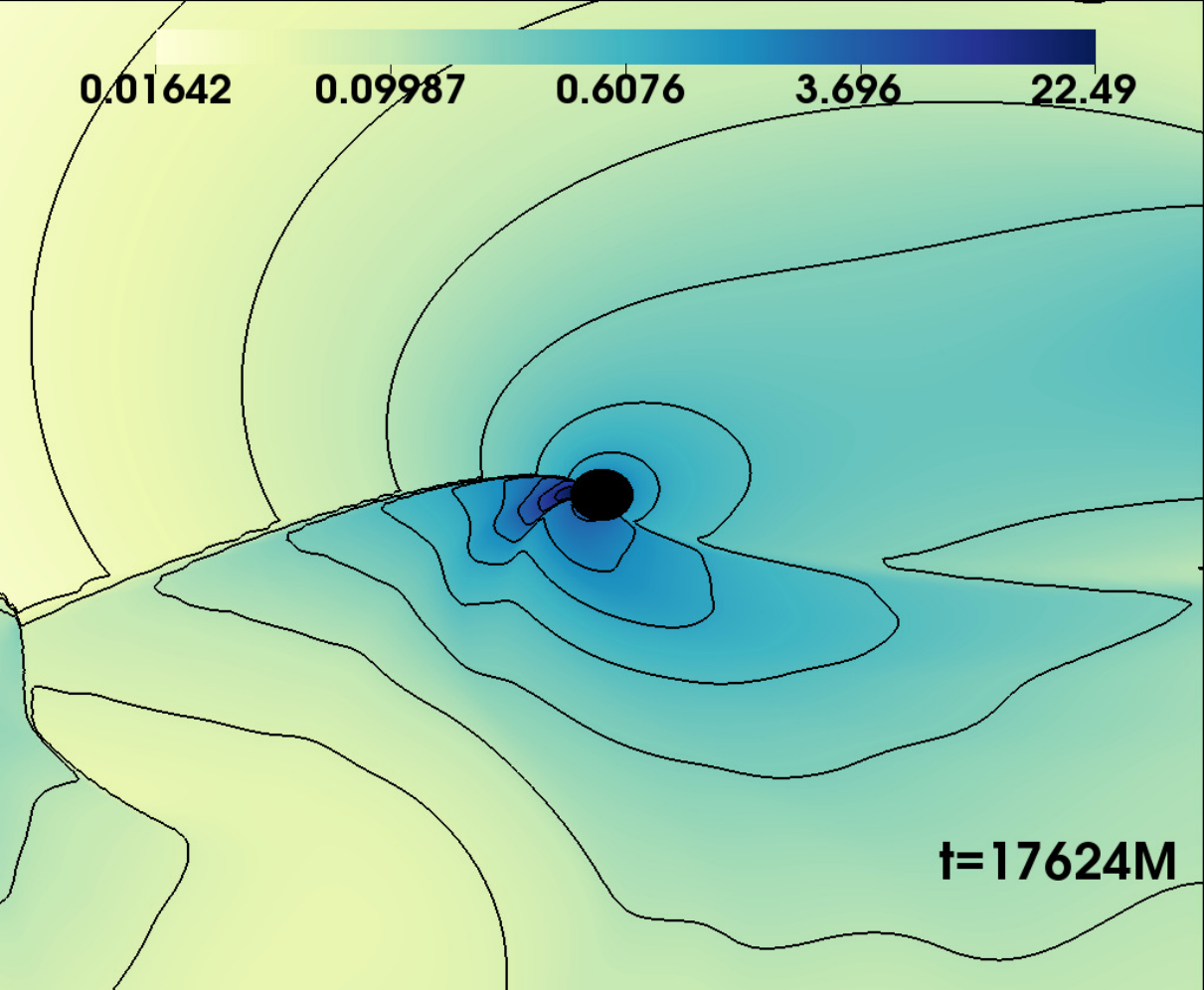,width=2.9cm,height=3.0cm}
  \psfig{file=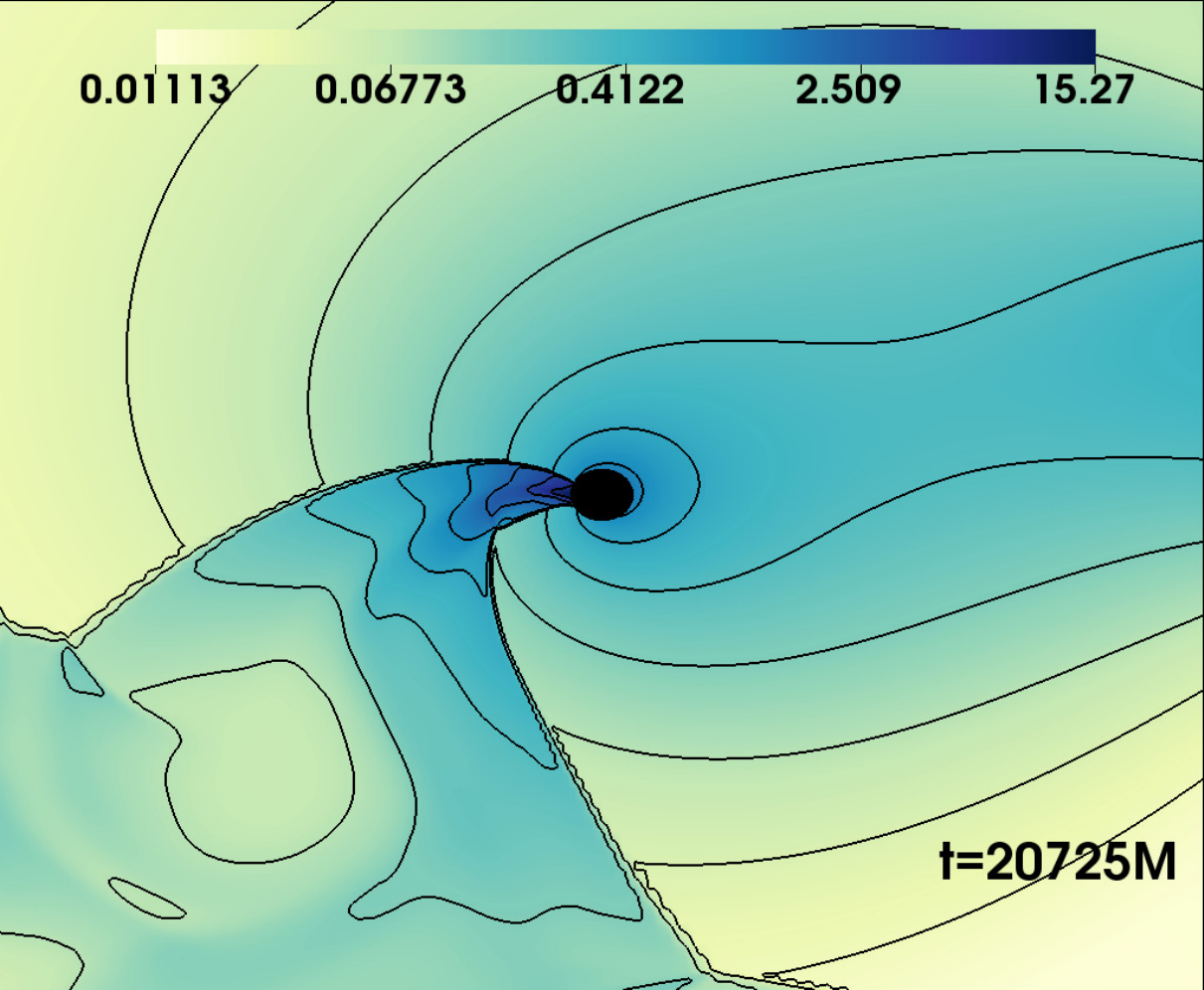,width=2.9cm,height=3.0cm}
  \psfig{file=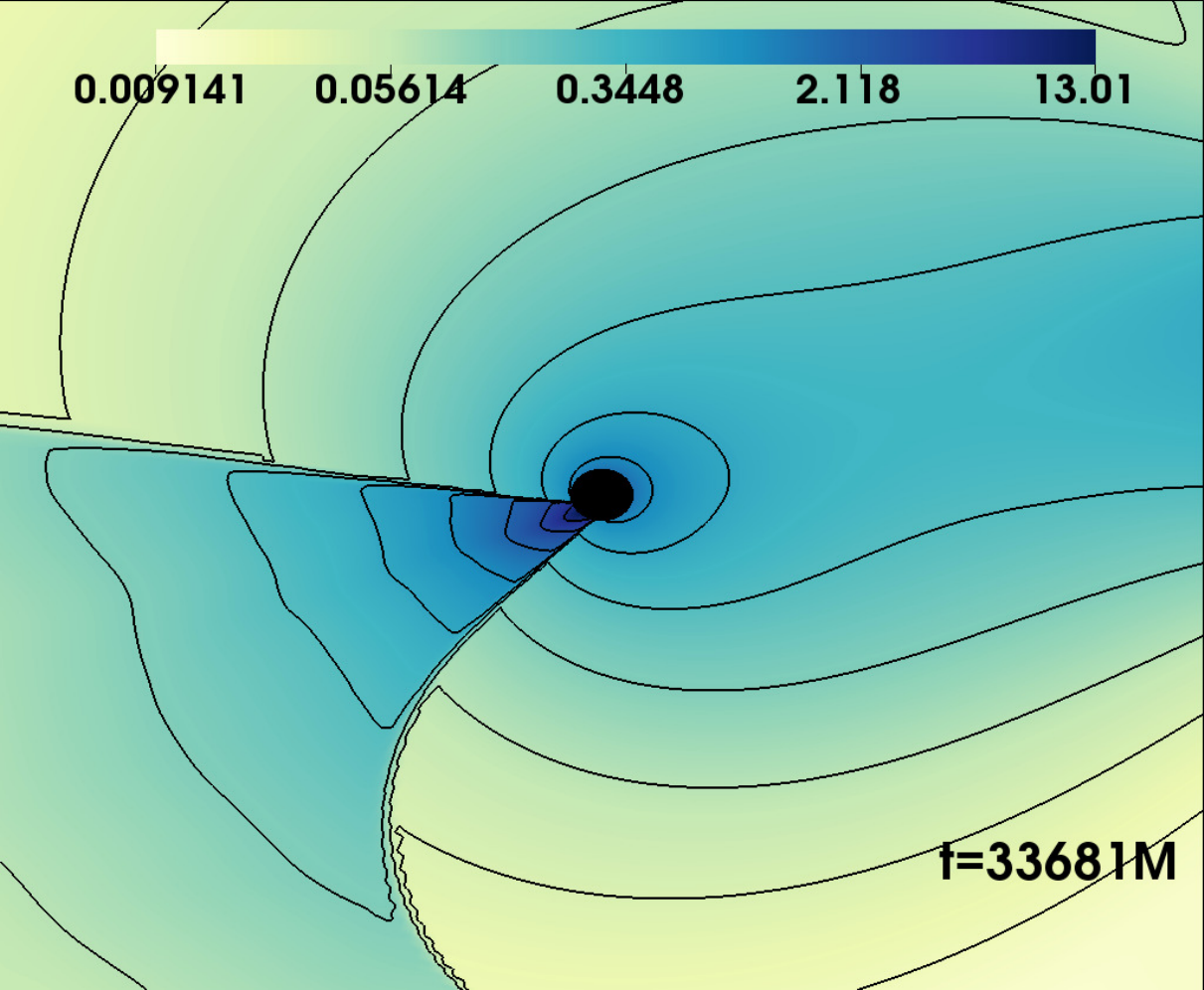,width=2.9cm,height=3.0cm}
  \psfig{file=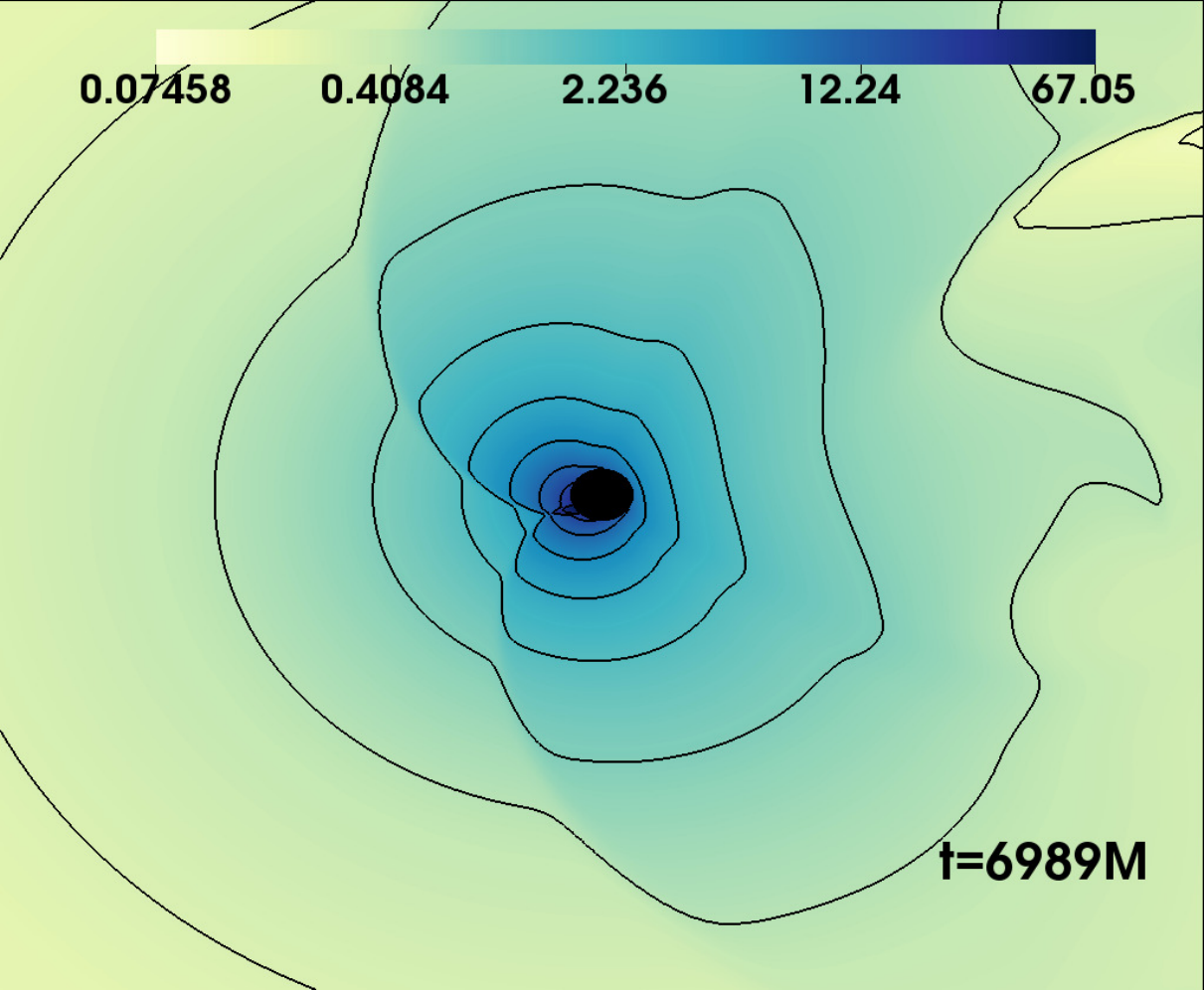,width=2.9cm,height=3.0cm}
  \psfig{file=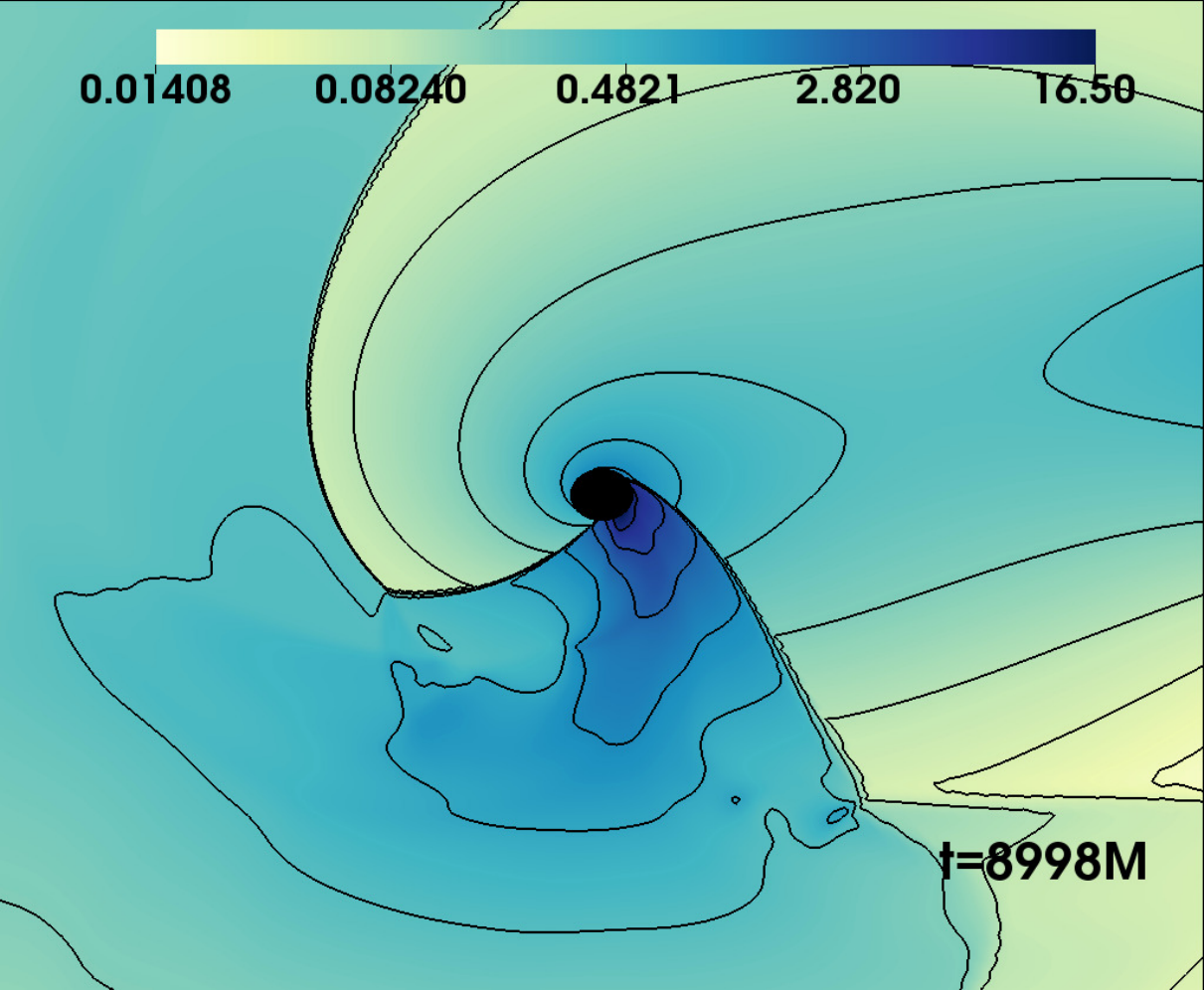,width=2.9cm,height=3.0cm}
  \psfig{file=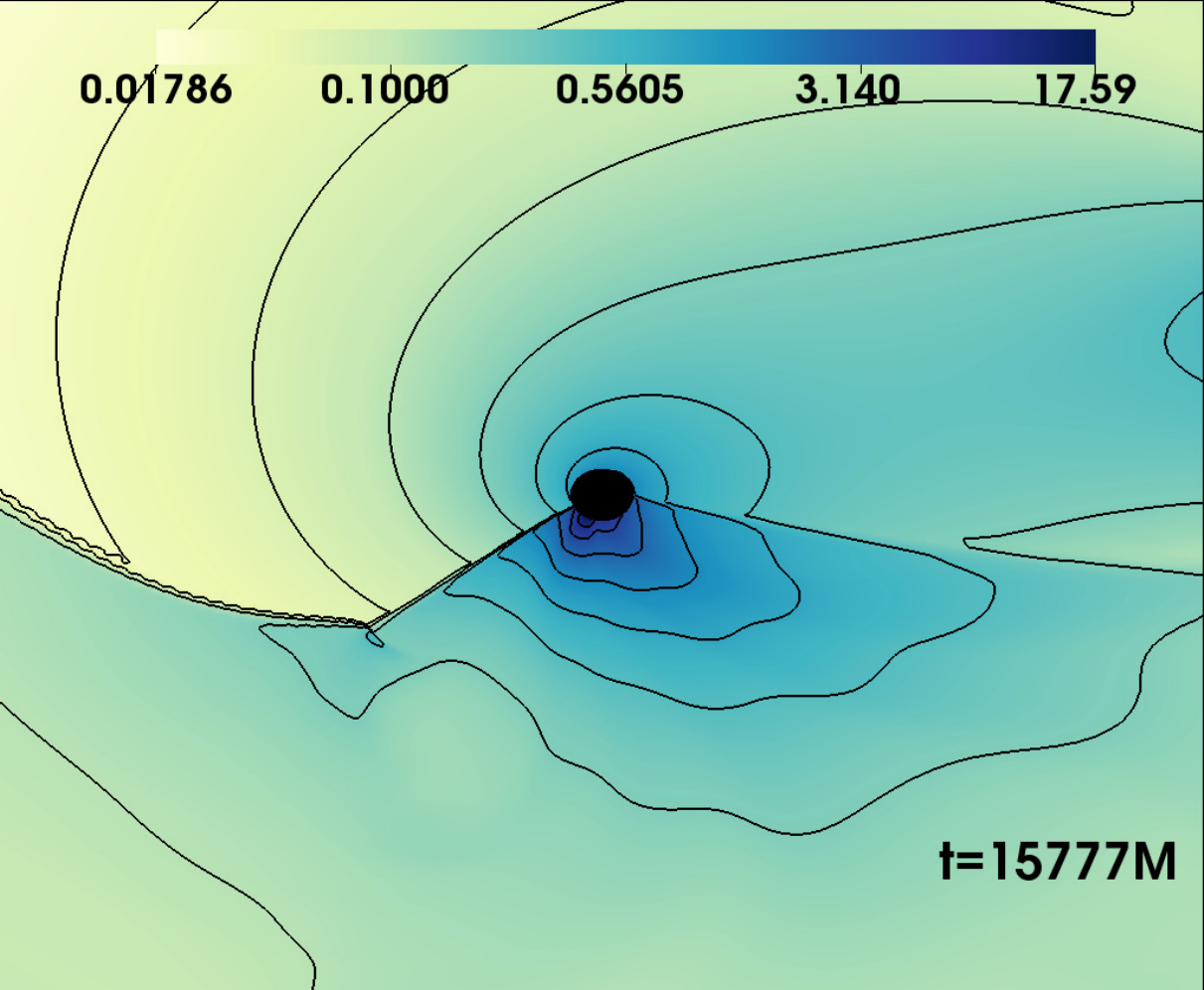,width=2.9cm,height=3.0cm}
  \psfig{file=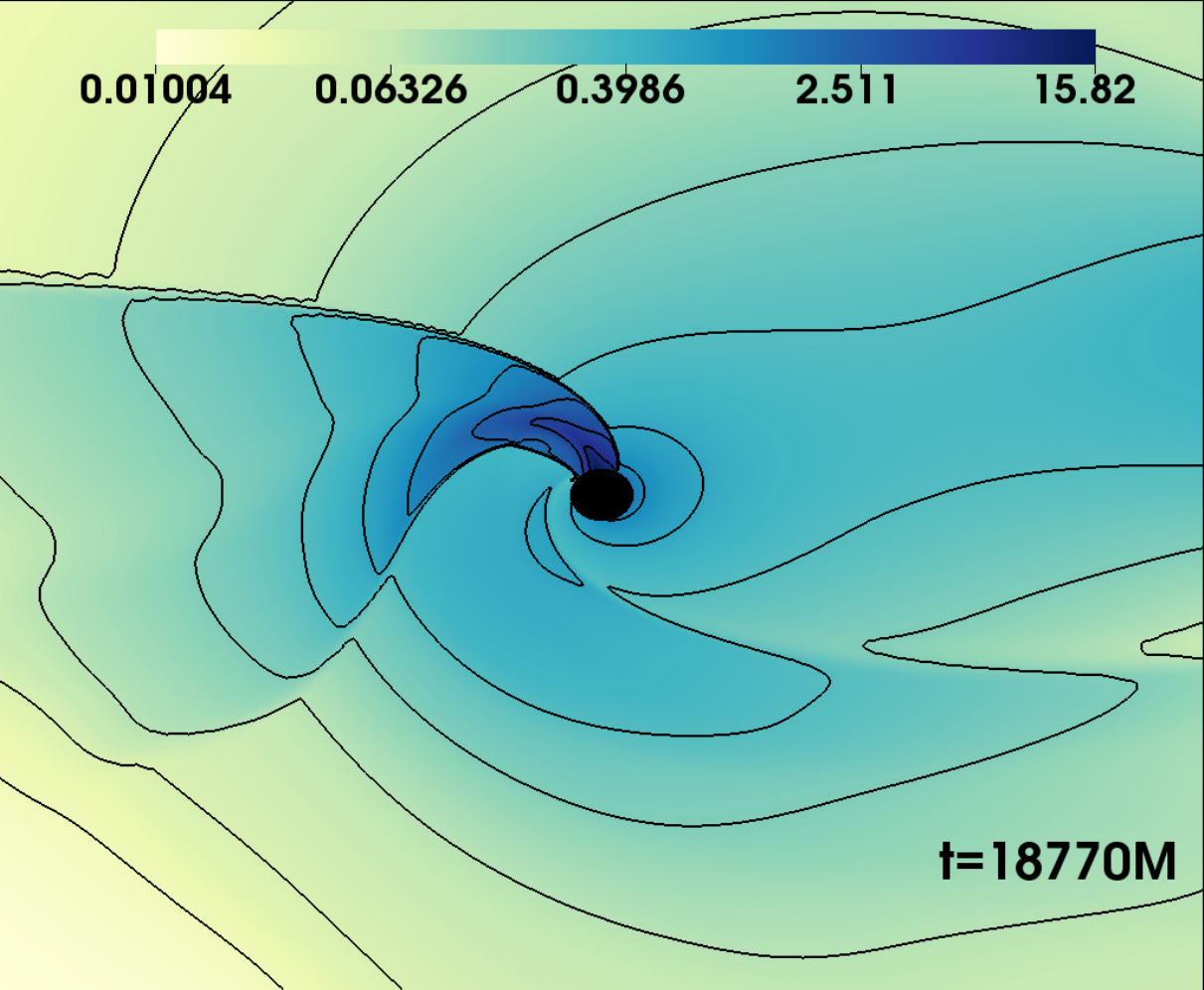,width=2.9cm,height=3.0cm}
  \psfig{file=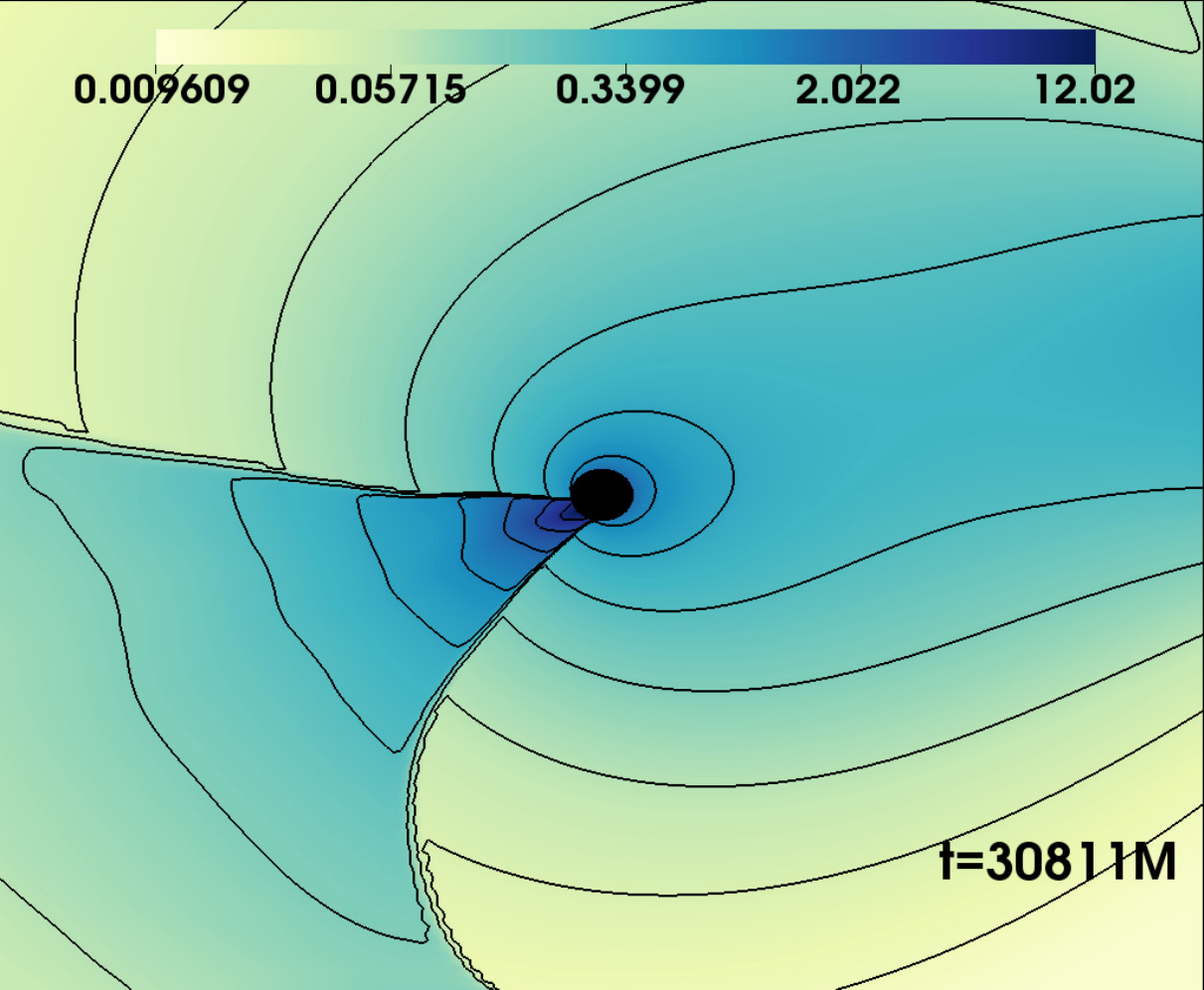,width=2.9cm,height=3.0cm}
  \psfig{file=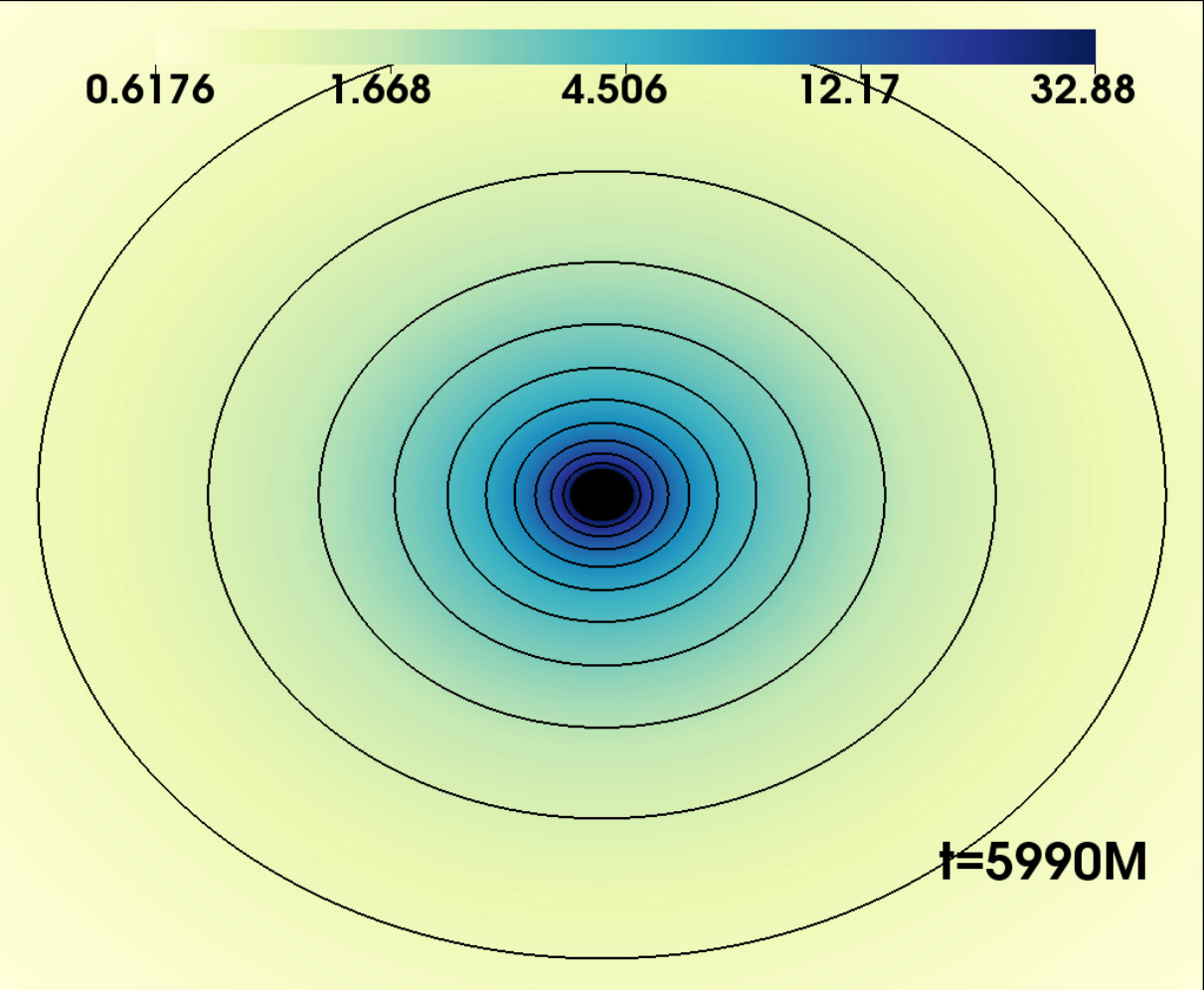,width=2.9cm,height=3.0cm}
  \psfig{file=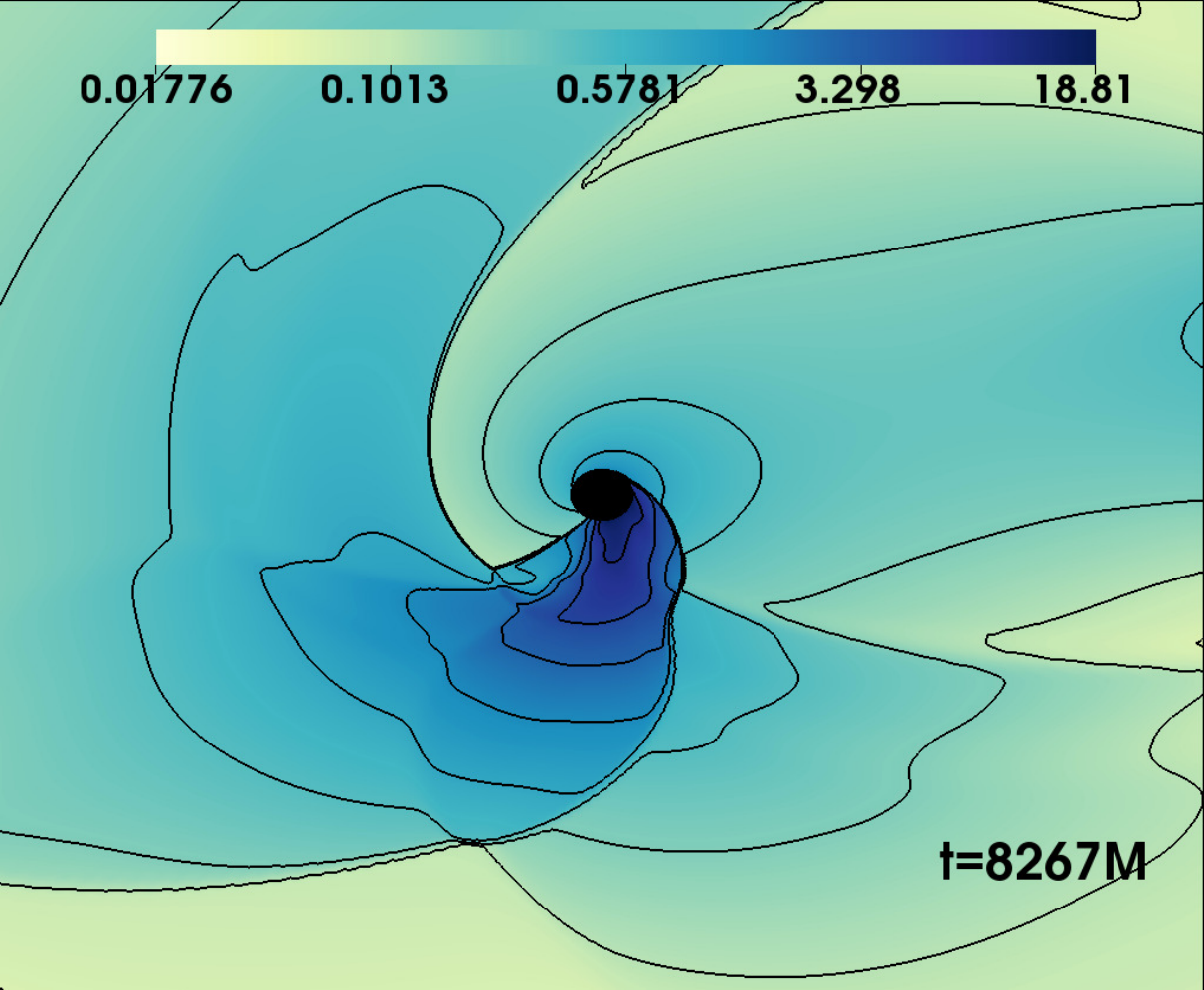,width=2.9cm,height=3.0cm}
  \psfig{file=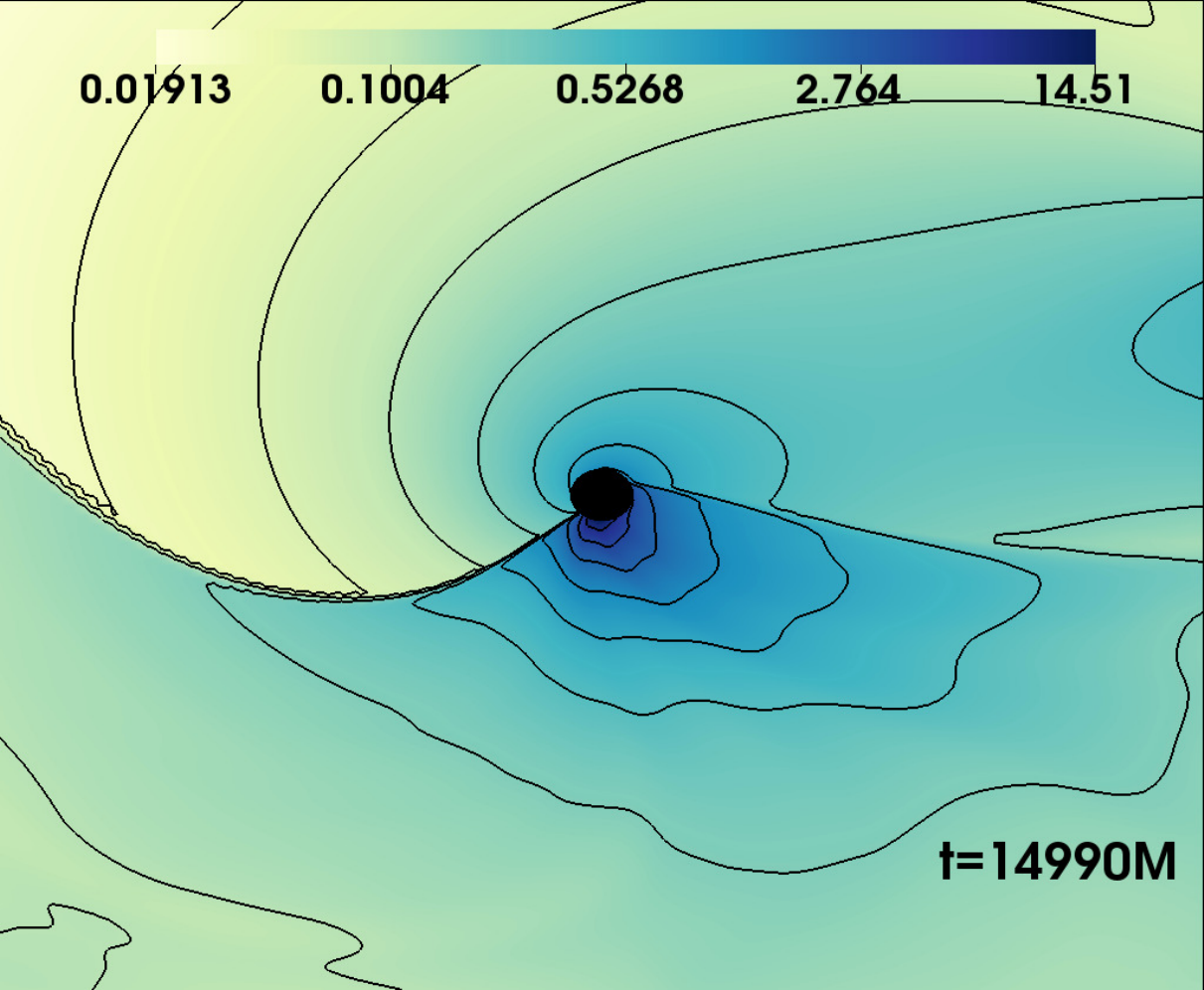,width=2.9cm,height=3.0cm}
  \psfig{file=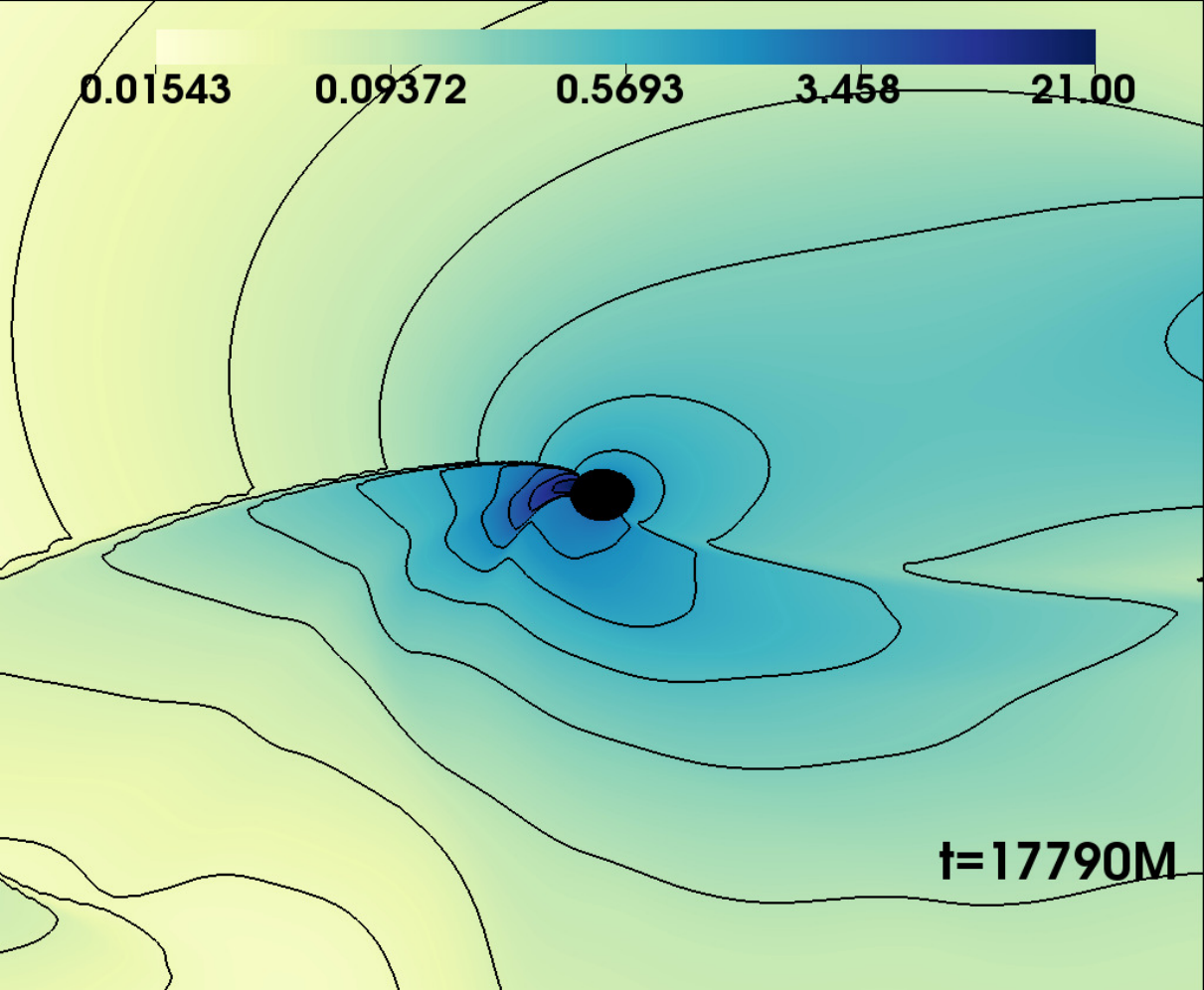,width=2.9cm,height=3.0cm}
  \psfig{file=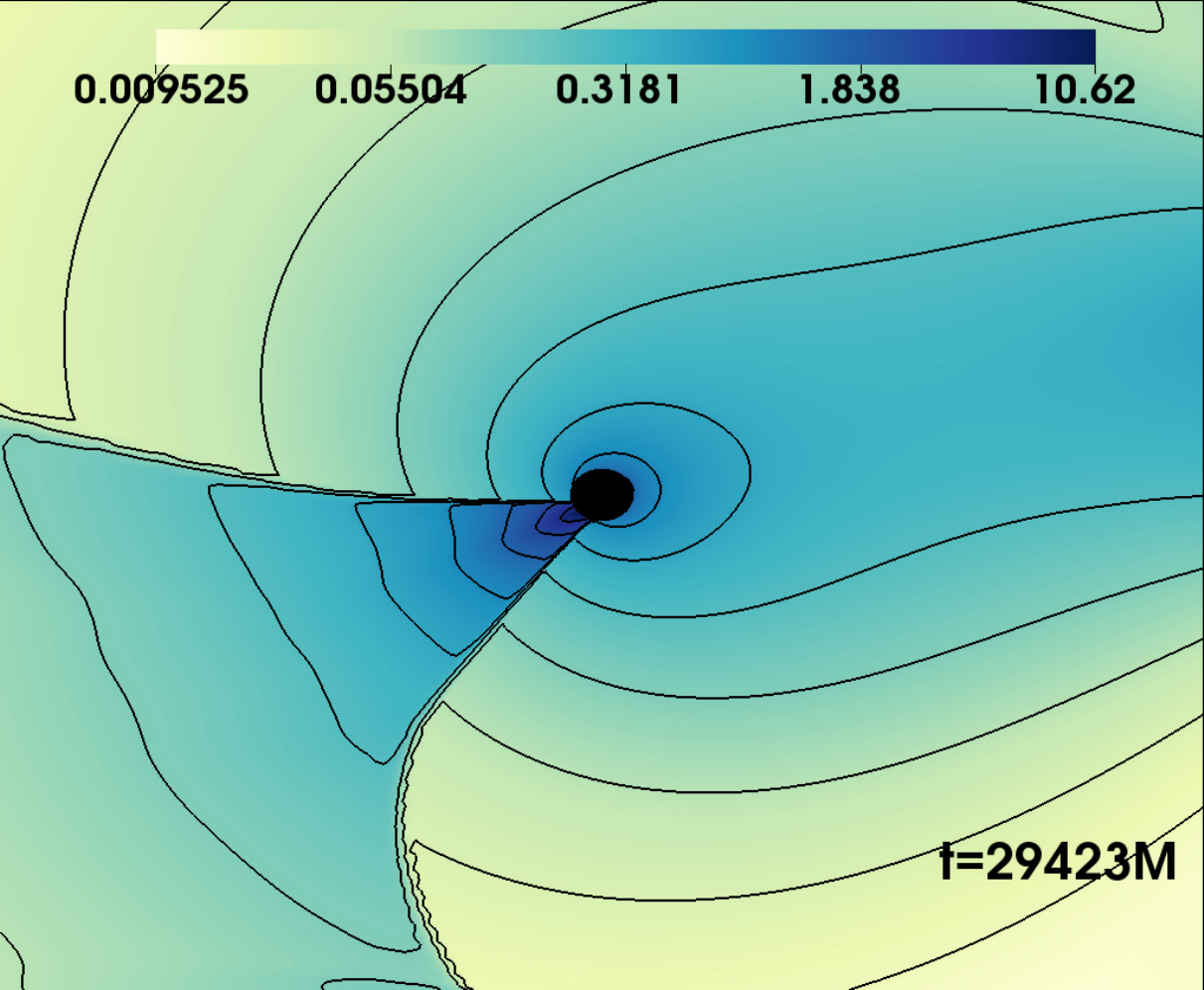,width=2.9cm,height=3.0cm}  
  \caption{Snapshots of the logarithmic rest-mass density are provided at different times for different values of the EGB coupling constant $\alpha$ for $a/M=0.9$. The first row shows how the dynamic structure of the accretion disk around the Kerr black hole changes over time during the perturbation process (model $K09$). The second row illustrates the dynamic changes in the disk for the largest $\alpha$ value used in Table \ref{Inital_Con} with $a/M=0.9$ (model $K09\_EGB8$), and the third row displays the dynamic changes in the disk corresponding to the largest value for negative $\alpha$ (model $K09\_EGB1$). In each column, the dynamic structure of the disk at times corresponding to the same time steps for different models is displayed, revealing the chaotic nature of the disk's dynamic structure. The dynamic boundary marked here extends from $[x, y]$ to $[-70M, 70M]$.}
\label{Color_log_a09}
\end{figure*}

\begin{figure*}
  \vspace{1cm}
  \center
  \psfig{file=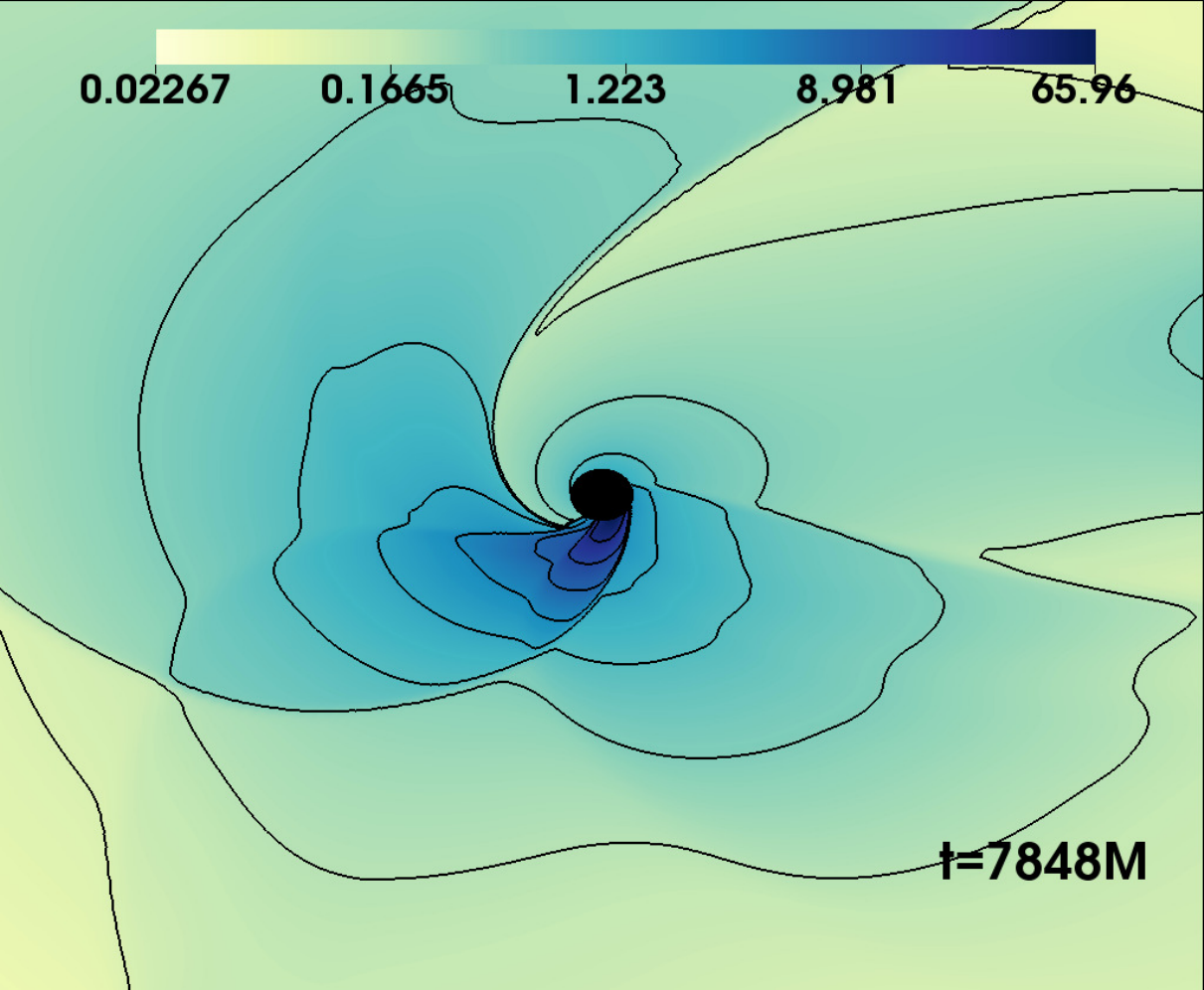,width=2.9cm,height=3.0cm}
  \psfig{file=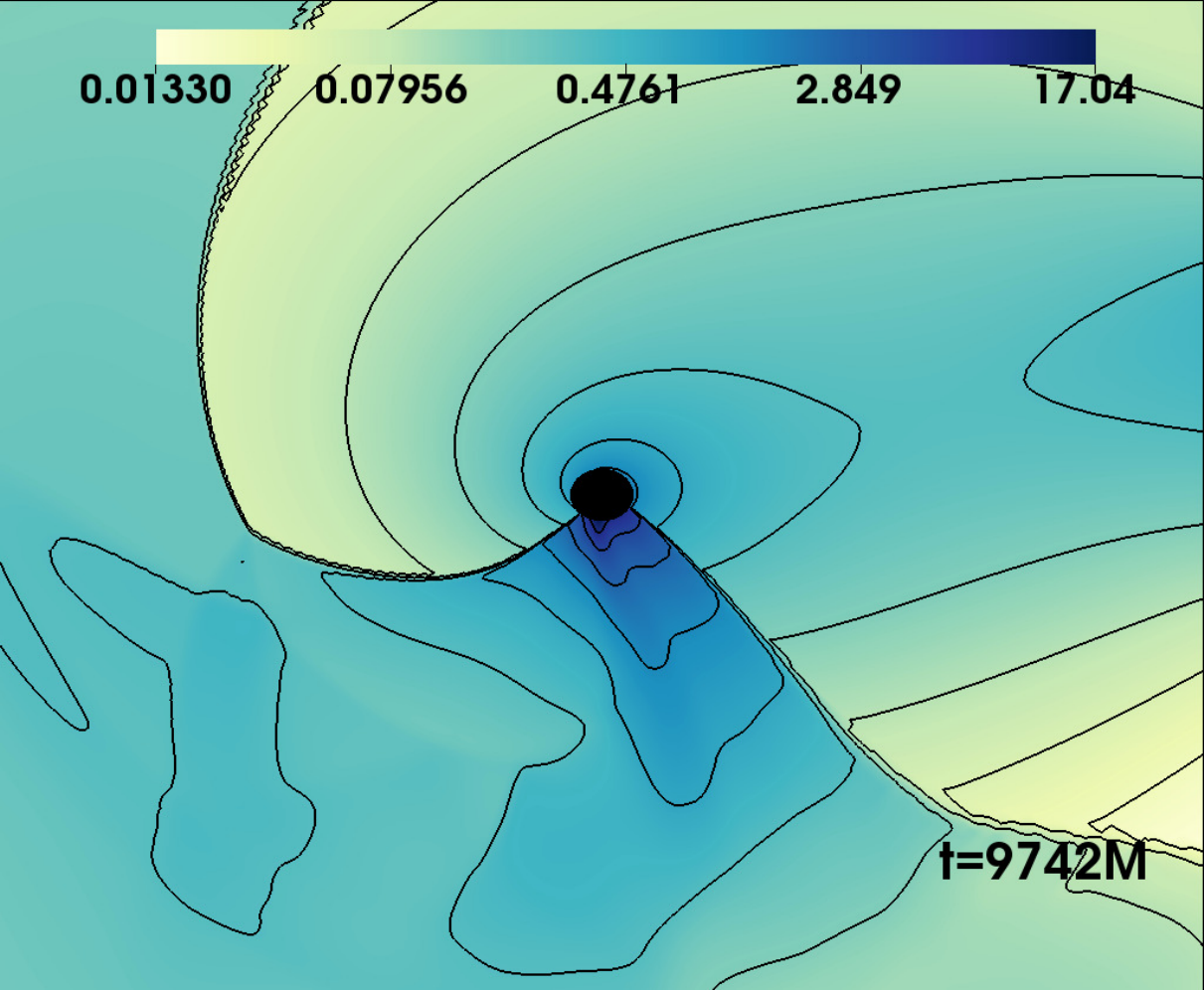,width=2.9cm,height=3.0cm}
  \psfig{file=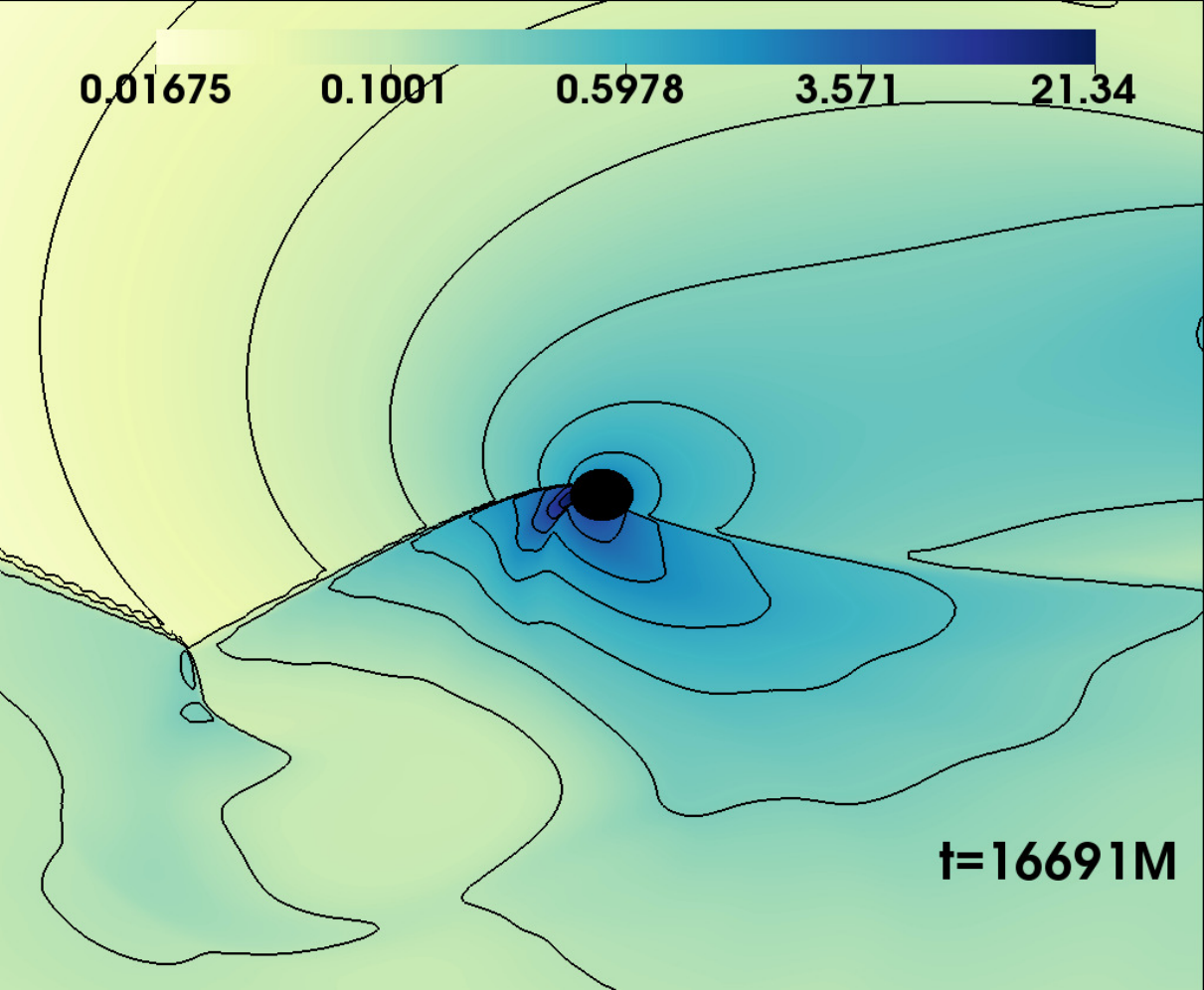,width=2.9cm,height=3.0cm}
  \psfig{file=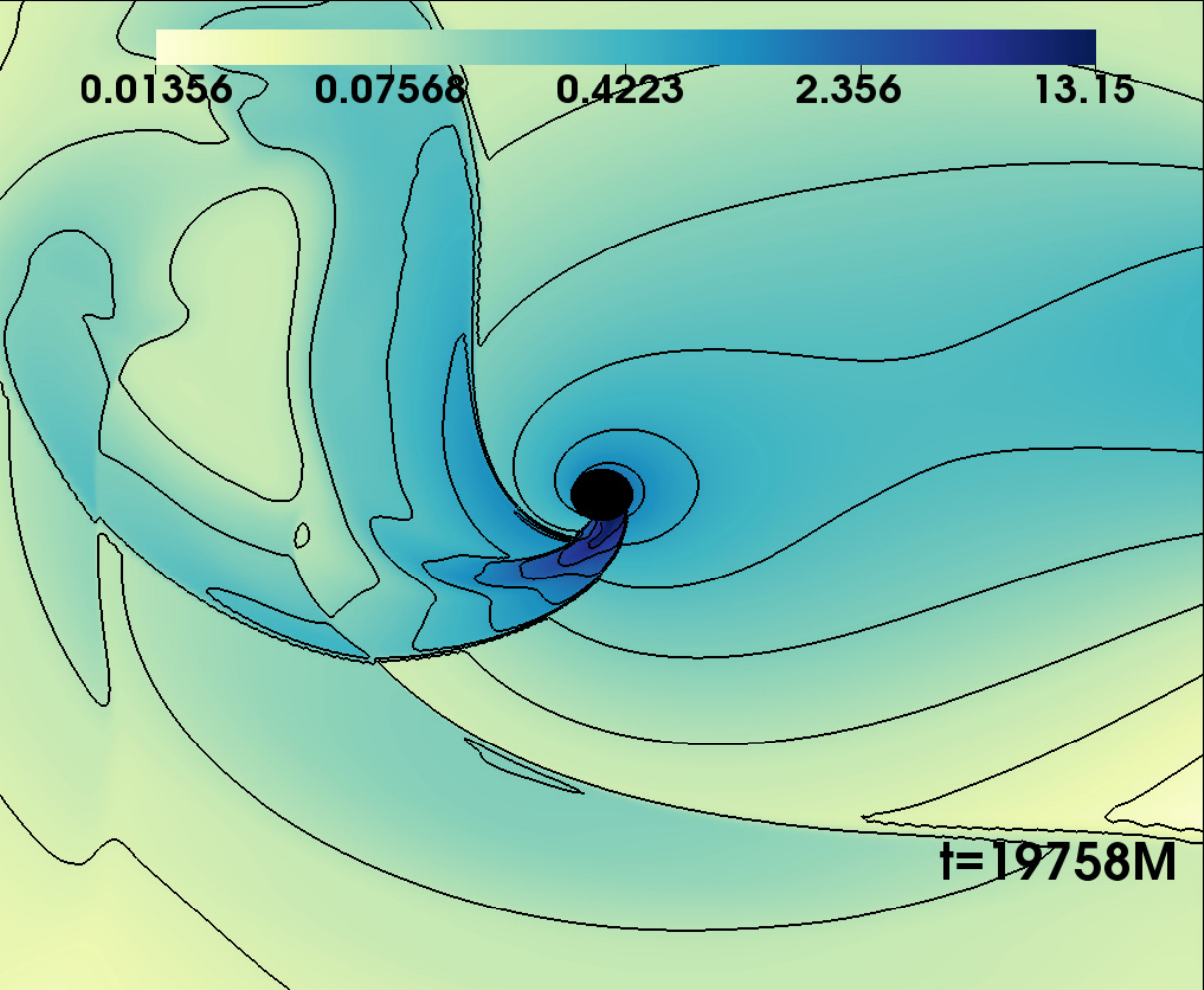,width=2.9cm,height=3.0cm}
  \psfig{file=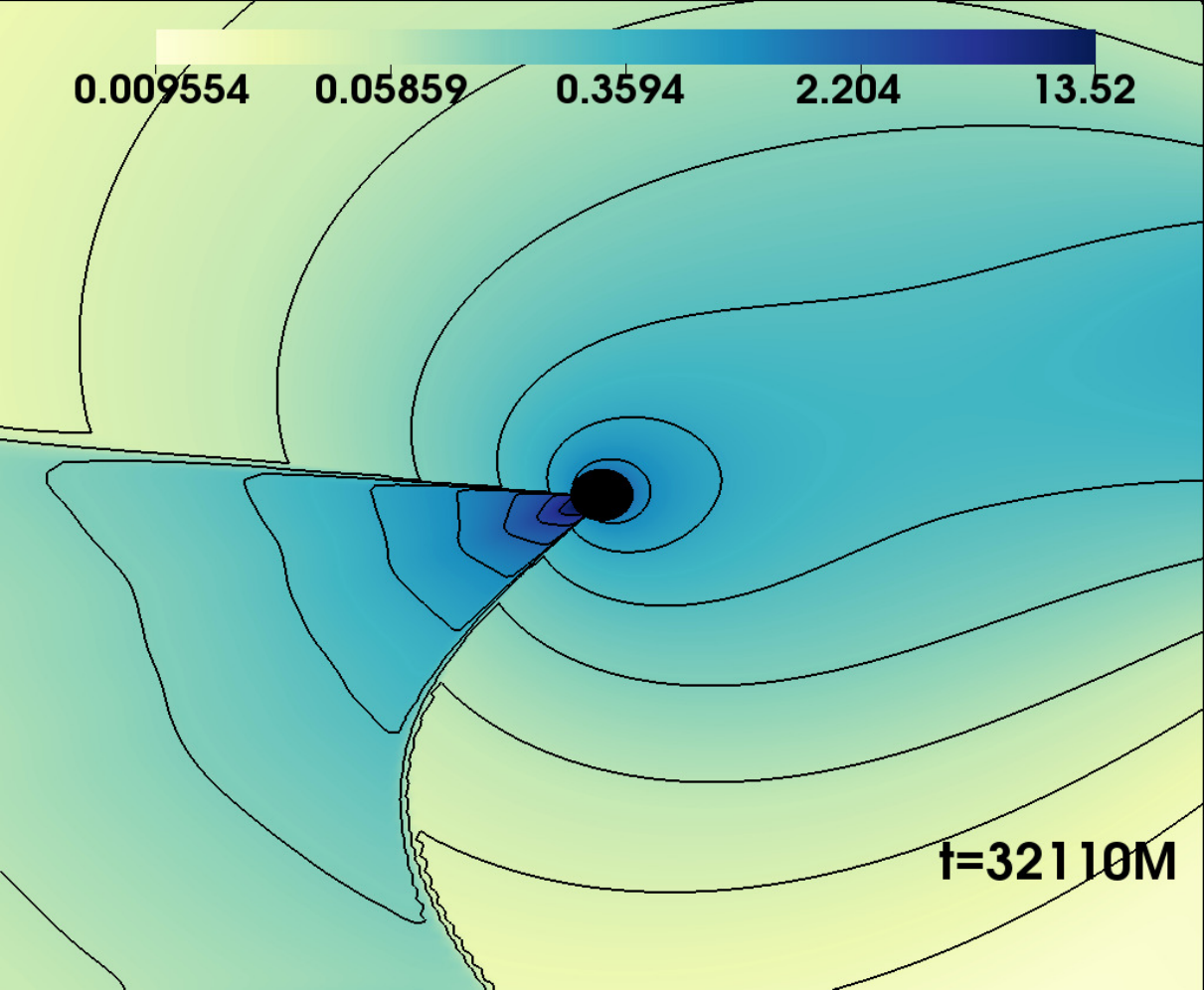,width=2.9cm,height=3.0cm}
  \psfig{file=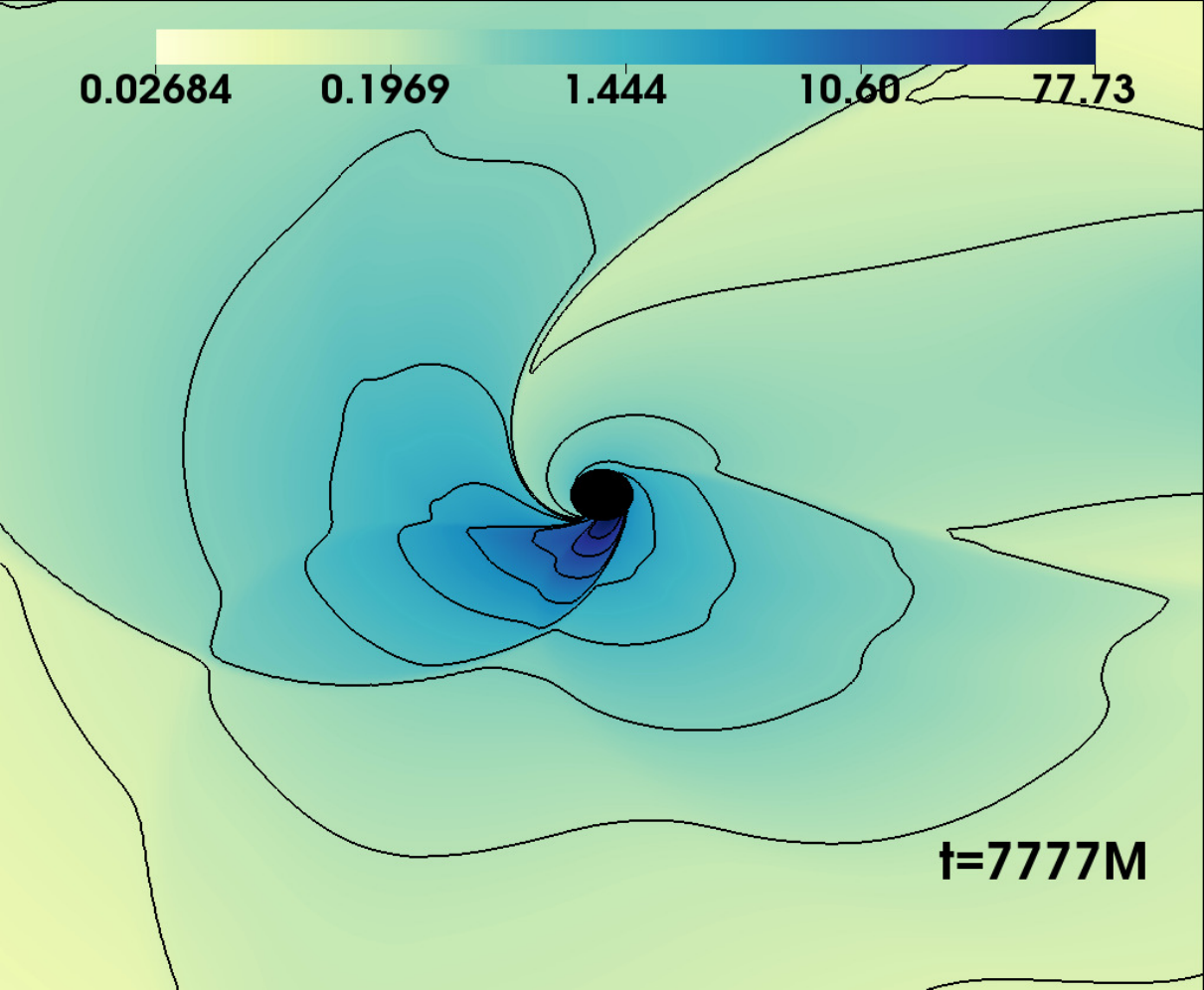,width=2.9cm,height=3.0cm}
  \psfig{file=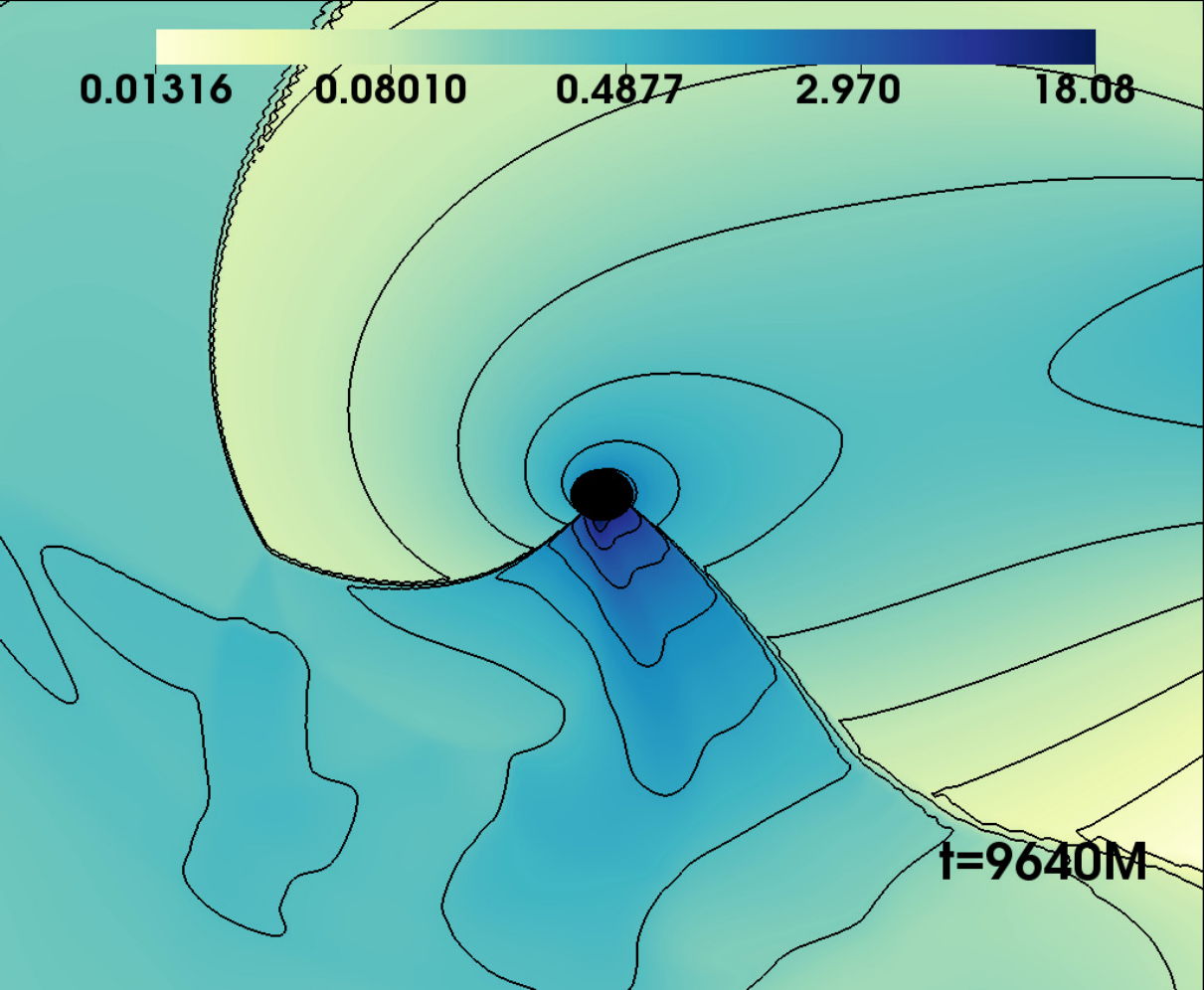,width=2.9cm,height=3.0cm}
  \psfig{file=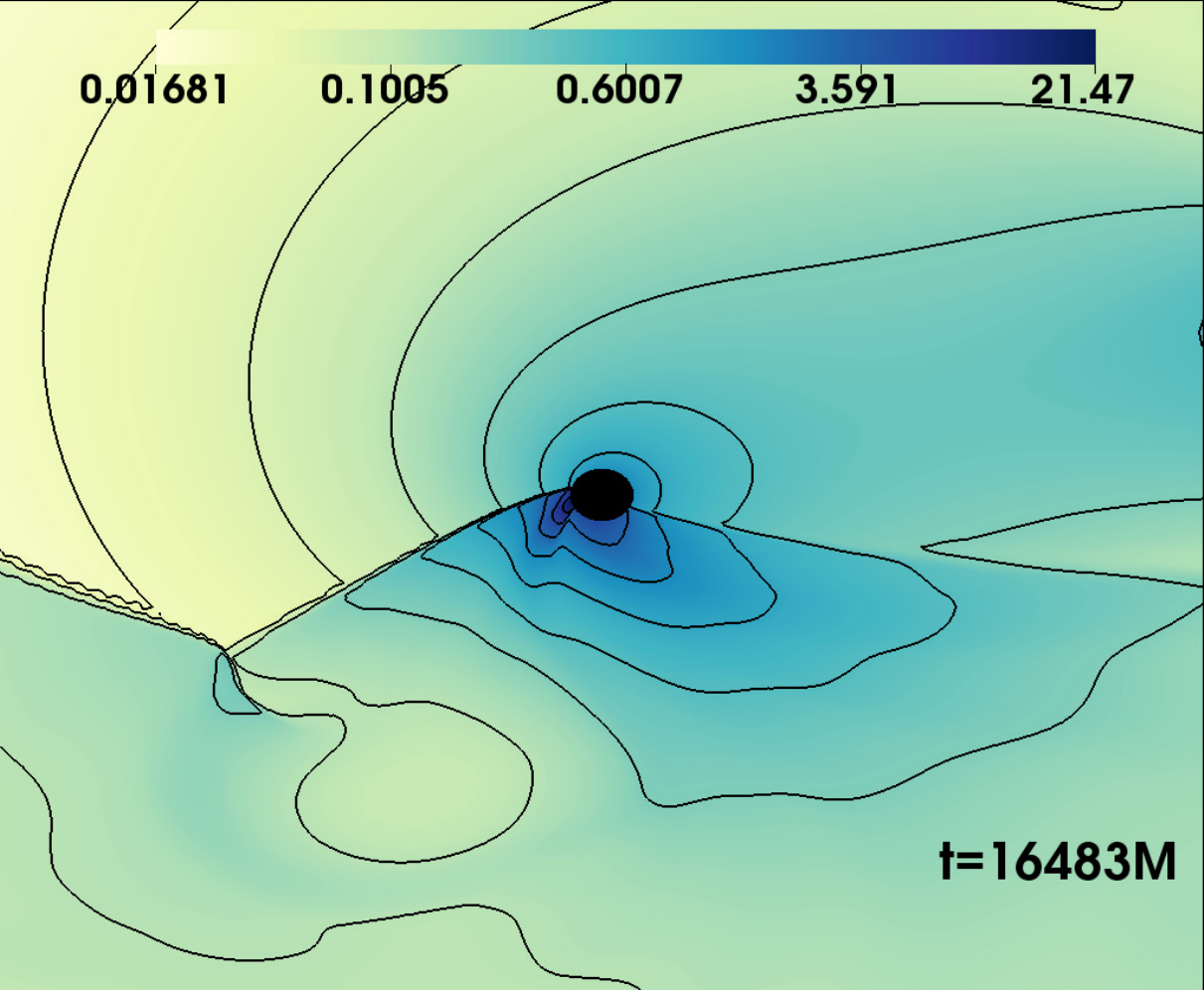,width=2.9cm,height=3.0cm}
  \psfig{file=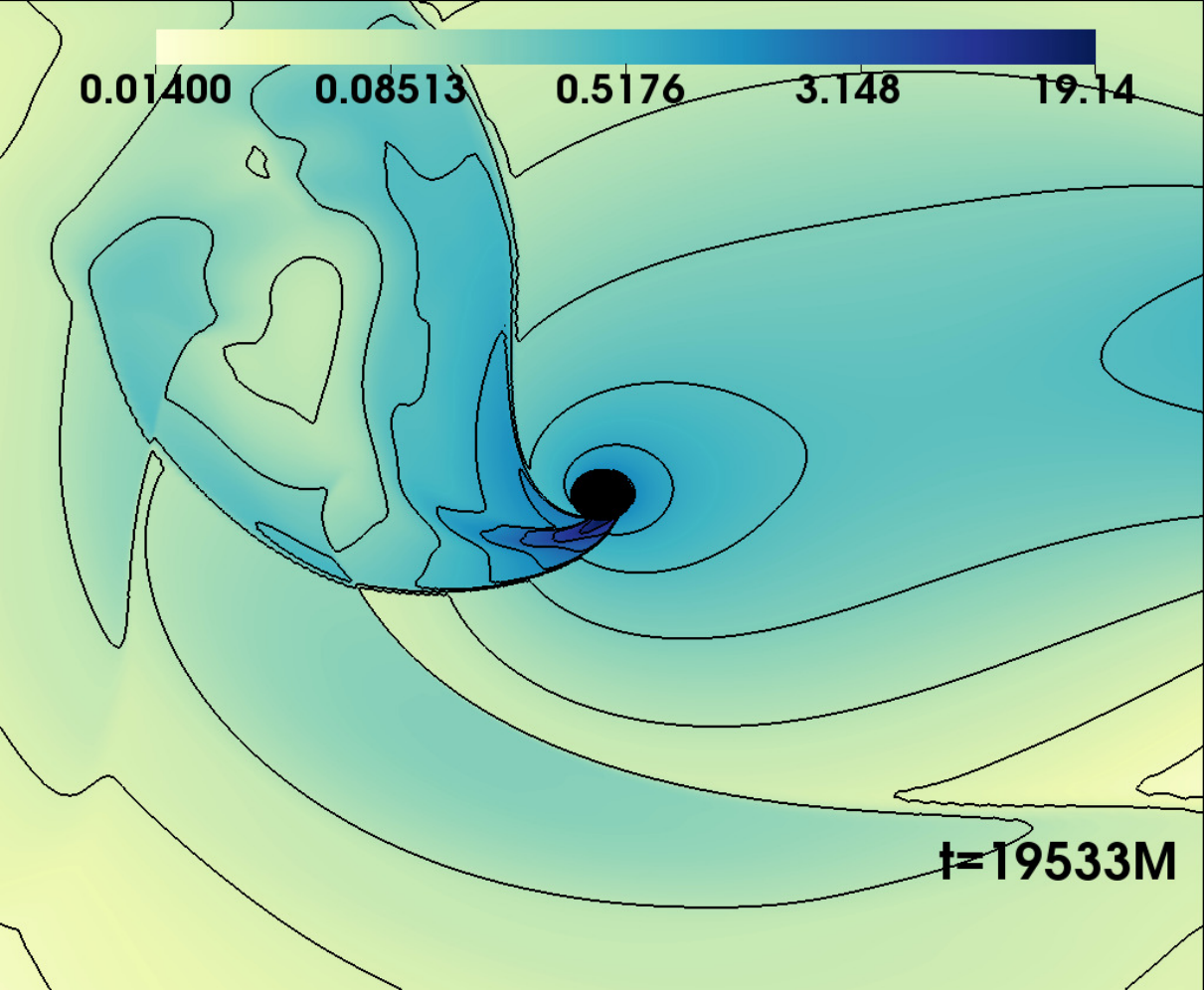,width=2.9cm,height=3.0cm}
  \psfig{file=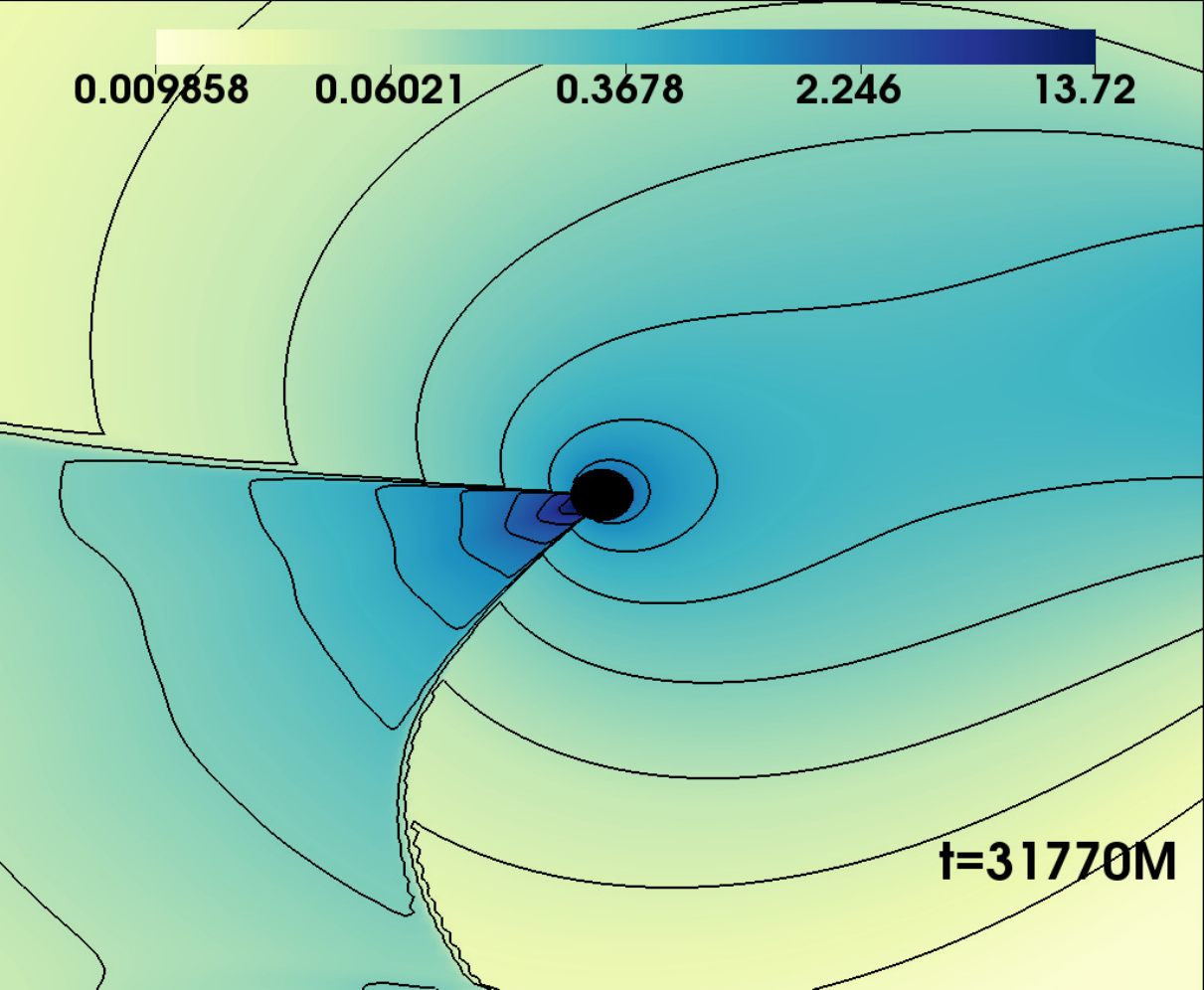,width=2.9cm,height=3.0cm}
  \psfig{file=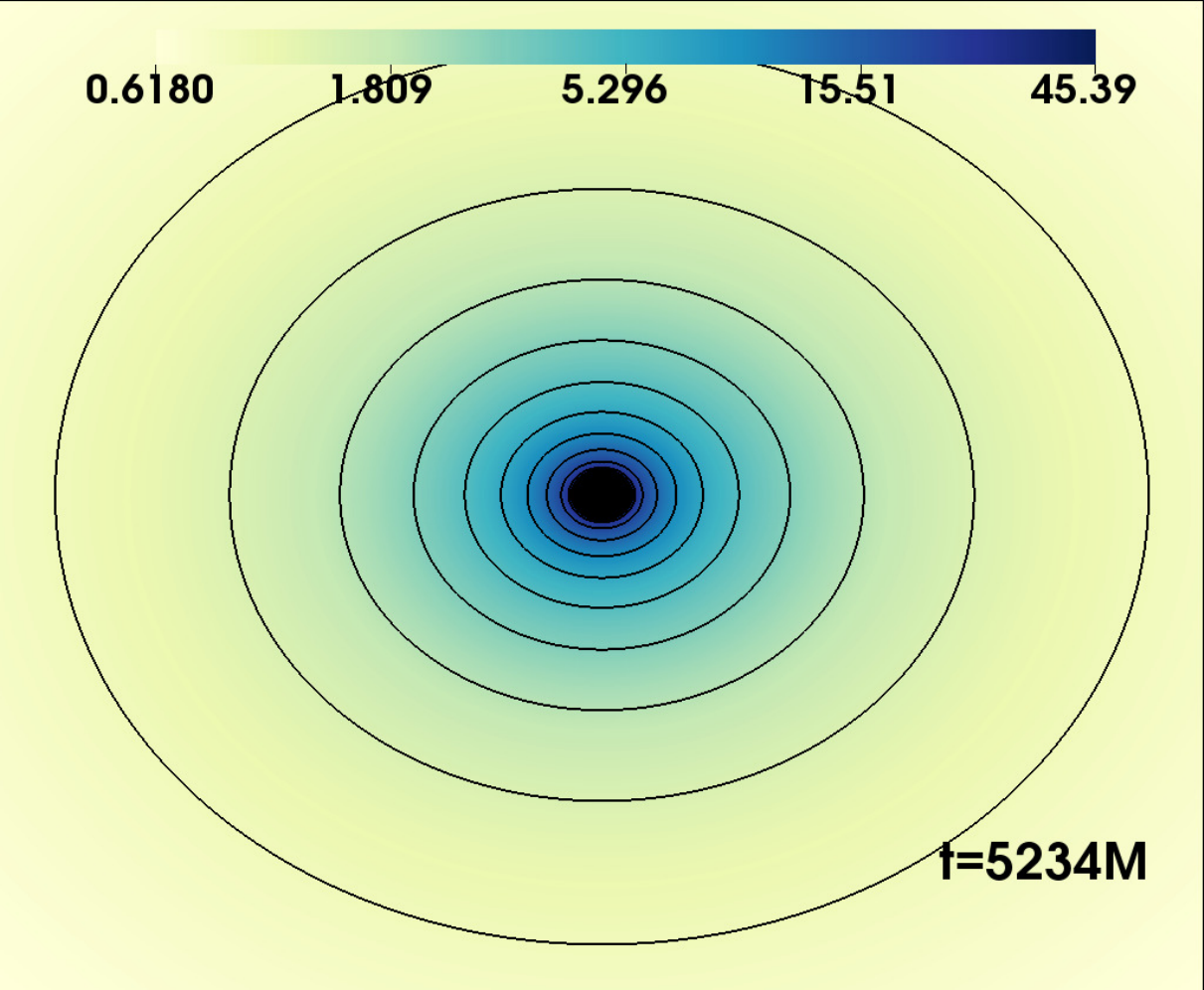,width=2.9cm,height=3.0cm}
  \psfig{file=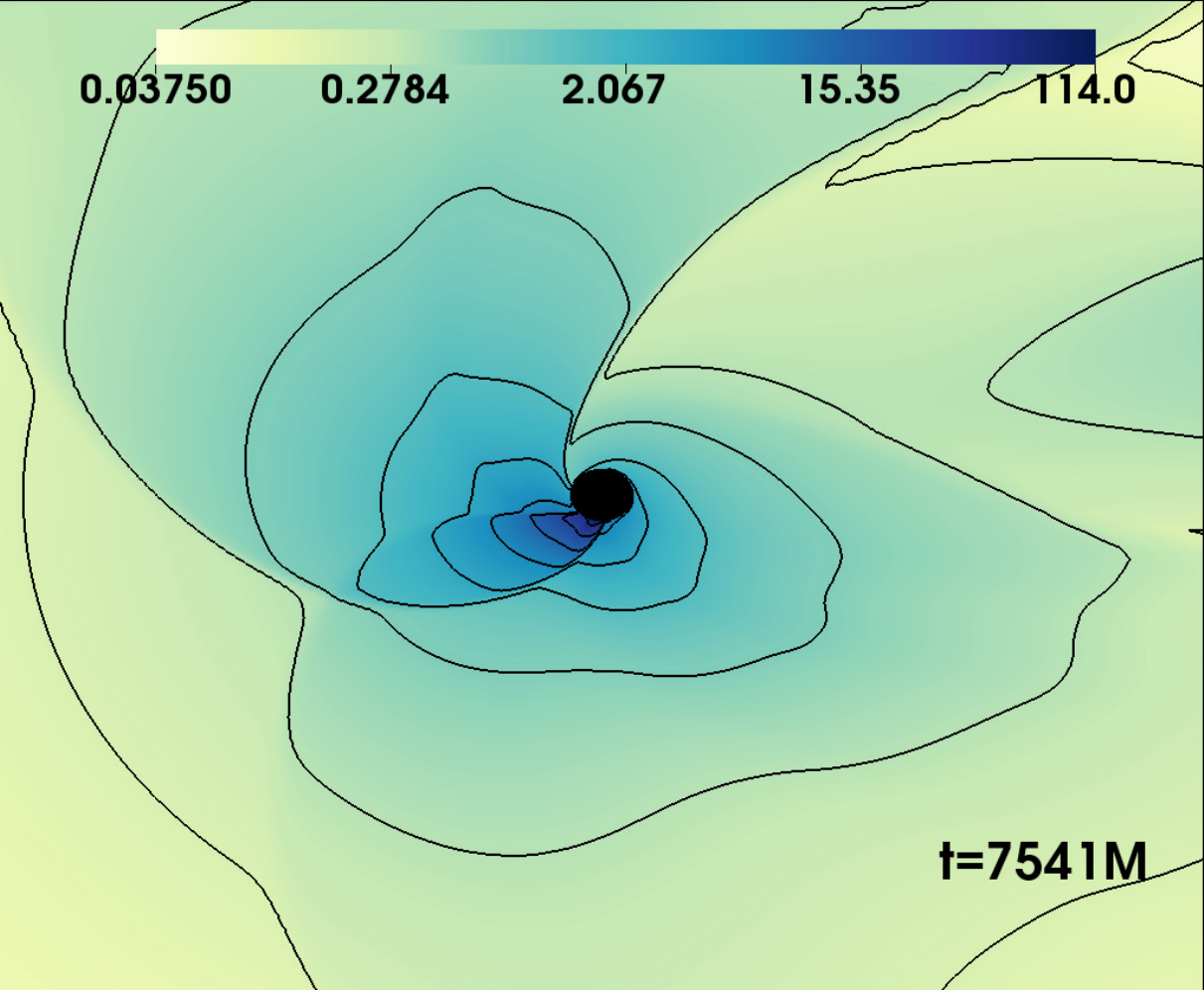,width=2.9cm,height=3.0cm}
  \psfig{file=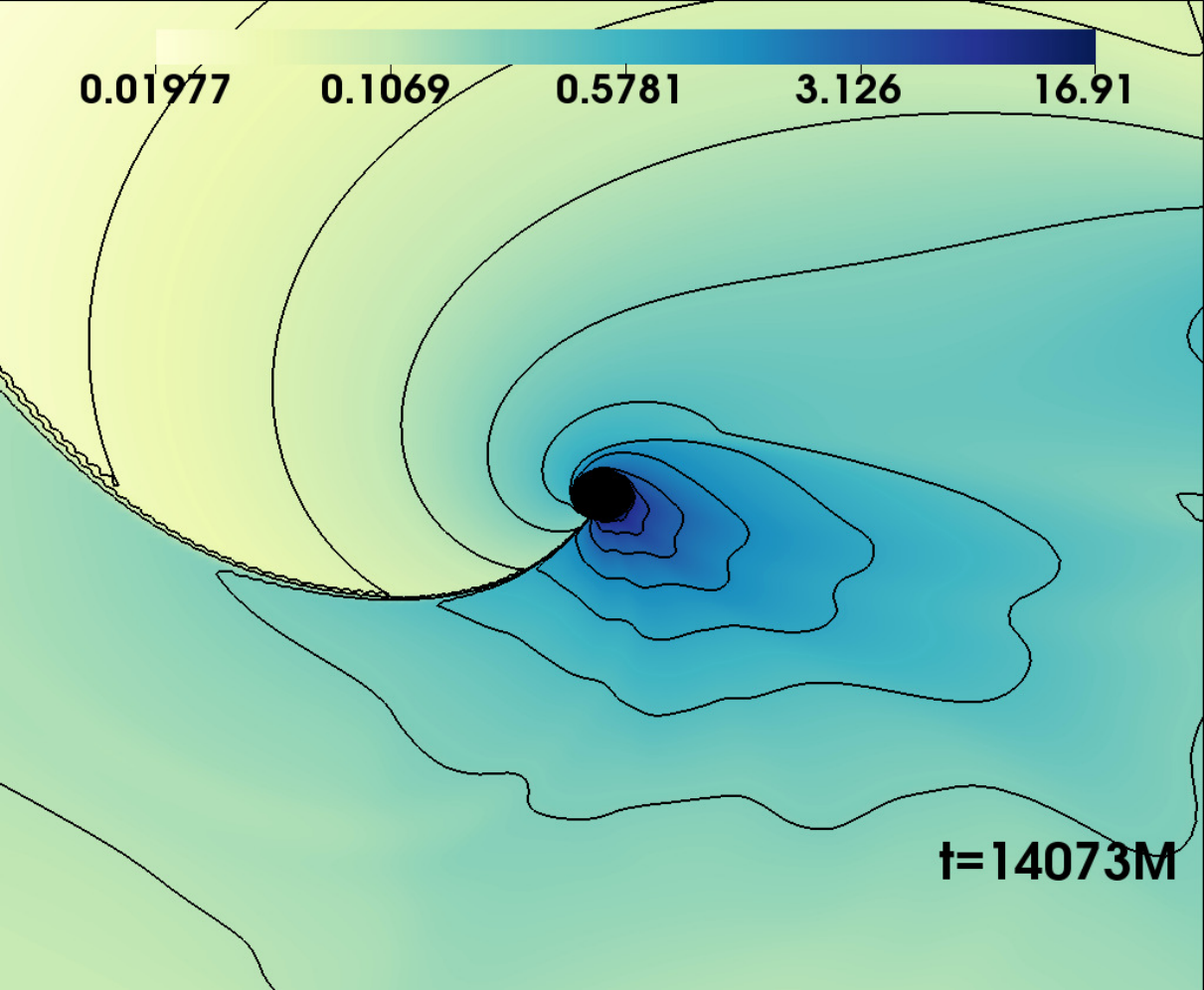,width=2.9cm,height=3.0cm}
  \psfig{file=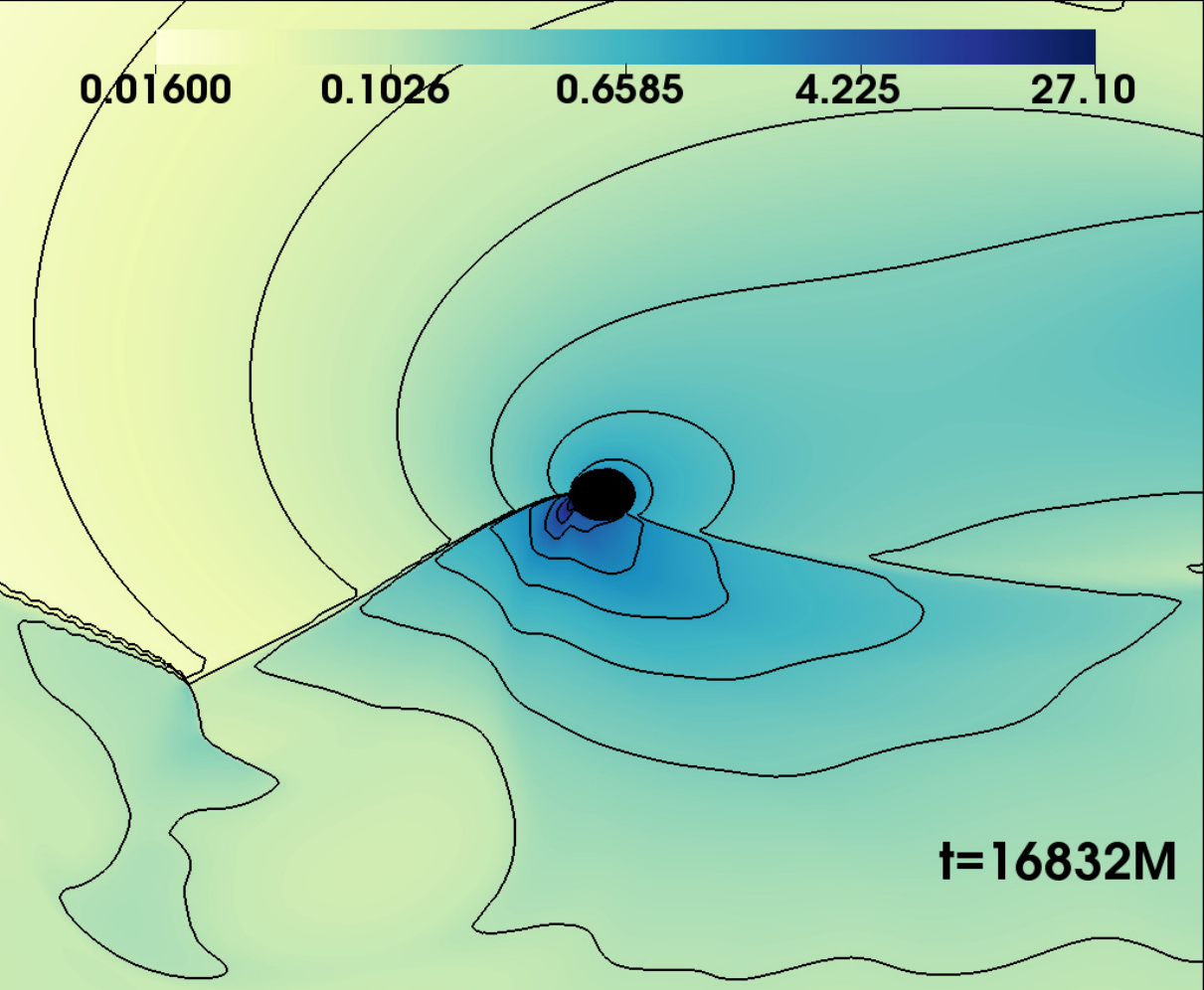,width=2.9cm,height=3.0cm}
  \psfig{file=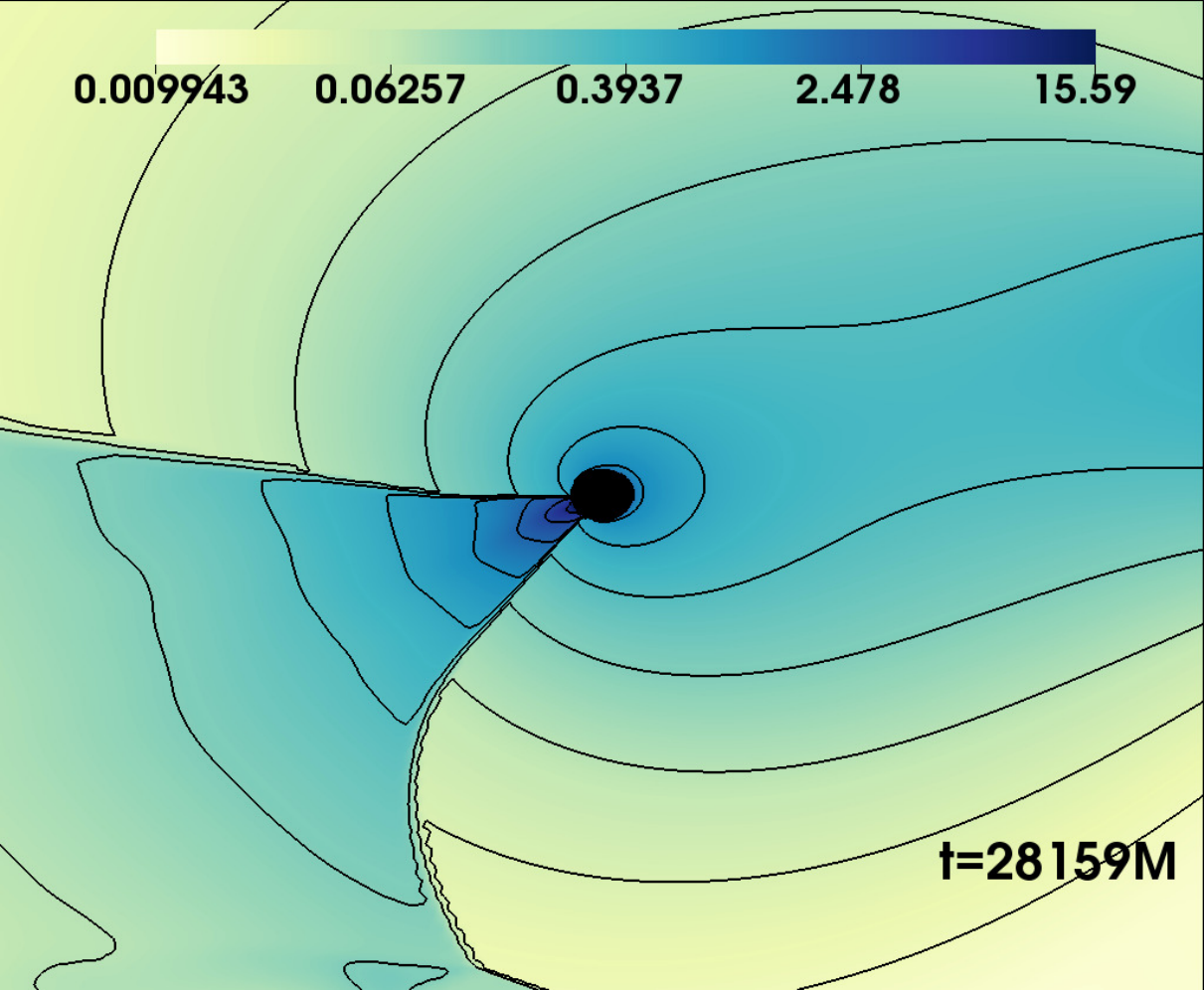,width=2.9cm,height=3.0cm}
  \caption{Same as Fig.\ref{Color_log_a09} but it is for $a/M=0.28$.
  They have been obtained using models $K028$, $K028\_EGB6$, and $K028\_EGB1$.}
\label{Color_log_a028}
\end{figure*}

Figures \ref{Color_log_a09} and \ref{Color_log_a028} illustrate how the mass density of the stable disk changes over time, either immediately before being perturbed or immediately after being perturbed. These snapshots are selected to correspond to the same time steps between $t=5000M$ and $t=30000M$ for each model. The steady-state disk is perturbed at $t=6000M$ in all models. As observed in the figures, the disk responds to the perturbation, initially exhibiting significant changes in its dynamic structure. These changes persist for a while before the disk reaches a steady state. Notably, a shock wave forms on the disk around $t=18000M$, initially exhibiting oscillations which transition into quasi-periodic oscillations at approximately $t=21000M$. This phenomenon persists throughout the entire simulation process. From the initial perturbation of the disk until it reaches a steady state, matter falls into the black hole. Simultaneously, the regularly fed disk initially creates a one-armed and then a two-armed shock wave. These arms appear to be mechanisms regularly feeding the black hole.

Figure \ref{Color_log_a09} illustrates the behavior of the accretion disk around a rapidly rotating black hole with a spin parameter $a/M=0.9$ under different gravities (Kerr and EGB), while Fig. \ref{Color_log_a028} demonstrates the dynamics of a perturbed disk around a slowly rotating black hole with a spin parameter $a/M=0.28$ under various gravities. In both figures, the first rows depict changes in the dynamics of the disk around the Kerr black hole, while the second and third rows illustrate how these disks respond to perturbations in EGB gravity for the positive and negative maximum values of $\alpha$. Overall, the disk response to perturbations exhibits similar dynamic structures and shock wave formation in all models, but significant differences are observed in the strength of oscillations, the amount of matter accreting onto the black hole, and the time it takes for these phenomena to occur. These differences lead to variations in the physical characteristics of the $QPOs$. As clearly seen in the figures, the times corresponding to the same time steps are shorter in EGB gravity compared to Kerr. This indicates that the disk structure and formation in EGB exhibit a more chaotic behavior. Additionally, the increase of $\alpha$ in the negative direction significantly intensifies the chaotic behavior.

Finally, when comparing Figs. \ref{Color_log_a09} and \ref{Color_log_a028}, it can be observed that the rotation parameter of the black hole affects the shock cone of the flow originating from the disk and its impact on oscillations. Especially in regions with strong gravity, such as near the black hole horizon, it has been observed that a rapidly rotating black hole curves the space-time, causing the shock cone attached to the surface to bend. This effect is particularly significant in the strong gravity region, leading to more matter falling into the black hole and resulting in $QPOs$.

As a result of the interaction between the black hole and the disk, matter falls towards the black hole under the influence of gravitational force. The rate at which this matter falls per unit time is known as the mass accretion rate. Mass accretion is typically observed close to the black hole, specifically at the boundary of the disk adjacent to the black hole. This provides insights into various features of the disk. The mass accretion rate indicates whether the disk oscillates over time, the rate of increase in the black hole mass, the frequency, and consequently the energy of electromagnetic radiation in regions with high gravitational pull, and the factors contributing to jet formation around the black hole. Therefore, calculating the mass accretion rate is essential for understanding these features. The general relativistic mass accretion rate on the equatorial plane is calculated using the following equation \citep{Petrich1989, Donmez6}:

\begin{eqnarray}
\frac{dM}{dt} = \int_0^{2\pi}\alpha \sqrt{\gamma}D\left(v^r-\frac{\beta^r}{\alpha}\right) d\phi
\label{MAR1}
\end{eqnarray}.

The mass accretion rate not only provides information about the behavior of the disk structure but also reveals the physical properties of electromagnetic radiation that may arise close to the black hole horizon due to black hole-disk interactions. Here, we compute the mass accretion rate around the black hole at $r=3.88 M$, very close to the black hole horizon at $r=3.7 M$. This proximity to the black hole horizon is crucial because the inner part of the disk, which experiences strong gravity near the black hole horizon, generates $QPO$ modes \citep{Ingram2019, Motta2023}. By doing so, we can compare the numerical results with observations, facilitating a deeper understanding of black hole-disk interactions.

\begin{figure*}
  \vspace{1.2cm}
  \center
  \psfig{file=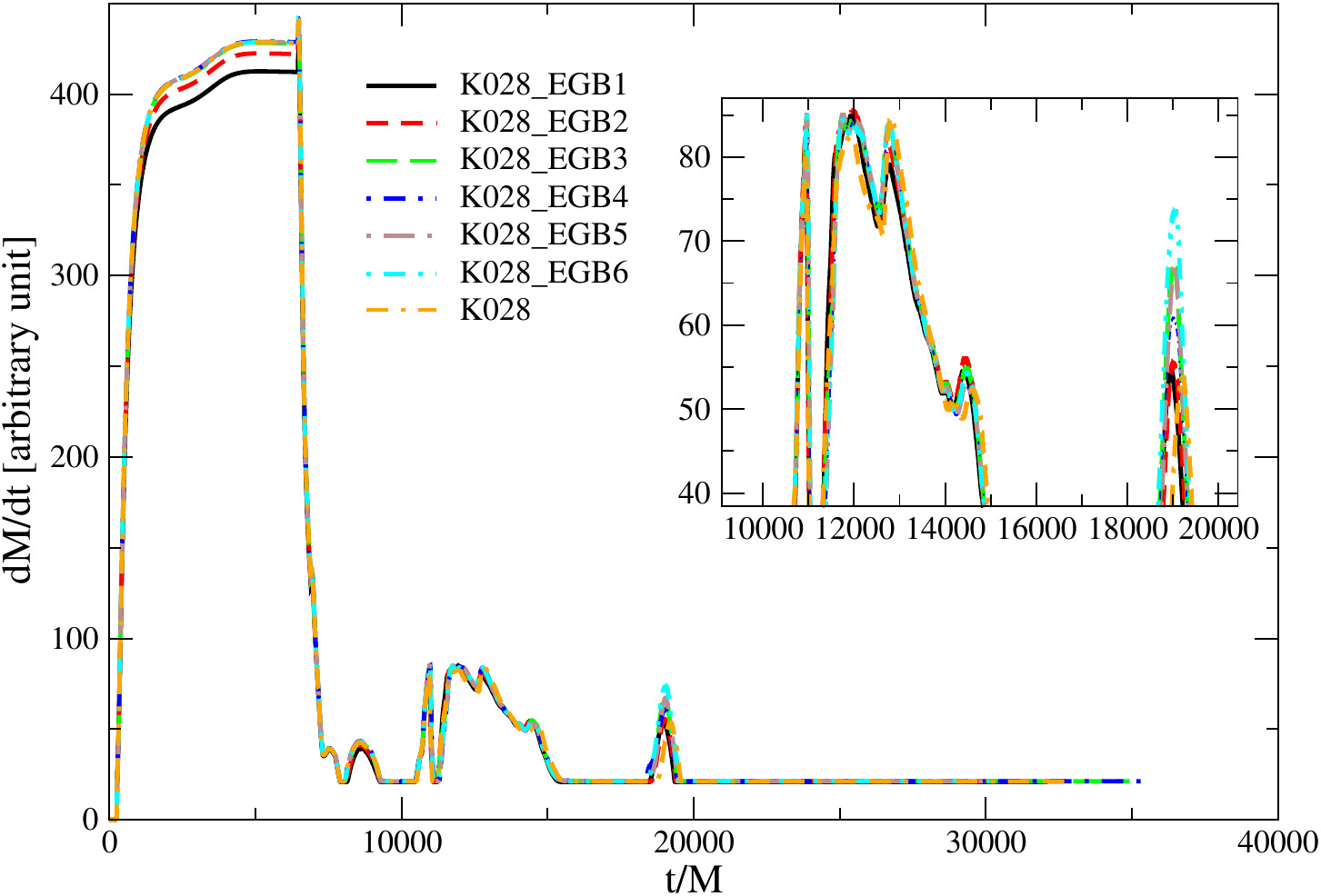,width=7cm,height=6.5cm}
 \hspace{0.5cm}    
  \psfig{file=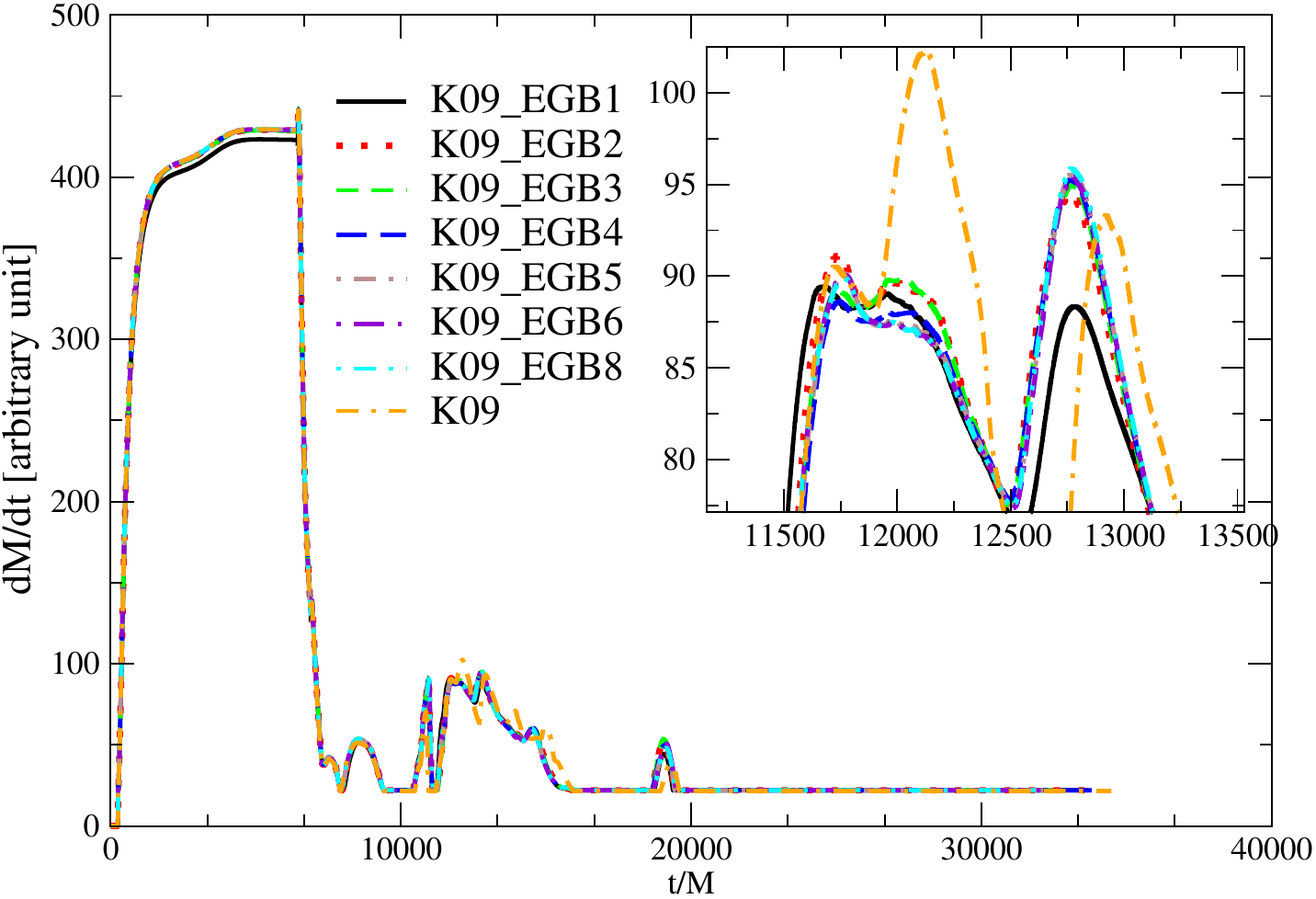,width=7cm,height=6.5cm}  
  \caption{
    The mass accretion rates at the point $r=3.88M$ near the black hole horizon are calculated for different values of the EGB coupling constant $\alpha$. These rates are plotted for the models with $a/M=0.28$ and $a/M=0.9$, as given in Table \ref{Inital_Con}, illustrating how the black hole spin parameter and $\alpha$ influence the mass accretion rate.
 }
\label{mass_acc}
\end{figure*}

In Fig. \ref{mass_acc}, the mass accretion rate is depicted for the cases of $a/M=0.28$ and $a/M=0.9$, illustrating how it changes over time in Kerr and EGB gravities. It is observed that as the initial stable accretion disk forms, the mass accretion rate exponentially increases over time, reaching a steady state around $t=4500M$. At the steady state, which occurs at $t=6000M$, the disk is perturbed. It is important to note that the same perturbation parameters are applied to all models depicted in Fig. \ref{mass_acc}. The perturbed disk rapidly loses mass, with most of the material falling into the black hole, while some is dragged beyond the computational domain. Around $t=7300M$, the mass accretion rate starts exhibiting non-linear oscillations around a specific value ($dM/dt \sim 30$). After undergoing strong oscillations until approximately $t=19500M$, it reaches a quasi-periodic phase with $dM/dt \sim 20$. In other words, the disk around the black hole reaches a steady state and maintains this phase throughout the simulation time ($t=32000M$).

\begin{figure*}
  \vspace{1.2cm}
  \center
  \psfig{file=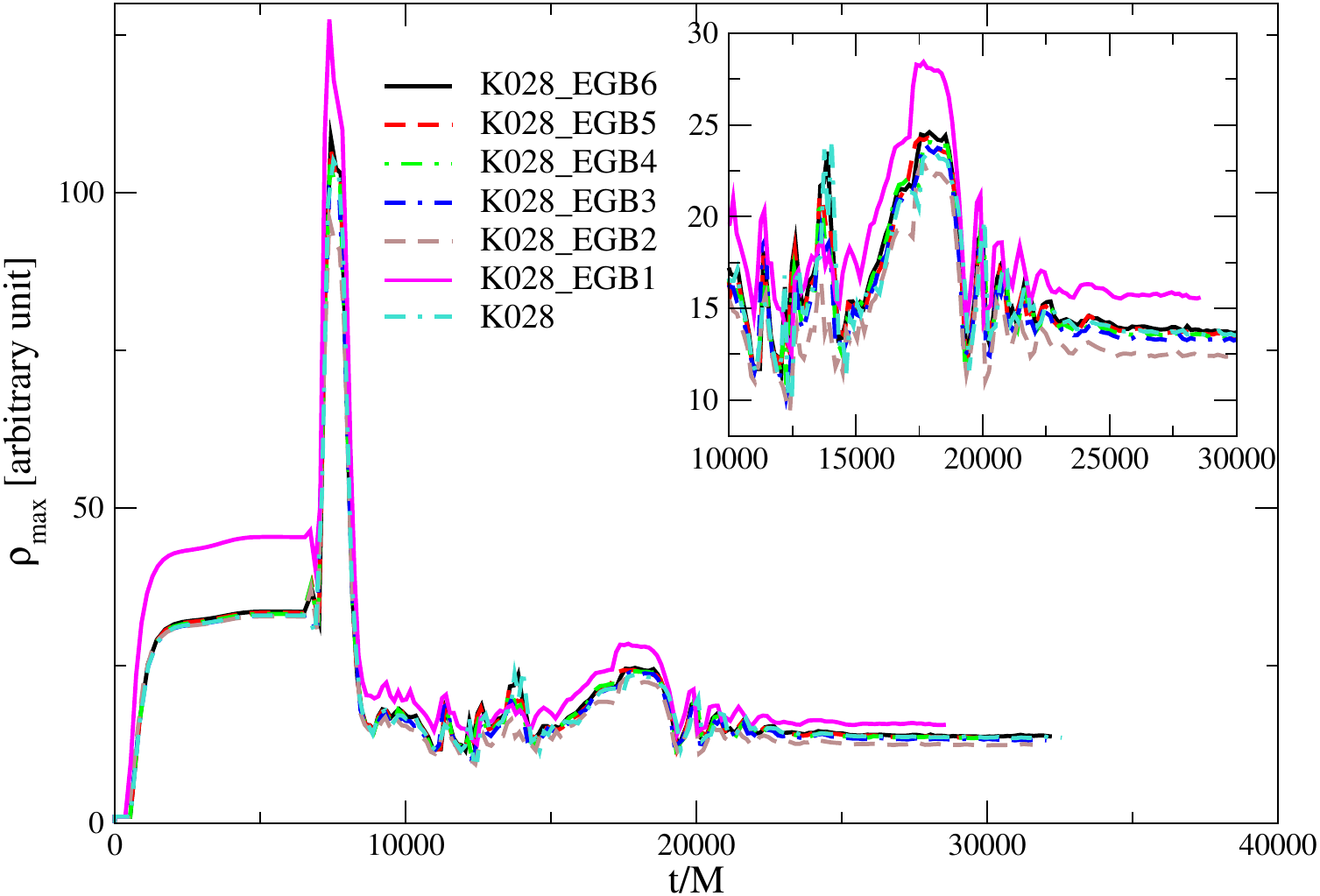,width=7cm,height=6.5cm}
 \hspace{0.5cm}    
  \psfig{file=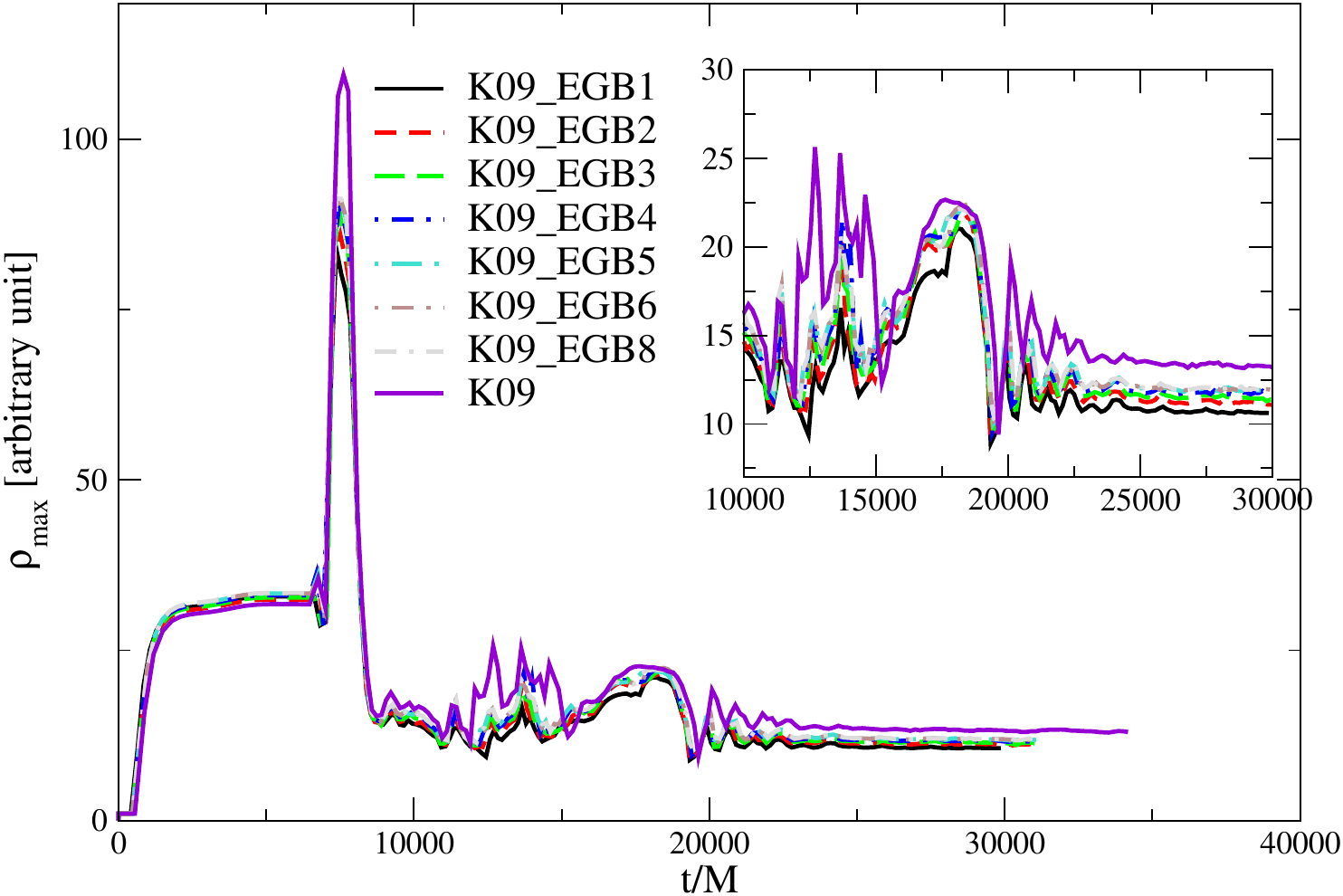,width=7cm,height=6.5cm}  
  \caption{
    The time-dependent variation of the maximum density of the accretion disk around the black holes with spin parameters $a/M=0.28$ and $a/M=0.9$ is depicted. The inner plot provides more details about how the disk behaves until it reaches a quasi-steady-state after being perturbed.
}
\label{max_den}
\end{figure*}

As seen in the highlighted graphic within Fig. \ref{mass_acc}, the effect of the EGB coupling constant, $\alpha$, significantly impacts the dynamic structure of the disk and the accretion of more matter during the stable disk formation process, especially in the case of a smaller black hole spin parameter, $a/M=0.28$. After perturbation, shock waves occur on the disk, and the disk returns to a steady state. In this scenario, the effect of $\alpha$ is more pronounced in the case of $a/M=0.9$, suggesting that $\alpha$ has a greater influence on mass accretion and oscillations within the disk when shock waves are present. It has been observed that as the value of $\alpha$ increases, the mass accretion rate also increases, causing a corresponding increase in the density of the accretion disk, as seen in Fig. \ref{max_den}. This finding has also been confirmed in \citet{FengLong2022}.

In order to compare the results obtained from Kerr and EGB gravities, both slowly and rapidly rotating black hole models are set with an inner radius of $r=3.7M$. The necessity for this choice is explained in Section \ref{InitialBC}. The inner radius of the computation domain is often slightly away from the black hole horizon in most models, so the distinct effect of the black hole spin parameter may not be very clear, especially in Figs. \ref{mass_acc}, \ref{max_den}, and \ref{Instability}.

However, upon closer examination, it becomes evident that the spin parameter affects the time it takes for the disk to reach the steady state, the amount of rest-mass density, and its influence varies under extreme conditions of $\alpha$. In addition, when the behavior of the rest-mass density is examined for different spin parameter scenarios, it is observed that the black hole's spin parameter has a significant impact, especially in the region below $r=6M$. The effect of the black hole spin parameter is further elaborated in detail in Fig. \ref{time_step}. However, the main goal of this article is not to reveal the effects of different spin parameters but to compare Kerr and EGB gravities, discussing the resulting shock wave structure and $QPOs$.

Figure \ref{max_den} illustrates how the maximum density of the accretion disk around the black hole changes throughout the entire simulation for both $a/M=0.28$ and $a/M=0.9$. The results presented here are consistent with the mass accretion results shown in Fig. \ref{mass_acc}. As mentioned earlier, and as can be seen more clearly here, $\alpha$ has a significant impact on the stable disk formation process. This behavior has also been observed in \citet{Donmez2023}. In particular, it is observed that the maximum density corresponding to the largest negative $\alpha$ values notably diverges from the others. However, although this divergence continues after perturbation, it is not as pronounced as before. For $a/M=0.9$, even though the higher values of $\alpha$ initially do not strongly affect the maximum density, we can observe this impact more distinctly during the $QPOs$ phase.

\subsection{Revealing the Instability of the Accretion Disk by Fourier Mode Analysis}
\label{Fourier_MA}

Various factors such as gravitational influences, thermal conditions, and magnetic fields can cause instabilities in accretion disks. These instabilities have profound effects on disk behavior, influencing the accretion rate onto the central object, radiation emission from the disk, and the generation of jets or outflows. Non-axisymmetric instabilities primarily arise in the equatorial plane due to gravitational influences and can be analyzed through linear perturbation studies of the black hole-accretion disk system \citep{Pierens2021}. To characterize these instabilities, we define the azimuthal wavenumber $m$ and calculate the saturation point using simulation data to determine the Fourier power in density. We employ equations provided by \citet{DeVilliers2002, Donmez4, Donmez2024Univ} to ascertain the Fourier modes $m = 1$ and $m = 2$, along with the growth rate at the equatorial plane (i.e., $\theta = \pi/2$). The imaginary and real parts of the Fourier mode are then analyzed.

\begin{eqnarray}
Im(w_m(r)) = \int_0^{2\pi} \rho(r,\phi) sin(m\phi)d\phi,
\label{FM1}
\end{eqnarray}  

\begin{eqnarray}
Re(w_m(r)) = \int_0^{2\pi} \rho(r,\phi) cos(m\phi)d\phi.
\label{FM2}
\end{eqnarray}

\noindent
The individual mode instability can be found by

\begin{eqnarray}
  P_m = \frac{1}{r_{out}-r_{in}}\int_{r_{in}}^{r_{out}} ln\left( \left[Re(w_m(r))\right]^2 +
  \left[Im(w_m(r))\right]^2\right)dr,
\label{FM3}
\end{eqnarray}

\noindent
where $r_{in}$ and $r_{out}$ represent the inner and outer radii of the computational domain, as described in Section \ref{InitialBC}. However, rather than using the outer boundary position of the disk for $r_{out}$, it is chosen to be approximately $35 M$ to encompass the region where the gravitational field is strong, and therefore, the instability is dominant.

\begin{figure*}
  \vspace{1.2cm}
  \center
  \psfig{file=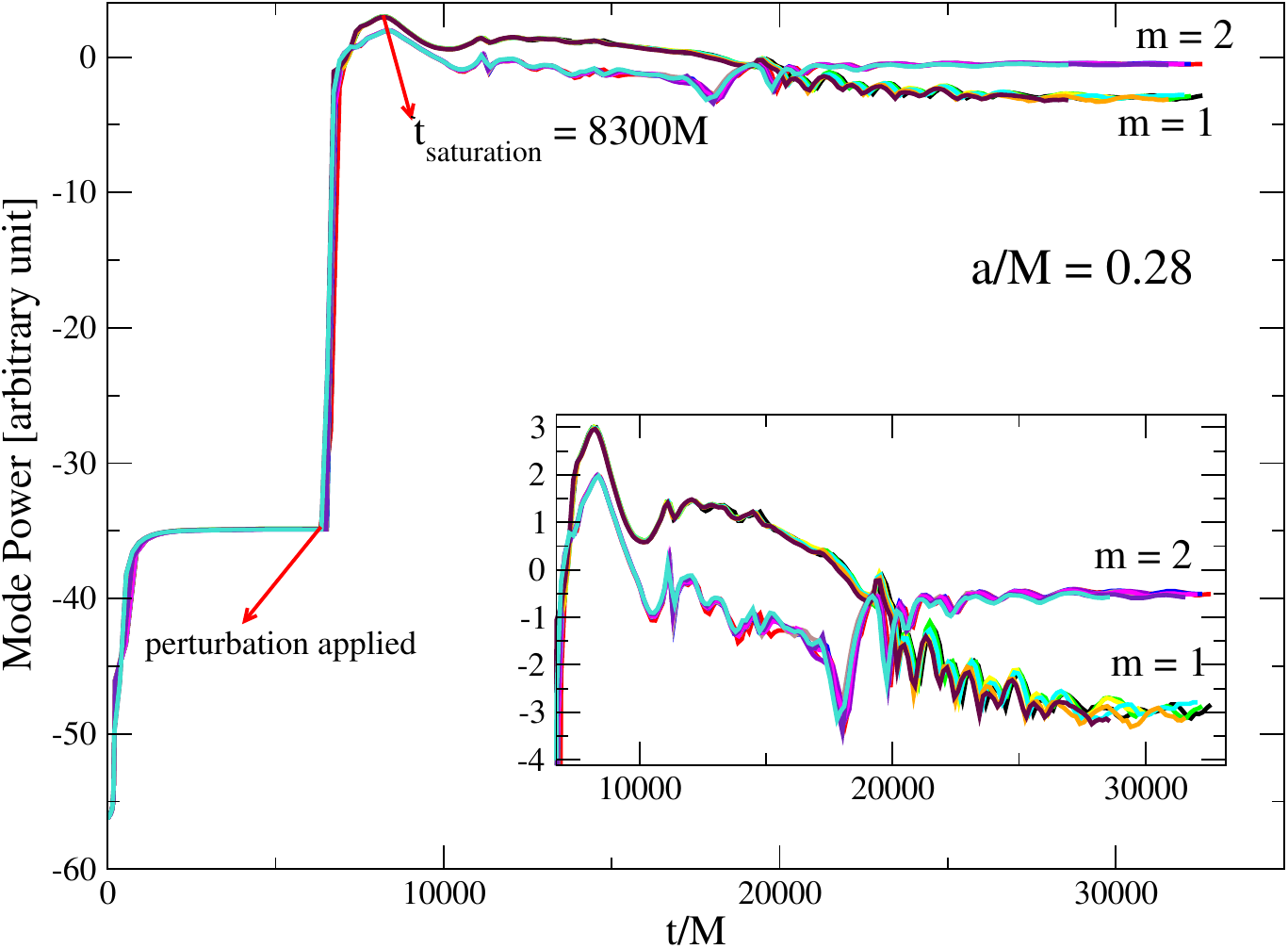,width=7cm,height=6.5cm}
 \hspace{0.5cm}    
  \psfig{file=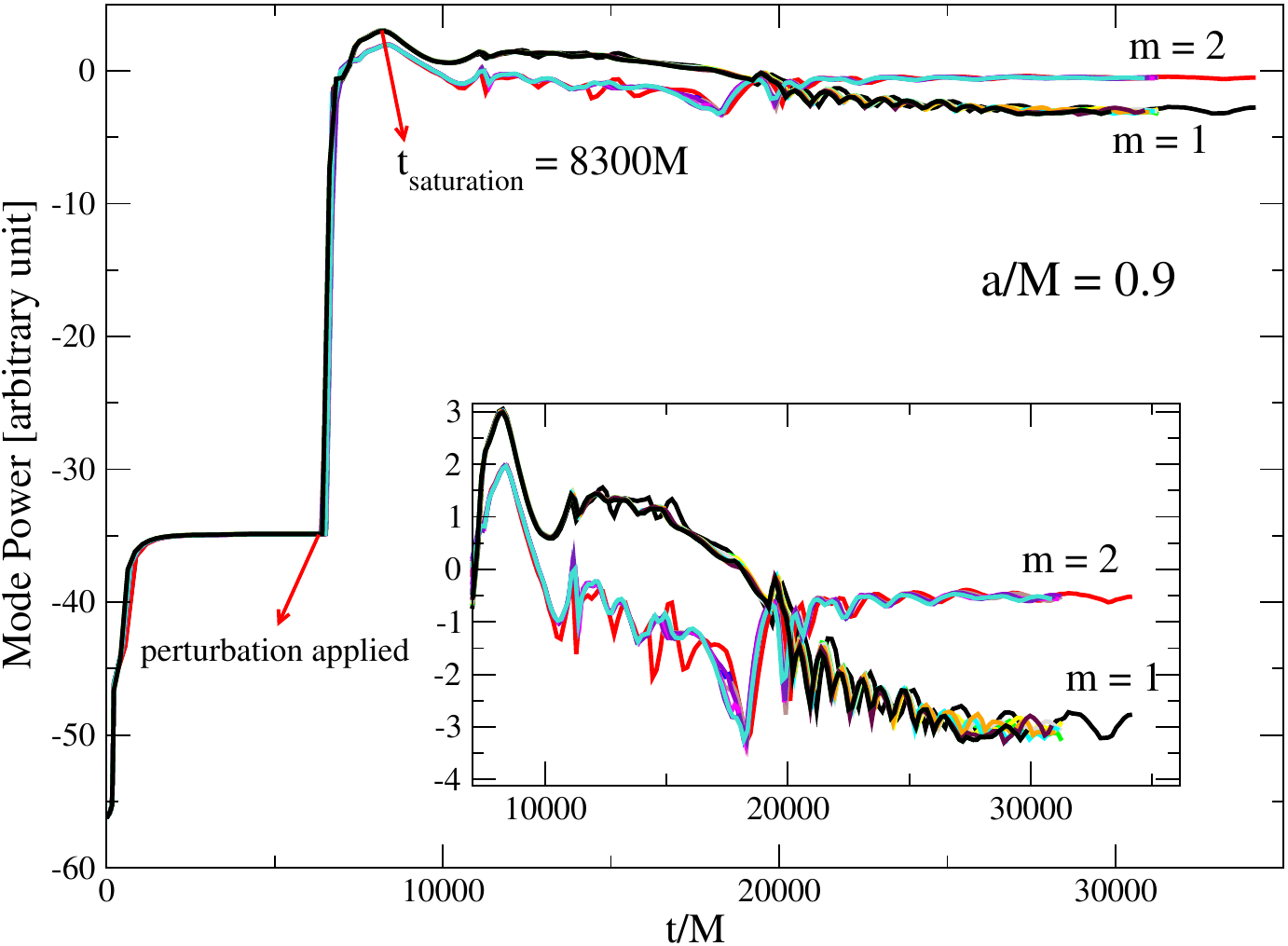,width=7cm,height=6.5cm}  
  \caption{
    The behavior of modes ($m=1$ and $m=2$) is observed for each of the two rotating black hole scenarios ($a/M = 0.28$ and $a/M = 0.9$) given in Table \ref{Inital_Con}. Both modes reach saturation, and it has been observed that the $m=2$ mode reaches a steady-state faster than $m=1$.
}
\label{Instability}
\end{figure*}

Here, we reveal how instability grows in the perturbed stable disk for the models described in Table \ref{Inital_Con}, determining whether it reaches saturation and if it enters a stable phase afterward. The numerical results in Fig. \ref{Instability} demonstrate that the behavior of the $m=1$ and $m=2$ modes is nearly identical for both the black hole spin parameter cases of $a/M=0.28$ and $a/M=0.9$. In the left and right panels of Fig. \ref{Instability}, initially stable disks are observed, reaching a steady state around $t=2000M$, as also depicted in Fig. \ref{Mass_acc}. These stable disks are perturbed using the physical parameters described in Section \ref{InitialBC} for the models listed in Table \ref{Inital_Con} at $t=6000M$. As depicted in Fig. \ref{Instability}, after perturbation, the stable disks become unstable, and for both models, the instability reaches saturation around $t=8300M$. Just before reaching saturation, it is observed that the $m=1$ and $m=2$ modes do not significantly deviate from each other.

Understanding the $m=1$ and $m=2$ modes is crucial not only for revealing the characteristics of the emitted $QPOs$ around a black hole but also for explaining observational results related to accretion disks \citep{ShapiroBook1985, LiNar2004}. The $m=1$ mode illustrates the oscillation caused by matter spiraling inward toward the black hole. On the other hand, the $m=2$ mode can be utilized to describe the long-term structure of the disk and the properties of matter falling into the black hole. These modes are valuable tools in interpreting the behavior of accretion disks around black holes.

Just before reaching the saturation point, both $m=1$ and $m=2$ modes start to separate from each other in all models. For a while, both modes exhibit unstable behavior, moving downwards from the saturation point. Around $t=19000M$, the $m=2$ mode displays stable oscillation, while instability continues in the $m=1$ mode. The $m=2$ mode maintains its stability until the end of the simulation, and towards the final stages of the calculation, the $m=1$ mode is also observed to transition to a steady state. These stable modes oscillating around a fixed value demonstrate the existence of a one-armed spiral shock wave on the disk and a two-armed oscillation. These modes indicate that the black hole is regularly fed by matter, thus showing an increase in the black hole mass over time. At the same time, they continuously lead to the formation of $QPO$ frequencies.

The results presented here contribute to the numerical understanding of the formation of massive black holes in the center of galaxies and active galactic nuclei (AGNs), as well as the physical mechanisms that can lead to observed $QPOs$. This may aid in providing a better explanation for observational data.

\subsection{QPOs in the Perturbed Disk}
\label{QPO}

The X-ray binary system GRS 1915+105, located in the Aquila constellation within our galaxy, is renowned for its core, which harbors a black hole being fed by a companion star. As matter falls from the star towards the black hole, it disrupts the pre-existing stable accretion disk, leading to the emission of high-energy X-rays and the presence of QPOs. The QPOs of GRS 1915+105 exhibit a wide frequency range, spanning from a few Hz to several hundred Hz, depending on their production location within the accretion disk \citep{Sreehari2020MNRAS, Majumder2022MNRAS}. The truncated radius for these QPOs varies, ranging from approximately 1.24 to about 19 gravitational radii ($R_g$) relative to the last stable orbit of an almost maximally spinning black hole. Notably, the frequencies of the C-type QPOs, falling within the range of 2 Hz to 6 Hz, closely conform to the pattern predicted by the relativistic dynamical frequency model. An intriguing observation is the presence of a high-frequency QPO in the range of 67.96 Hz to 72.32 Hz \citep{Sreehari2020MNRAS, Majumder2022MNRAS}, which exhibits similar behavior, implying that both types of QPOs likely stem from the innermost stable circular orbit, sharing a common underlying mechanism. This phenomenon aligns with what is observed in the more frequently observed C-type QPOs \citep{Dhaka2023}. The estimated mass of GRS 1915+105 is around $12.4(+2.0,-1.8)M_\odot$ \citep{Reid2014}.

We conducted a fast Fourier transform to compare the results obtained from numerical modeling with observations of the source GRS 1915+105. Oscillation frequencies were obtained by either sampling the density values at a fixed point inside the shock cone, where oscillations were prominent over time, or by utilizing the time-dependent mass accretion rate measured at the inner boundary of the disk (near the black hole) in the Fourier transformation. Subsequently, to represent the obtained results in the frequency domain, we converted from the geometrized unit to the frequency domain using Equation \ref{FT1}.

\begin{eqnarray}
  f(Hz) = f(M) \times 2.03 \times 10^5 \times \left(\frac{M\odot}{M}\right),
\label{FT1}
\end{eqnarray}

\noindent
where $M$ represents the mass of the black hole. It has been chosen as $M = 12.4M_{\odot}$ to facilitate comparison between the results obtained from the fast Fourier transform of numerical data and the observational results provided in the above sources. The numerical data is sampled around $r=1.9R_g$ (where $R_g$ denotes gravitational radii), in a region where the gravitational field is particularly strong.

\begin{figure*}
  \vspace{1.2cm}
  \center
  \psfig{file=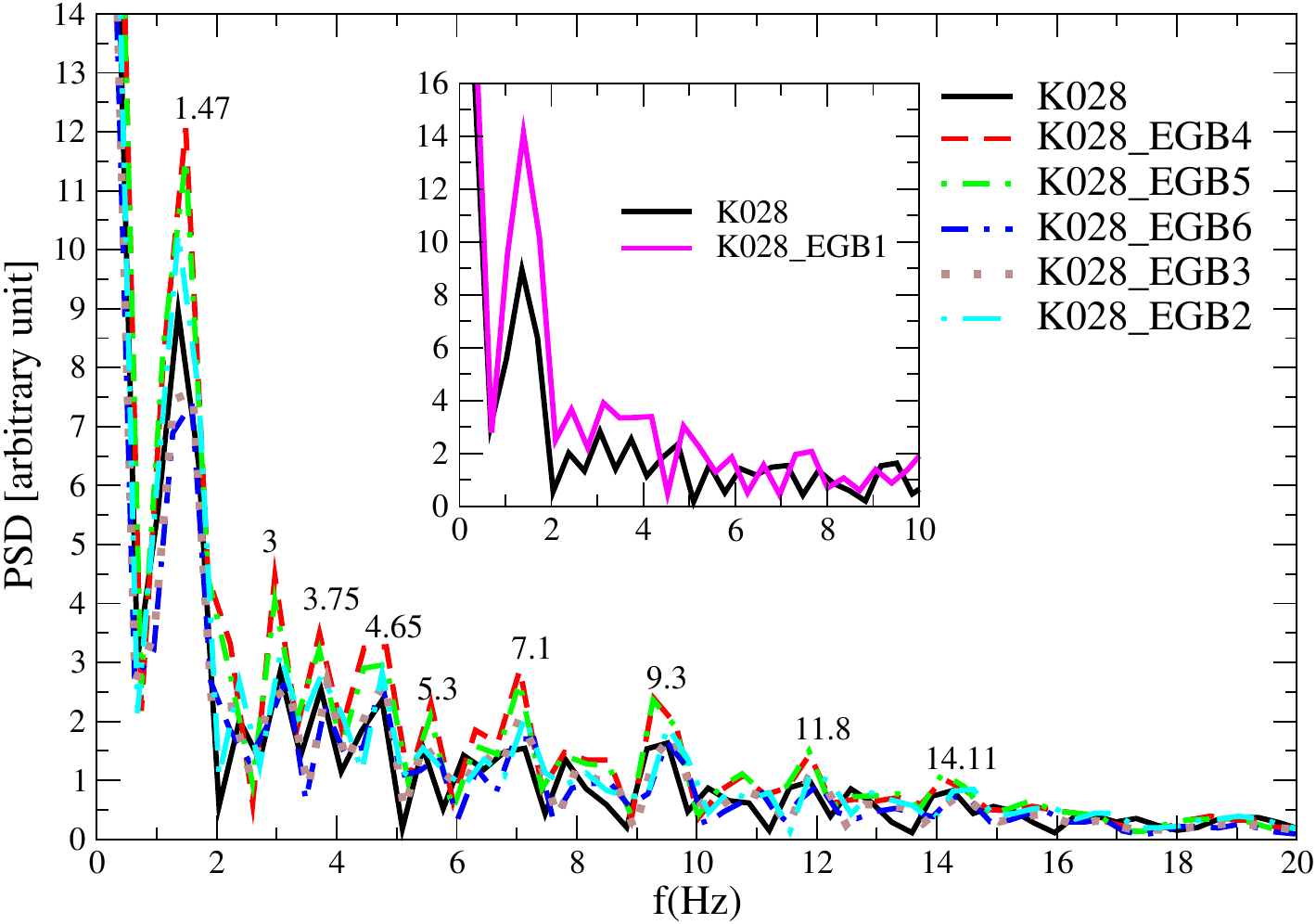,width=7cm,height=6.5cm}
  \hspace{0.4cm}    
   \psfig{file=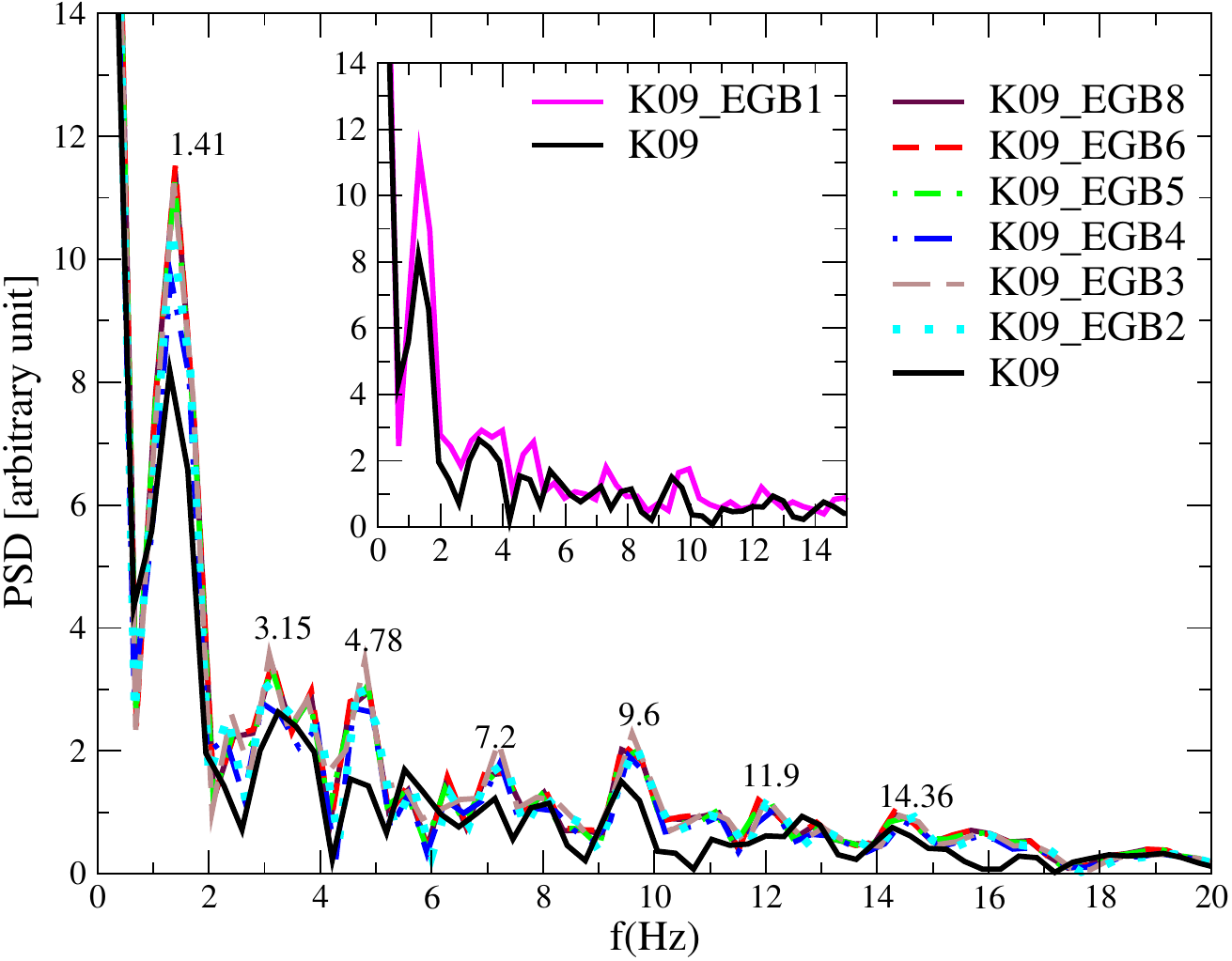,width=7cm,height=6.5cm} 
   \caption{The power spectrum density (PSD) is calculated using the mass-accretion rate at $r=3.88M$ for a black hole with a mass $M=12.4M_{\odot}$, which is the approximate mass of GRS 1915+105. The graph on the left shows the frequencies of QPOs in the region near a rapidly rotating black hole with a spin parameter $a=0.9$, for both Kerr and EGB black holes with different values of the coupling constant $\alpha$. Conversely, the graph on the right displays the QPO frequencies in the vicinity of a slowly rotating black hole with a spin parameter $a=0.28$.
}
\label{QPOs1}
\end{figure*}

The observation of frequencies generated by oscillations due to shock waves in the dynamics of the disk and their comparison with observed sources are essential for understanding the physical mechanisms that cause this emission in observed sources. In this context, the power spectrum of the time-dependent mass accretion rate is calculated to identify the eigenmodes of oscillation during the formation, perturbation, and subsequent processes of the accretion disk. These are computed for different black hole spin parameters and EGB coupling constants, as shown in Fig. \ref{QPOs1}. The left and right panels illustrate how the oscillation frequencies of the disk around slowly and rapidly rotating black holes vary with different EGB coupling constants $\alpha$. For $a/M = 0.28$ and $a/M = 0.9$, it is observed that the oscillation frequencies on the disk are nearly the same for different $\alpha$ values given in Table \ref{Inital_Con}. Only for the highest negative value of $\alpha$, the frequencies are slightly different, believed to be due to the highly chaotic motion discussed in Section \ref{Comparision}.

On the other hand, as observed in Fig. \ref{QPOs1}, the calculated frequencies vary in magnitude depending on the value of $\alpha$. It can be noted that the magnitude is larger for positive values of $\alpha$ compared to negative values. Given that high-magnitude frequencies are easier to observe with telescopes, observation results are more likely to align with positive values of $\alpha$. The oscillation frequencies generated by negative values may not be distinguishable from the background noise frequency.

\begin{table}
\footnotesize
\caption{
Frequencies of the first two genuine modes are measured close to the black hole horizon from the evolution of the mass accretion rate. Here, $a/M$ represents the black hole spin, $f_1$ corresponds to the first genuine mode, $f_2$ corresponds to the second genuine mode, and $Remarks$ indicates for which $\alpha$ values these are valid.
}
 \label{geniue_mode}
\begin{center}
  \begin{tabular}{cccc}
    \hline
    \hline
    $a/M$  & $f_1(Hz)$ & $f_2(Hz)$ & $Remarks$ \\
    \hline
    $0.28$ &  $1.47$   & $3$   & $the \; same \;  for \;  all \;  \alpha \;  values$  \\
    $0.9$  &  $1.41$   & $3.5$   & $the \; same \;  for \;  all \;  \alpha \;  values$   \\
    \hline
    \hline
  \end{tabular}
\end{center}
\end{table}

Nonlinear coupling frequencies occur when multiple fundamental modes combine, resulting in their superposition \citep{Landau1976, Zanotti2005MNRAS, Donmez6}. The first two peaks seen in Fig. \ref{QPOs1} and the modes listed in Table \ref{geniue_mode} are fundamental modes. Various combinations of these modes, such as $f_1 + f_2$, $3f_2$, $3f_1$, $3f_2 + f_1$, and others, generate additional oscillation modes around the black hole. The frequencies resulting from nonlinear coupling stem from nonlinear behaviors in small oscillation cases \citep{Landau1976}. These behaviors have also been numerically demonstrated in simulations by \citet{Zanotti2005MNRAS, Donmez2023arXiv231108388D, Donmez2024Univ,Donmez2024arXiv240216707D} for thick accretion disks around black holes. Conversely, as observed in Fig. \ref{QPOs1}, these couplings give rise to numerous new modes. However, as the frequency value increases, both the amplitude decreases and distinguishing between them becomes more difficult.

Figure \ref{QPOs1} displays the power spectrum frequencies resulting from genuine modes and their nonlinear couplings. Table \ref{geniue_mode} provides genuine modes for each case, observed for every value of $\alpha$. However, for the largest negative values of $\alpha$, while the genuine modes remain the same, the frequencies resulting from nonlinear coupling differ from other $\alpha$ cases. Additionally, the separation of frequencies following the genuine modes for the largest negative $\alpha$ value from those for other $\alpha$ values provides evidence of the accretion disk exhibiting a more chaotic behavior.

As evident from the numerical results, both genuine modes and their nonlinear couplings can provide insights into the origin of the observed low-frequency $QPOs$ in the source GRS 1915+105. More specifically, the numerically observed genuine modes and their several nonlinear couplings fall within the observed range of C-type low-frequency $QPOs$. These findings offer predictions regarding the nature of perturbations that could lead to the formation of these $QPOs$, as well as the potential physical characteristics of such perturbations.

According to observations given in \citet{Sreehari2020MNRAS, Majumder2022MNRAS}, it is known that the GRS 1915+105 source produces high-frequency QPOs. To compare the numerical results with high-frequency observation results, we examined whether shock waves formed around Kerr and EGB black holes generate high-frequency modes. However, no peak was found in the 67-72 Hz region in the PSD analysis. As previously mentioned, only background oscillations have been observed. This suggests that the spiral shock wave structure obtained in the numerical modeling can only explain the low-frequency $QPOs$ of the source GRS 1915+105.

As mentioned before, in Figure \ref{QPOs1} and Table \ref{geniue_mode}, LFQPOs have been observed in both slow and rapidly rotating black hole models. As seen in this table and figure, the influence of the black hole spin parameter on QPOs is minimal. We believe there could be three different reasons for this. Firstly, one reason could be the inner radius of the accretion disk, which we have numerically defined to be at $r=3.7M$. Although the inner radius of the disk is in a region where the gravitational field is very strong, in many cases, the black hole horizon occurs at $2M$ or smaller, which may reduce the impact of the black hole spin parameter on the resulting QPOs. The second and most important reason could be, as seen in \citet{Liu_2021, Dhaka2023}, that observed LFQPOs from the GRS 1915+105 source can occur in a region between the black hole horizon and $r=36.3M$. If the LFQPOs we found here occur far from the horizon and within this region, the influence of the black hole spin parameter on the QPOs could significantly diminish. Lastly, considering that the shock waves we found in numerical simulations are very strong, these waves could play a significant role in the formation and excitation of QPOs, thereby suppressing the influence of the spin parameter. Due to these reasons, I believe the influence of the spin parameter on the resulting QPOs is approximately around $5\%$. Of course, this effect may be more pronounced in HFQPOs, as these frequencies typically occur around $4M$ for the GRS 1915+105.

One of the other important conclusions drawn here is that similar frequencies occur at different values of the EGB coupling constant ($\alpha$) under the same spin parameter condition. This observation is explained both in Table \ref{geniue_mode} and observed in Fig.\ref{QPOs1}. As the value of $\alpha$ increases in the negative direction, it amplifies the resulting frequency, enhancing their observability. On the other hand, the existence of the same fundamental frequencies and their nonlinear couplings at different $\alpha$ values can be attributed to the similarity in the physical structure of shock waves formed around the black hole for all $\alpha$.


\subsection{Effect of Angular Momentum of the Perturbation on  the  Disk Dynamics}
\label{Angular_Vel_effect}

The stable accretion disk around the black hole experiences perturbation at different initial angular speeds (rotational angular momenta) in the vicinity of the black hole, leading to variations in the disk dynamics, the rate of change in the accretion rate, instabilities characteristic of strong shock waves, and consequently, a change in the frequency of the emitted electromagnetic radiation \citep{FengLong2022}.

\begin{figure*}
  \vspace{1.2cm}
  \center
  \psfig{file=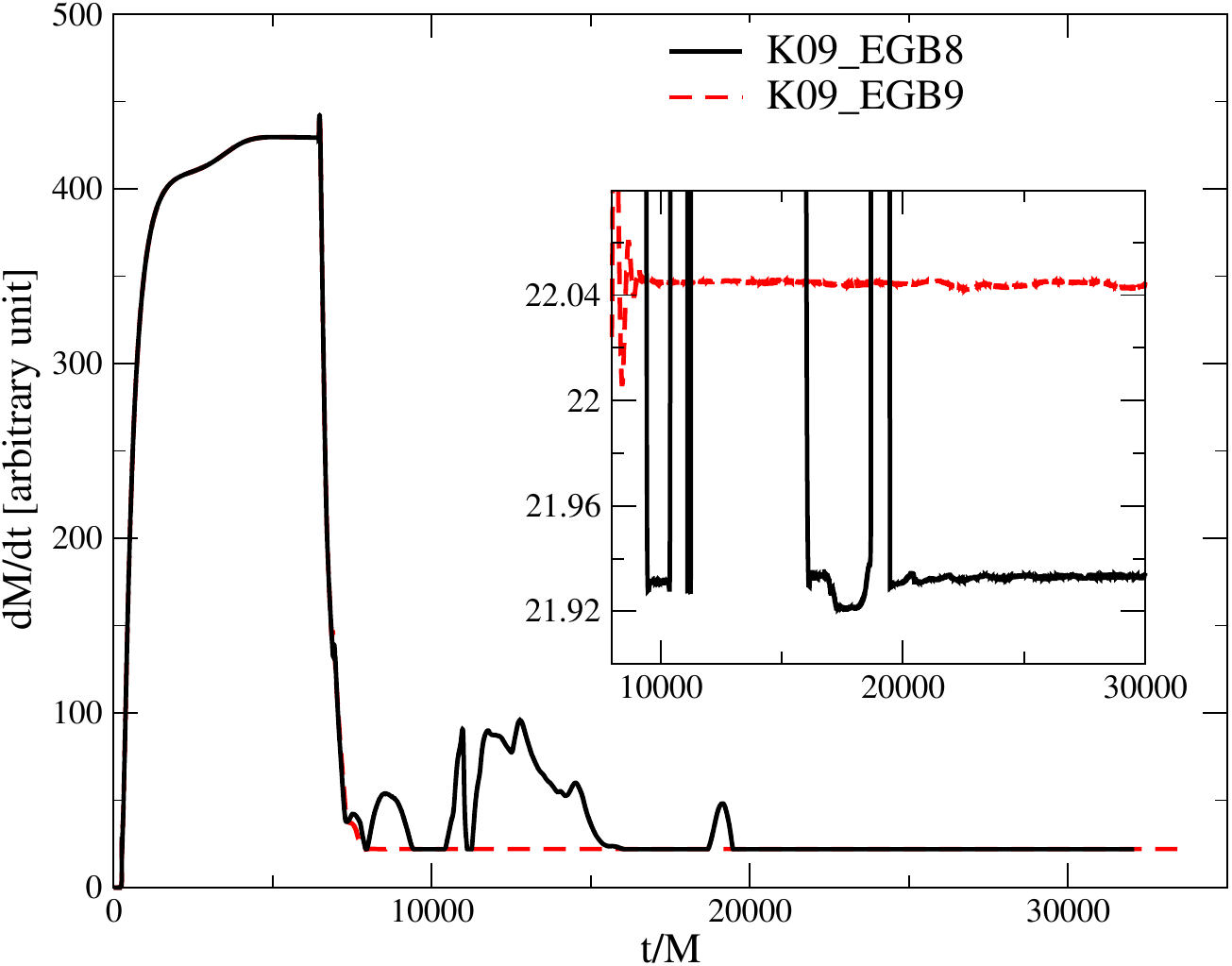,width=7cm,height=6.5cm}
 \hspace{0.5cm}    
  \psfig{file=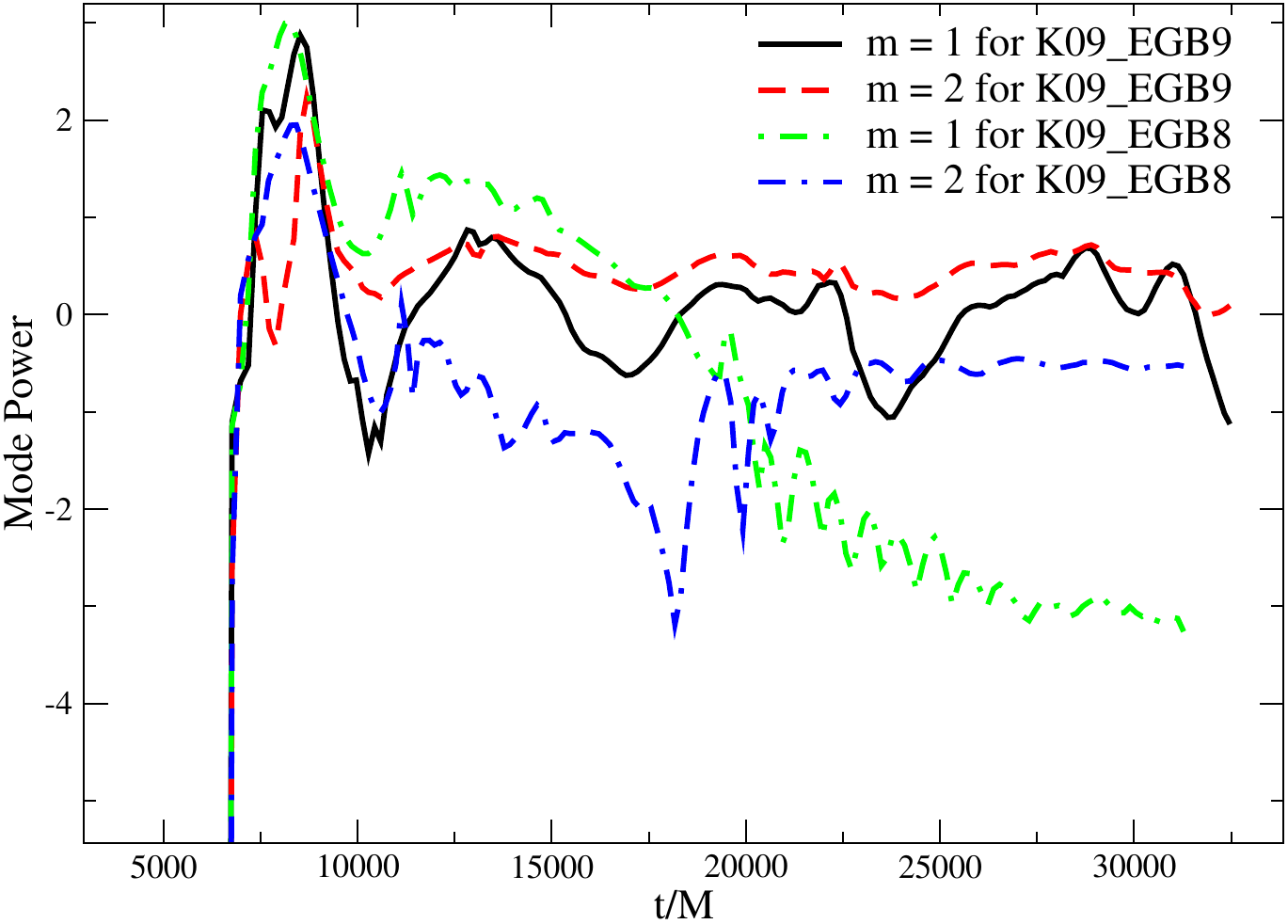,width=7cm,height=6.5cm}  
  \caption{
The models $K09\_EGB8$ and $K09\_EGB9$ are compared by calculating the mass-accretion rate and the power of the disk oscillation modes. It appears that the behavior of the perturbation with angular velocity is more complex in relation to the physical parameters, and it seems that the accretion disk exhibits stronger non-linear oscillations compared to the non-rotating perturbation case.
}
\label{Effect_Angular_Momentum1}
\end{figure*}

The left panel of Fig. \ref{Effect_Angular_Momentum1} illustrates how the change in mass-accretion rate over time significantly influences the formation and stabilization of the disk when considering the perturbation angular velocity. After perturbation of the stable disk at $t=6000M$, it is observed that matter around the black hole quickly either falls into the black hole or is ejected from the computational domain. This process continues until $t=8000M$. Subsequently, the two models exhibit distinct behaviors. The perturbation with zero angular velocity ($K09\_EGB9$) quickly reaches a steady state at $t=8500M$, whereas the other one ($K09\_EGB8$) oscillates until $t=19500M$. This illustrates the significant impact of the physical properties of the perturbation on the formation, accretion rate, and physical parameters of the disk around the black hole.

As observed in the right panel of Fig. \ref{Effect_Angular_Momentum1}, both systems quickly become unstable after perturbation. This instability escalates rapidly in the early stages of perturbation, reaching saturation when the mode value is approximately 3. Subsequently, in the non-rotating perturbation case ($K09\_EGB9$), the $m=1$ and $m=2$ modes oscillate around a certain value with a high amplitude, indicating a steady state. However, the scenario is markedly different in the case of perturbation with angular velocity ($K09\_EGB8$). Despite both modes of $K09\_EGB8$ displaying unstable behavior after the saturation point, the $m=2$ mode begins to exhibit stable oscillations at approximately $t=20000M$, maintaining this condition throughout the simulation. Conversely, it is observed that $m=1$ reaches an approximate steady state towards the end of the calculation time, oscillating around a specific value. This suggests that the presence of a single dominant mode of $m=1$ is insufficient to maintain the stable structure of the disk. In contrast, in the case of perturbation with angular velocity, the existence of two dominant modes, as in $m=2$, appears to ensure the stability of the disk, enabling it to reach a steady state.

\begin{figure*}
  \vspace{1cm}
  \center
  \psfig{file=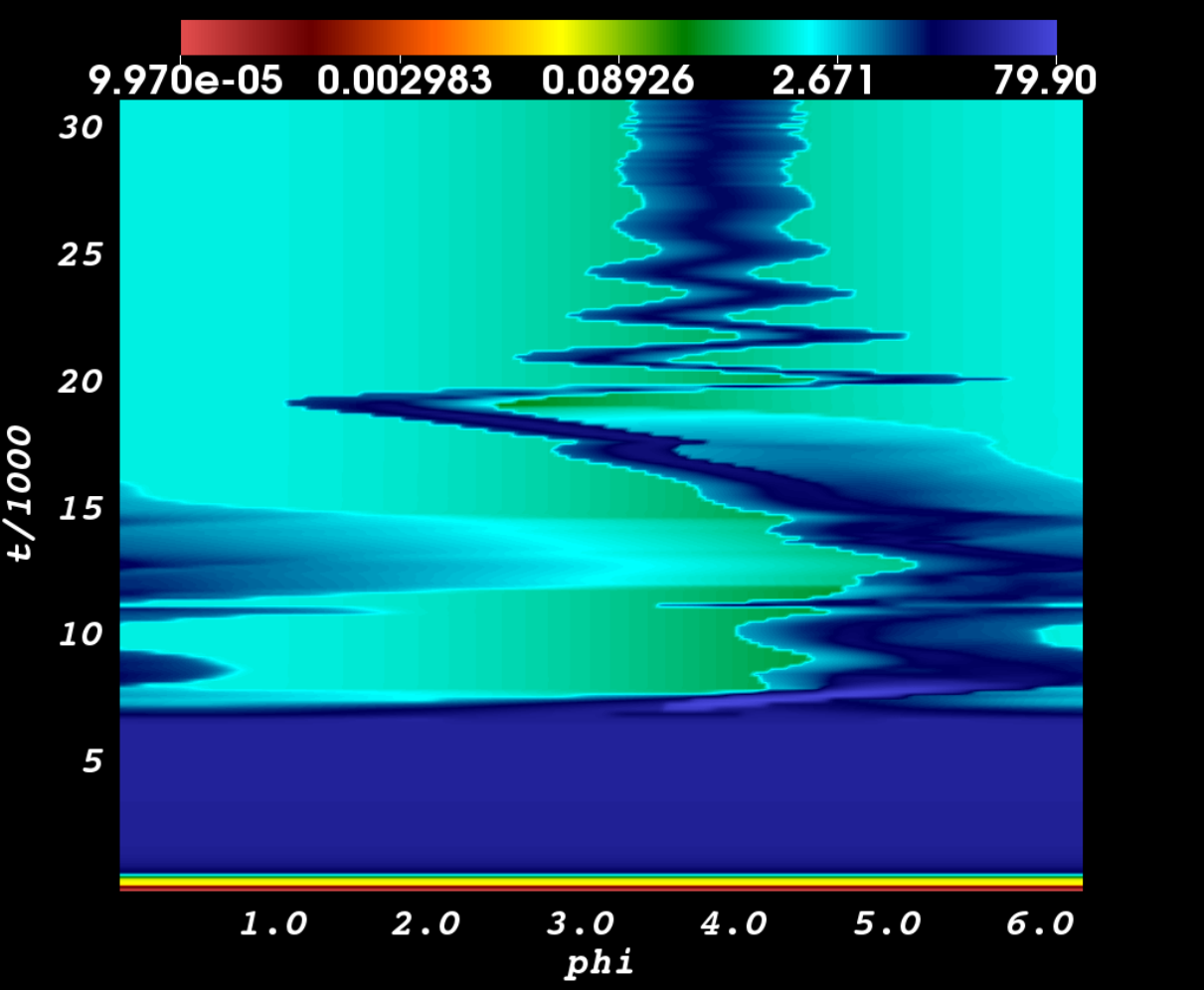,width=7cm,height=6.5cm}
  \psfig{file=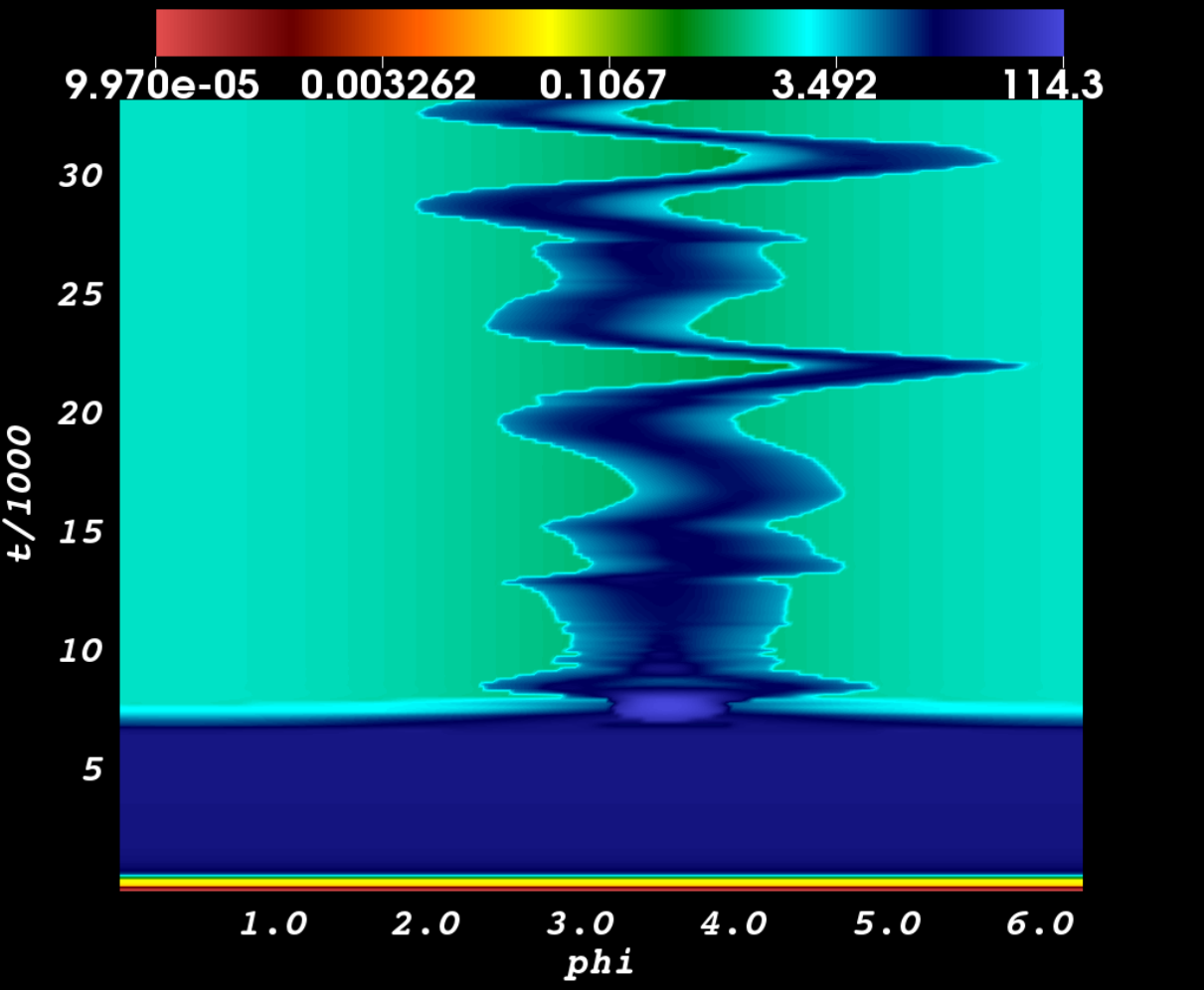,width=7cm,height=6.5cm}
  \caption{
The time-dependent variation of density as a function of $\phi$ ($\phi$). The left panel is for model $K09\_EGB8$, and the right panel is for $K09\_EGB9$ at a constant $r=3.88M$. The effect of the angular velocity of the perturbation on the disk structure and the resulting shock wave oscillation is clearly observed.
}
\label{Effect_Angular_Momentum2}
\end{figure*}

Figure \ref{Effect_Angular_Momentum2} illustrates how the density of the disk oscillates in the azimuthal direction ($\phi$) over time, from the initial formation of the disk to the end of the simulation, at $r=3.88M$ for the $K09\_EGB8$ (left panel) and $K09\_EGB9$ (right panel) models. It is evident that the disk forms through spherical accretion up to $t=6000M$. After applying two different perturbations at $t=6000M$, the evolving behavior of the stable disk over time becomes apparent. The left panel displays the oscillation of the disk resulting from the perturbation with angular velocity, while the right panel shows the disk oscillation resulting from the perturbation with zero angular velocity. It is clearly visible that the angular velocity of the matter sent as perturbation significantly affects the disk oscillation. This, in turn, leads to changes in various important physical parameters, such as oscillation modes, the physical characteristics of X-rays emitted in the region near the black hole, and the amount of matter falling into the black hole.

\begin{figure*}
  \vspace{1cm}
  \center 
  \psfig{file=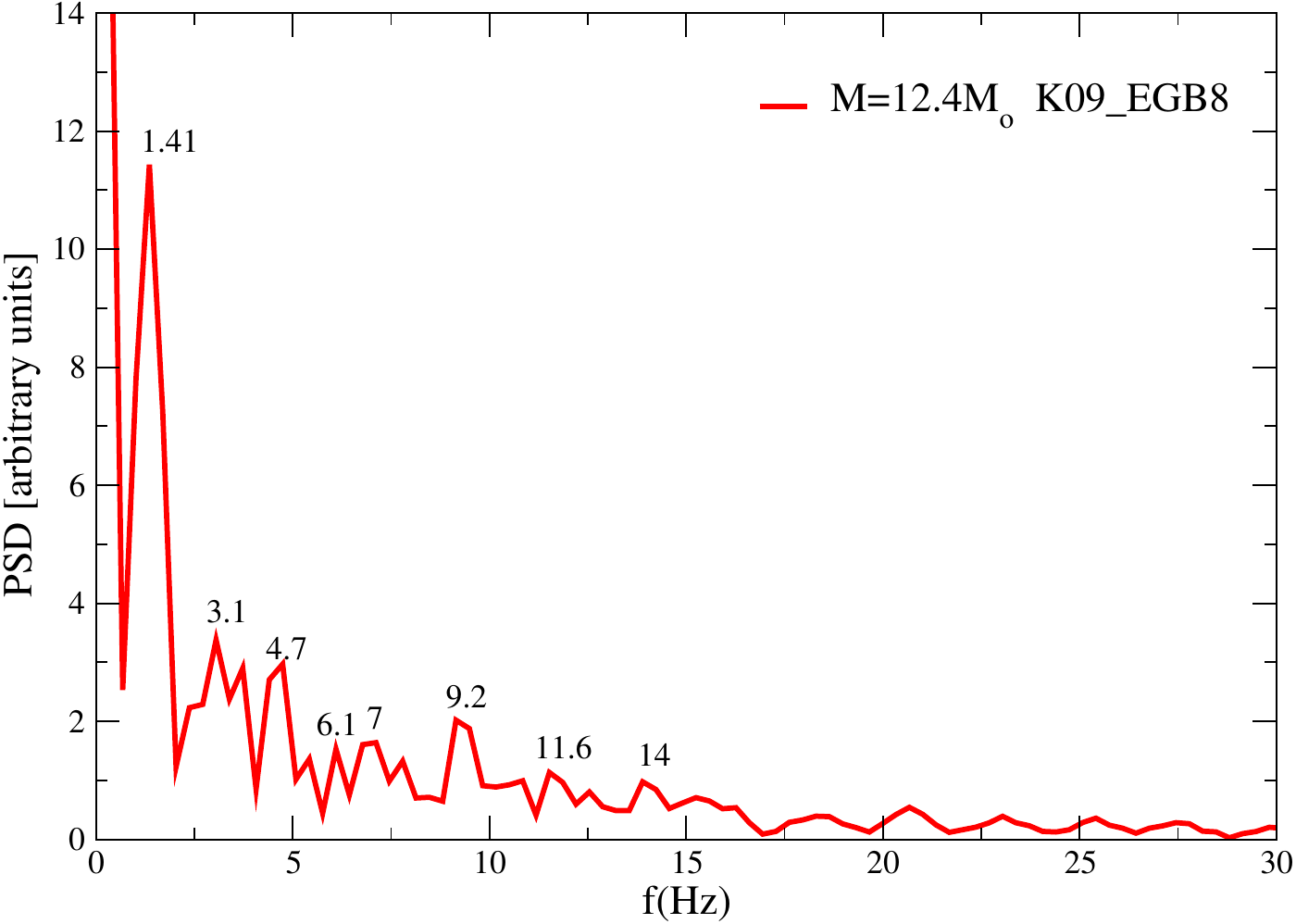,width=7cm,height=6.5cm}
 \hspace{0.3cm}  
  \psfig{file=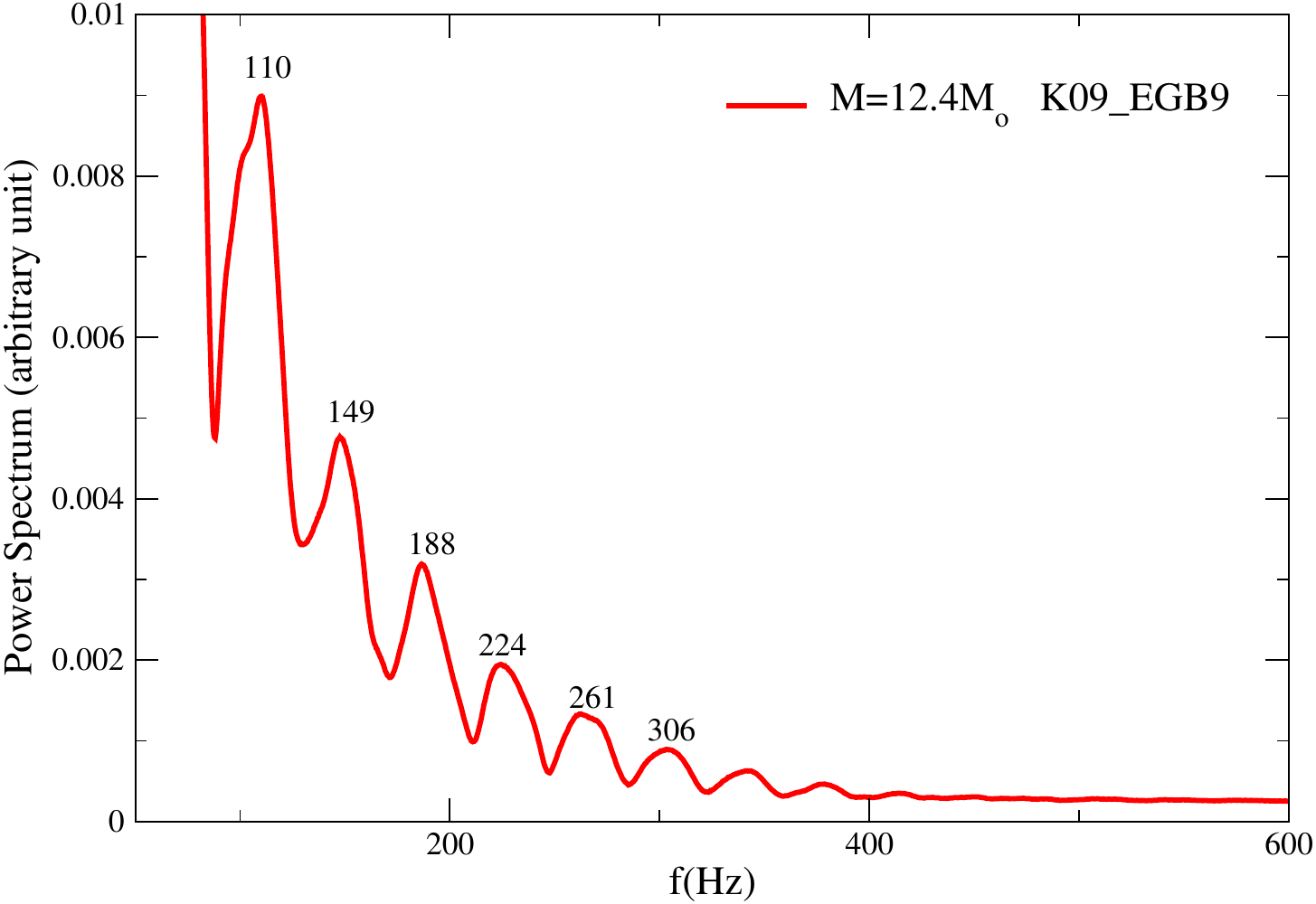,width=7cm,height=6.5cm}
  \caption{
    The same models as seen in Fig.\ref{Effect_Angular_Momentum2} are considered, but this time, the oscillation frequencies on the disk are revealed using the mass-accretion rate at $r=3.88M$. The frequency axis is obtained using the mass of the black hole with $M=12.4M_{\odot}$.
  }
\label{Effect_Angular_Momentum3}
\end{figure*}

Furthermore, Fig. \ref{Effect_Angular_Momentum3} demonstrates that different initial perturbation angular velocities lead to significant differences in observable oscillation frequencies. As expected, the frequencies obtained from the $K09\_EGB8$ model, which exhibits more chaotic behavior, reveal a more chaotic structure, while the oscillation frequencies of the disk resulting from the non-rotating perturbation matter exhibit a more linear behavior. Additionally, the amplitudes of the frequencies obtained from the $K09\_EGB8$ model are nearly $10^3$ times stronger than the amplitudes of the genuine mode frequencies obtained from the other model. This facilitates the observation of these emissions by detectors. Many telescopes can easily observe high-amplitude oscillations because they can distinguish them from background noise. Additionally, these types of oscillations carry more information about the black hole-disk system.

The oscillation frequencies obtained here have been compared with observational results in Sections \ref{Astro_Mot} and \ref{QPO}. To facilitate this comparison, we used $M=12.4M_{\odot}$, which approximates the observed mass of the GRS 1915+105 black hole. The comparison revealed that the results obtained from the perturbation with angular momentum align well with the observed results for the GRS 1915+105 source. Specifically, the genuine modes observed through numerical analysis, along with their numerous nonlinear interactions, correspond closely to the observed spectrum of C-type QPO frequencies. This indicates that the frequency observed from this source is attributable to a perturbation with angular velocity of the disk. Therefore, the disk structure, the behavior of oscillation modes, and the physical characteristics of the disk obtained for the $K09\_EGB8$ model can help explain the GRS 1915+105 source.

The first two frequencies, $1.41 \text{ Hz}$ and $3.1 \text{ Hz}$, obtained from the Fourier spectrum of model $K09\_EGB8$ shown on the left side of Fig. \ref{Effect_Angular_Momentum3}, are genuine eigenmodes. The others are modes resulting from their nonlinear couplings. For example, the frequency $4.7 \text{ Hz}$ is almost equal to the sum of the first two genuine eigenmodes. These genuine eigenmodes and their nonlinear coupling often produce ratios, such as $1:2:3$, and they fall within the observed frequency range of GRS 1915+105 and C-type QPOs.

\subsection{Exploring the Differences: Disk Dynamics under Kerr and EGB Gravities}
\label{Comparision}

It is evident that the different curvatures around the rotating black hole affect the dynamic structure of the forming disk and the physical properties of the emitted X-rays. In this section, the influence of Kerr and EGB gravities on the disk structure has been elucidated.

The impact of the black hole spin parameter and the EGB coupling constants on the mass accretion rate of the disk is shown in Fig. \ref{mass_acc}. As observed in the figure, the amount of matter falling from the disk towards the black hole is greater for non-zero values of $\alpha$ in EGB gravity compared to Kerr gravity. This implies that for different values of $\alpha$, the disk feeds the black hole more, leading to the black holes having larger masses than those in Kerr gravity. When Fig. \ref{mass_acc} is examined in detail, it can be seen that this difference is very small for values of $\alpha$ close to zero, which is an expected result. On the other hand, it is evident that a higher mass accretion rate leads to a lower maximum density in the disk, as seen in Fig. \ref{max_den}.

Figure \ref{QPOs1} displays the oscillation frequencies of QPOs around Kerr black holes and in EGB gravity with different coupling constants. It is evident that, although the same frequencies are observed in both EGB and Kerr gravity scenarios, the amplitudes of the frequencies are greater in the EGB gravity cases. This enhances the detectability of these frequencies. Large-amplitude QPO oscillations are more likely to be detected by detectors due to their amplitude exceeding the background noise modes. At the same time, these oscillations help establish that their origin is not the result of numerical artifacts. On the other hand, for the most negative value of $\alpha$ (see the internal graphics in Fig. \ref{QPOs1}), while the first genuine mode has the same frequency, subsequent modes differ from Kerr gravity.

The occurrence of shock waves in the dynamic structure of the accretion disk determines the severity of disk instability. Such systems may exhibit consistent behavior at the same frequency, making them valuable sources for observations. To quantify the severity of disk instability, we compute the ratio of the total time step at $t=30000M$ for each model to the total time step in the Kerr solution. This allows us to determine which systems require more time for calculations. According to the Courant condition, the time step varies depending on the instability occurring in the disk. In cases of severe instability, the time step $dt$ decreases, leading to slower progress in time and differences in time steps within the same total time.

As seen in Table \ref{Inital_Con}, changes in the ratio of time steps can be observed depending on the chaos exhibited by the disk. The reasons for these changes include the fact that instability occurs only near the black hole throughout the entire simulation (which causes a decrease in the total time step) or that the modeling of this instability affects the entire disk and is also influenced by the alternative gravity.

\begin{figure*}
  \vspace{1.5cm}
  \center 
  \psfig{file=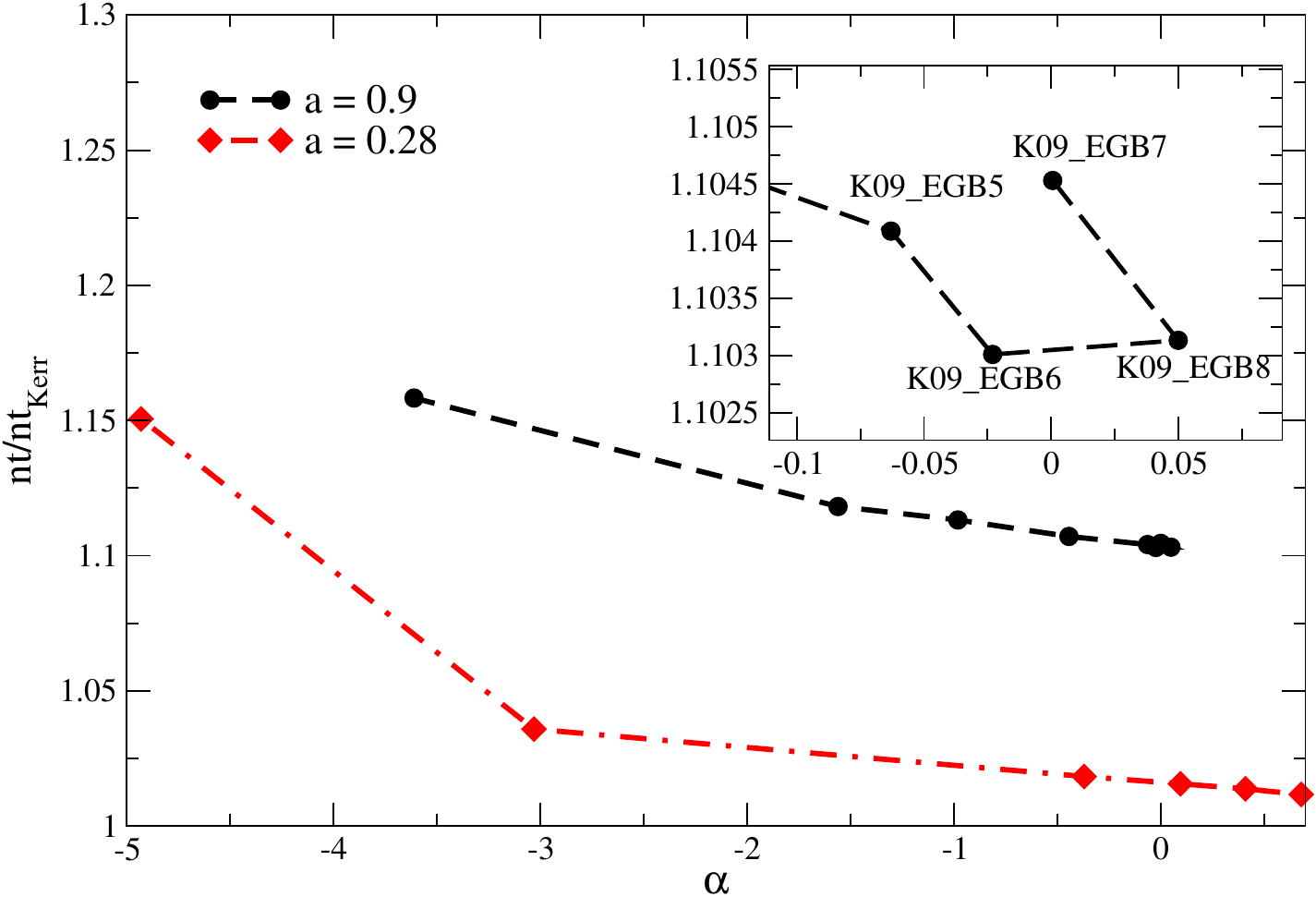,width=10cm,height=9cm}
  \caption{
    The ratio of the required time steps for each model to reach $t=30000M$
    to the required time steps for a Kerr black hole at that time.
    The necessary time steps for each model have revealed the chaotic behavior of
    the accretion disk.}
\label{time_step}
\end{figure*}

All models shown in Fig. \ref{time_step} indicate that an increase in the negative direction of the EGB parameter $\alpha$ leads to the disk exhibiting more chaotic behavior. As observed in the same figure, the black hole rotation parameter $a$ affects the accretion of matter close to the black hole horizon, and therefore, $a$ influences the chaotic structure of the disk.

Furthermore, as observed in Fig. \ref{time_step}, it is evident that $\alpha$ is not significantly effective in determining the chaotic behavior of a rapidly rotating black hole, particularly for values close to zero (both negative and positive). Therefore, for $a/M=0.9$, unexpected behavior has been observed for these $\alpha$ values.

These chaotic structures can be used to explain intense continuous emissions from black hole-disk systems and provide a wide frequency range in the emitted radiation \citep{Misra2004, Guo2020}. The amplitude of the radiation changes over time, and these systems may also be sources of sudden bursts \citep{Cao2023MNRAS}. As seen in Section \ref{QPO}, fundamental modes of oscillation create new frequencies by nonlinear coupling, thus suggesting them as a physical mechanism for the observed sources given in Section \ref{Astro_Mot}.

\section{Discussion and Conclusion}
\label{Conclusion}
In this study, we investigated the time variation of accretion disks around rotating black holes formed by Kerr and EGB gravities, considering the black hole spin parameter ($a/M$) and the EGB coupling constant ($\alpha$). We examined the effects of both positive and negative values of $\alpha$ for both slowly rotating ($a/M=0.28$) and rapidly rotating ($a/M=0.9$) black hole scenarios. This involved modeling the shock waves generated around the black hole, analyzing variations in shock intensities, changes in QPO frequencies, and the chaotic structures they exhibited, all dependent on these parameters.

A stable disk undergoes perturbation through the introduction of matter from outside the computational domain, akin to an X-ray binary system. When the perturbation is directed towards the black hole with specific angular and radial velocities, it initiates the formation of a one-armed shock wave around the black hole, succeeded by the emergence of a two-armed shock wave. Concurrently, the disk begins to exhibit nonlinear oscillations. These shock waves persist beyond their initial formation, and the disk demonstrates QPOs before reaching the simulation's maximum time.

Conversely, in scenarios where the perturbation has zero angular velocity, one- and two-armed shock waves still form, but the oscillations display different characteristic behaviors. These distinct disk structures and oscillation conditions notably influence the QPO frequencies of the disk. In cases with zero angular velocity, oscillation frequencies are observed in the hundreds to a few hundred Hz range. However, when angular velocity is present at a specific value, these frequencies are found in the several Hz range.

These frequencies arise from two genuine modes and their nonlinear couplings, often manifesting in ratios like 1:2:3. Some observed sources exhibit these frequency ratios \citep{LiNar2004}. Comparing the obtained QPO frequencies with observations suggests that the observed low-frequency QPOs from the GRS 1915+105 black hole could be explained by the disk structure and shock waves generated by perturbations with angular velocity. The frequencies of QPOs triggered by one- or two-armed spiral shock waves are consistent with the observed low-frequency QPOs for this source \citep{Ingram2019}.

Numerical calculations have strongly demonstrated the close relationship between the mass accretion rate generated by perturbations with angular velocity and the black hole spin parameter ($a/M$) and EGB coupling constant ($\alpha$). It has been observed that the mass accretion rate decreases, leading to an increase in the amount of matter falling into the black hole, due to positive and negative variations in the $\alpha$ constant. Particularly significant increases in this mass accretion rate have been clearly observed for large negative values of $\alpha$. Consequently, the disk density around the black hole and the intensity of the shock wave forming on it increase for both positive and negative $\alpha$ values. As a result, the QPO frequencies around the black holes exhibit a wider range, with larger amplitudes for increasing $\alpha$. This, in turn, increases the likelihood of these frequencies being observed by telescopes.

The calculations conducted here have demonstrated that, as expected, instabilities grow shortly after the perturbation and reach saturation. When describing these instabilities, both $m=1$ and $m=2$ modes are examined. Both modes exhibit instability for a certain period after reaching saturation \citep{ShapiroBook1985, LiNar2004}. Specifically, the $m=1$ mode, which characterizes a one-armed shock wave, showed instability for an extended period but eventually stabilized towards the end of the simulation. On the other hand, the $m=2$ mode, characterizing a two-armed shock wave, exhibited QPOs shortly after saturation and displayed stable behavior.

These modes serve as evidence for the formation of shock waves in the numerically calculated disk, which are the physical mechanisms behind the observed QPOs. Utilizing the disk structure presented here, it becomes possible to explain the causes of QPOs in strong gravitational fields around different types of black holes, such as stellar and massive ones, that have not been discussed in this article.

In summary, the numerical results provided here can offer insights into understanding the QPOs observed in X-ray binaries, galaxies, and AGNs, where the physical origins are yet to be fully understood or remain subject to ongoing debates \citep{Ingram2019, Song2020, Heena2023}. These QPOs stem from the interaction between black holes and disks at the centers of these systems and can be elucidated using the findings presented in this study.

Finally, in future studies, the physical characteristics of perturbations provided in the $K09\_EGB8$ and $K09\_EGB9$ models will be explored over broader parameter ranges, particularly around moderately rotating black holes. The impact of angular velocity on disk dynamics and QPOs will be presented in greater detail in forthcoming research.

\normalem
\section*{Acknowledgments}
All simulations were performed using the Phoenix  High
Performance Computing facility at the American University of the Middle East
(AUM), Kuwait. The authors express gratitude to the anonymous reviewer for their
valuable comments and suggestions, which have significantly enhanced the clarity of the manuscript.\\

\bibliography{paper.bib}

\end{document}